\documentclass[journal]{IEEEtran}
\ifCLASSINFOpdf
\else
\fi
\usepackage{comment}
\usepackage{cite}
\usepackage{indentfirst}
\usepackage{graphicx}
\usepackage{changepage}
\usepackage{stfloats}
\usepackage{amsfonts,amssymb}
\usepackage{bm}
\usepackage{diagbox}
\usepackage{algorithm}
\usepackage{algorithmic}
\usepackage{amsmath}
\usepackage{amsmath,amssymb,amsfonts}
\usepackage{times}
\usepackage{url}
\usepackage{textcomp}
\usepackage{xcolor}
\usepackage{color}
\usepackage{setspace}
\usepackage{enumitem}
\usepackage{float}
\usepackage{diagbox}
\usepackage[T1]{fontenc}
\usepackage[utf8]{inputenc}
\usepackage{authblk}
\usepackage{threeparttable}
\usepackage{booktabs}
\usepackage{footnote}
\usepackage{graphicx}
\usepackage{pifont}
\usepackage{subfigure}
\usepackage{mathrsfs}
\usepackage{makecell}
\usepackage{array}
\usepackage{booktabs}
\usepackage{lipsum}
\usepackage{multirow}
\usepackage{adjustbox}
\usepackage{bbding}
\usepackage{tabularx}
\usepackage{colortbl}
\newcolumntype{Y}{>{\centering\arraybackslash}X}
\begin{document}
\abovedisplayshortskip=2pt
\belowdisplayshortskip=2pt
\abovedisplayskip=2pt
\belowdisplayskip=2pt
\textfloatsep=4pt
\floatsep=4pt
\intextsep=4pt
%
\title{Next-Generation MIMO Transceivers for Integrated Sensing and Communications: Unique Security Vulnerabilities and Solutions}
%
%
%

\author{Kawon~Han,~\IEEEmembership{Member,~IEEE,}
        Christos~Masouros,~\IEEEmembership{Fellow,~IEEE,}
        Taneli~Riihonen,~\IEEEmembership{Senior Member,~IEEE,}
        and~Moeness~G.~Amin,~\IEEEmembership{Life Fellow,~IEEE}

\thanks{Kawon Han and Christos Masouros are with the Department of Electronic and Electrical Engineering, University College London, London, UK (E-mail: {kawon.han, c.masouros}@ucl.ac.uk).}
\thanks{Taneli Riihonen is with the Faculty of Information Technology and Communication Sciences, Tampere University, 33720 Tampere, Finland}
\thanks{Moeness G. Amin is with the Center for Advanced Communications, Villanova University, Villanova, PA 19085 USA}}
\maketitle

\begin{abstract}
Integrated sensing and communications (ISAC), which is recognized as a key enabler for sixth generation (6G), has brought new opportunities for intelligent, sustainable, and connected wireless networks. Multiple-input multiple-output (MIMO) transceiver technology lies at the core of this paradigm, providing the degrees of freedom required for simultaneous data transmission and accurate radar sensing. The tight integration of sensing and communication introduces unique security vulnerabilities that extend beyond conventional physical-layer security (PLS). In particular, high-power transmissions directed at sensing targets may empower adversarial eavesdroppers, whereas passive interception of ISAC echoes can reveal sensitive information such as target locations and mobility patterns. This article presents an overview of recent advances in MIMO ISAC transceiver design, considering transmitter perspectives, receiver architectures, and full-duplex implementations. We examine MIMO transceiver designs under unique security threats specific to ISAC and highlight emerging countermeasures, including secure signaling design, interference exploitation, and transceiver optimization under adversarial conditions. Finally, we discuss challenges and research opportunities for developing secure ISAC systems in next-generation wireless networks. 
\end{abstract}

%
\IEEEpeerreviewmaketitle

\section{Introduction}
Radar and communication are two fundamental applications of radio frequency systems that have profoundly shaped modern society. Although both rely on electromagnetic (EM) waves, they have traditionally been developed in isolation, following independent design principles and operating on separate hardware platforms. As a result, the two technologies often compete for scarce spectrum resources rather than cooperating. Elements toward integrating radar and communication can be traced in the literature since as early as the 1960s \cite{mealey2007method}, but for decades the concept remained largely unexplored due to technological and practical barriers, as well as the absence of driving commercial or defense applications. This landscape is now changing. Advances in millimeter-wave (mmWave) systems and the widespread adoption of multiple-input multiple-output (MIMO) architectures have revealed strong commonalities between radar and communication \cite{larsson2014massive, li2007mimo}, including shared transceiver hardware, common antenna and array architectures, and overlapping channel characteristics. These developments are transforming integration and dual-functionality from a long-standing vision into a practical opportunity, marking a paradigm shift from co-existence to co-design, now unified under the concept of Integrated Sensing and Communication (ISAC).

The emergence of sixth-generation (6G) networks further amplifies this need, demanding technologies that provide both ubiquitous connectivity and high-resolution situational awareness. Embedding sensing functionality into communication signals or reusing radar waveforms for data transmission, ISAC enhances spectral and energy efficiency, reduces hardware redundancy, and supports a wide range of emerging applications \cite{liu2022integrated}. This convergence is also driven by spectrum scarcity and escalating demands on throughput, reliability, and latency, which make separate spectrum allocation increasingly impractical. Finally, the recent chip-crisis has created a drive for efficient hardware reuse and the development of multi-functional radio frequency platforms that provide sensing and communication capabilities without the need for hardware duplication across separate sensing and communication systems. Advances in waveform design, transceiver architectures, and especially MIMO techniques now enable ISAC systems to achieve high data rates, accurate sensing, and robust interference management within shared spectral and hardware resources through joint optimization.

At the same time, this new opportunity brings new challenges, as integration introduces security and privacy risks that are far less pronounced in conventional wireless systems. Because ISAC operates over shared spectrum and often uses common waveforms, any compromise of the transmitted signal can simultaneously jeopardize both data exchange and radar sensing. Adversaries may exploit ISAC illumination to intercept or manipulate confidential information, while unauthorized receivers can passively reconstruct sensitive environmental details, such as user/target movements, locations, or object dynamics, without the need to access the communication payload. In short, the fusion of sensing and communication multiplies the potential benefits but also expands the attack surface, making security a critical concern for practical ISAC deployment.

These challenges in ISAC call for a holistic security perspective that goes beyond classical physical layer security (PLS) in wireless communications. Future ISAC systems must embed security properties directly into waveform and transceiver design, leverage physical-layer characteristics to impair adversarial observations, and adopt cross-layer protocols that jointly safeguard communication integrity and sensing privacy. Without such measures, large-scale ISAC deployment may be jeopardized not by technical feasibility, but by the inability to guarantee trust, resilience, and privacy in real-world environments. 

This article presents a comprehensive overview of the latest MIMO ISAC transceiver design techniques, including radar- and communication-centric approaches, joint signaling strategies, and interference-exploitation techniques, which bring advanced sensing and communication (S\&C) trade-off performance. In addition, we highlight the emerging concept of secure ISAC transceivers, which are tailored to simultaneously safeguard communication data eavesdropping and sensing target information. By bridging transceiver design and security, our objective is to examine the state-of-the-art approaches and provide a forward-looking perspective on how ISAC can be realized in practice while remaining resilient to new classes of threats; ultimately, a practical and safe-for-use technology. 

The overall organization of this article is as follows: Section \ref{Sec::TX_A} discusses transmitter-side ISAC designs focusing on radar- and communication-centric approaches, and Section \ref{Sec::TX_B} focuses on the recent advances on joint MIMO precoding and Section \ref{Sec::SLP} investigates interference exploitation for ISAC. Section \ref{Sec::RX} turns to receiver-side processing, highlighting both a unique ISAC receiver architecture and joint receiver designs, while Section \ref{Sec::FD-ISAC} examines full-duplex ISAC transceivers with an emphasis on self-interference cancellation. Section \ref{Sec::PLS} introduces the emerging dimension of ISAC physical-layer security, focusing on transceiver designs tailored to protect both data, and Section \ref{Sec::SensingSecure} for sensing security. Section \ref{Sec::PoC} highlights the ISAC proof-of-concept demonstration. Finally, Section \ref{Sec::Conclusion} concludes the article with a summary of key insights and future outlooks.

\begin{figure}[t!]
    \centering
    {\includegraphics[width=0.49\textwidth]{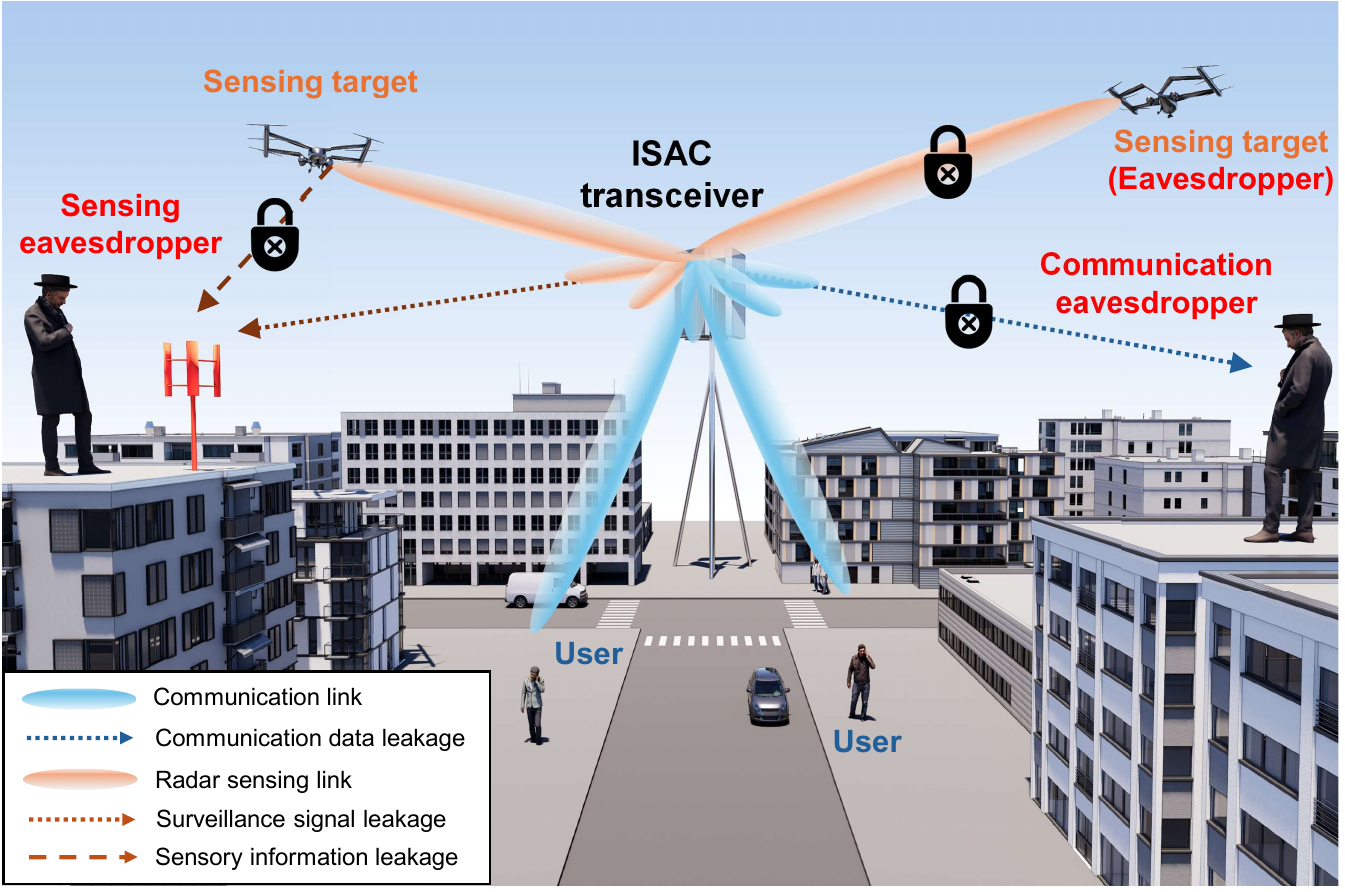}}
    \caption{Illustration of MIMO-enabled ISAC operation and new physical layer security vulnerabilities in ISAC systems. }
    \label{Fig::1}
\end{figure}

\section{The ISAC Transmitter: Modulation and Constellation Design} \label{Sec::TX_A}
This section provides an overview of ISAC transmitter design methodologies, focusing on the dual-functionality perspectives, namely radar-centric and communication-centric approaches. 

\subsection{Radar-Centric Design: Direct Data Modulation on Radar Pulses} 
A straightforward realization of radar-centric dual-functional radar-communication (DFRC) systems is to convey communication data by directly modulating radar pulses. In this approach, traditional radar probing waveforms, such as linear frequency modulation (LFM), frequency-modulated continuous wave (FMCW), or phase-modulated continuous wave (PMCW), are preserved, while data symbols are embedded through slow-time or fast-time coding. These systems maintain radar compatibility with strong target detection and parameter estimation capability but typically offer only moderate communication throughput~\cite{de2021joint}. Representative examples include intentional pulse modulation schemes, where the radar pulse serves as the carrier and the communication message or symbol sequence acts as the modulating signal~\cite{nowak2016co, sahin2017novel, hassanien2017dual}.

Slow-time coding (or phase modulation), also known as complex scaling, often used for waveform diversity in MIMO radar, conveys data bits across pulses without compromising sensing performance. While this approach is highly radar-compatible, its communication rate is fundamentally limited by the pulse repetition interval~\cite{zhang2021overview}. In contrast, fast-time coding increases the communication data rate by modulating symbols within a pulse, but it alters the radar waveform structure, potentially causing spectral spreading and out-of-band leakage~\cite{uysal2019phase}. These schemes allow direct symbol recovery at the communication receiver without requiring inverse dictionaries, yet remain suboptimal for both sensing and communication due to the lack of dual-function co-optimization. Therefore, direct data modulation on radar waveforms can be regarded as a baseline DFRC approach, providing compatibility but limited joint performance.

\subsection{Radar-Centric Design: Conveying Data Bits Over Legacy Radar Systems via Index Modulation} \label{Sec::Radar-centric}
By the virtue of system co-design, maximizing the performance of one function should meet satisfactory performance constraints for the other. Just like communication waveforms are modulated in amplitude and phase to convey data bits, radar waveforms can, in principle, be modulated as well. To date, a particular modulation format, index modulation (IM), is considered an effective approach to expand the degree of freedom (DoF) available to both functions, thereby easing co-design tradeoffs and enhancing overall system performance.

IM has been examined in DFRC systems in which digital communications are achieved using legacy radar platforms. In these DFRC systems, also referred to as a radar-centric approach, the radar is the primary function \cite{blunt2010embedding,hassanien2016signaling,hassanien2019dual,zheng2019radar,liu2020joint0,martone2021view}. The communication function treats the radar as a system of opportunity. This concept implies that system resources or features of one function, including the signal waveforms, can be utilized by the other function. The type of resources employed as well as the extent of their utilization define the underlying DFRC system. From the authors’ point of view, and for the purpose of this paper organization, we consider legacy radar-centric DFRC systems are those radar systems that hold on to their legacy waveforms, bandwidths, and beams without significant alterations that stem from accommodating the communications function. As such, involving orthogonal frequency division multiplexing (OFDM) as the transmit waveforms is not typically considered a radar-centric approach. 

In IM, the communication symbols, drawn from a given signal constellation, are not necessarily transmitted to the receivers as in-phase and quadrature components. Rather, each symbol can be additionally or solely represented by different radar parameter set values.  In essence, to communicate different communication symbols, referred to as IM symbols, radar parameters, independently or in combinations, would assume different values, allowing for different transmit waveform characteristics and beamforming. The communication receiver, being aware of the indexing, which is the dictionary mapping between the radar parameter values and the corresponding symbols, seeks to optimally decipher the transmitted signal and retrieves the information.  Radar parameters, proposed to implement IM for radar-centric DFRC systems, include signal processing level parameters, like the array weights and the pulse waveform shapes, and system-level parameters, like central frequencies, signal bandwidth, and the array aperture and configuration. It is important to note that if changing the radar parameters leads to the transmission of the exact amplitude- and phase-based communication symbols, then it is no longer an indexing and lies outside the realm of IM.

\subsubsection{IM involving Radar Beam Sidelobes}
\begin{figure}[t!]
    \centering
    {\includegraphics[width=0.375\textwidth]{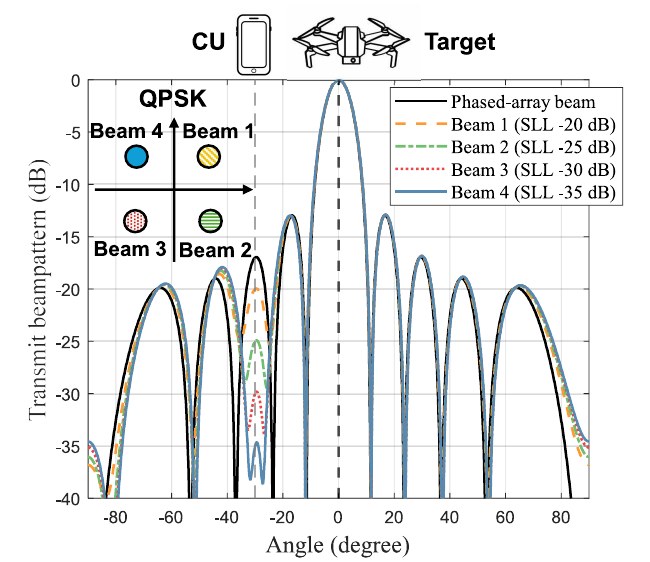}}
    \caption{QPSK signal is indexed by four sidelobe levels in the radar beam.}
    \label{Fig::IM1}
\end{figure}
The real and complex sidelobes levels, acting alone or in conjunction with multiple radar waveforms, can be used to represent the communications symbols without significant alterations to the main radar beam \cite{hassanien2016signaling, euziere2014dual, hassanien2016phase, hassanien2015dual, ahmed2018dual}. Special cases of sidelobe variations are amplitude-shift keying (ASK) \cite{hassanien2016signaling} and phase-shift keying (PSK) \cite{hassanien2016phase}. In this type of IM, the array weights change with communication symbol, resulting in corresponding changes in sidelobe levels towards the intended communication receiver. In this regard, the indexing of the array weights morphs into indexing of the sidelobe levels via Fourier transform and beamforming. A simple case is demonstrated in Fig. \ref{Fig::IM1}, where a quadrature phase-shift keying (QPSK) signal is indexed by four sidelobe levels. We maintain that if the sidelobe level values are the same as the communication constellation values, then this type of transmission is considered directional modulation, in lieu of IM.

\subsubsection{IM Involving Radar Waveforms}
The radar waveforms themselves can be considered an index with which to create a communication constellation, if allowed to change from one pulse repetition period to another \cite{blunt2010embedding, tedesso2018code}. In this case, the size of the signal constellation is dictated by the radar waveform diversity. Up and down chirps, discussed in \cite{chalise2023information}, represent a simple waveform diversity for binary-phase-shift keying (BPSK) constellation. It is worth noting that, in selecting the transmit radar waveforms, the constant-modulus property should be maintained to enable the transmit power amplifiers to operate in saturation, as typical for radar transmission. In addition, it is known that waveform variations over slow-time cause undesirable range sidelobe modulation, hindering target detection and resolution. This problem can be avoided or mitigated by applying mismatched filters \cite{sahin2017filter}.

\subsubsection{IM Involving Radar Antennas and Array Configuration}
\begin{figure}[t!]
    \centering
    {\includegraphics[width=0.40\textwidth]{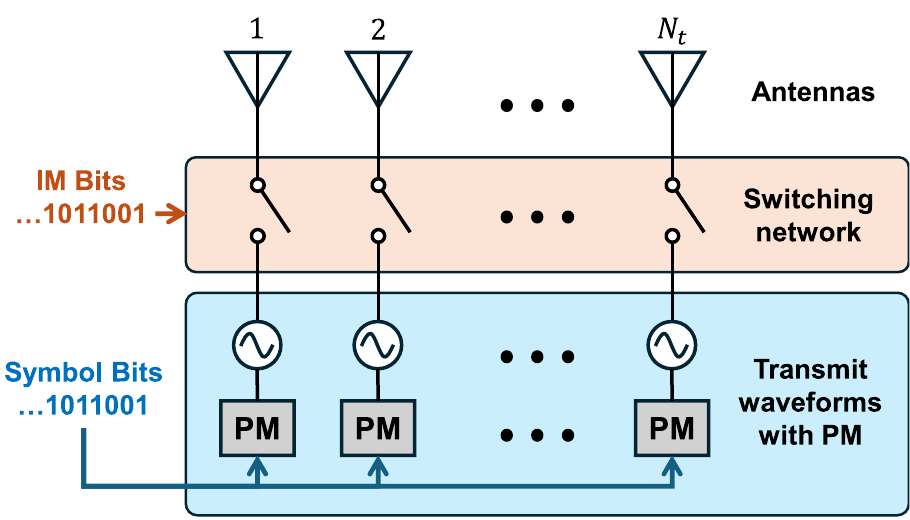}}
    \caption{Antenna-selection IM combined with signal phase modulation in MIMO radar \cite{elbir2024index}.}
    \label{Fig::IM2}
\end{figure}
For MIMO radar platforms, different antennas emit orthogonal waveforms, thereby providing more indexing opportunities and higher data rates compared to phased arrays. Since multiplication of each waveform by a complex value does not change the waveform orthogonality, IM in MIMO radar can be combined with concurrent transmission of phase modulated signal. Antenna selections over the radar aperture pattern generate different sparse array configurations, in which case the selection matrix serves as an index for communication symbols, with the radar waveforms kept intact \cite{wang2018dual, ma2021spatial}. This embedding strategy is shown in Fig. \ref{Fig::IM2}. Another strategy for indexing in MIMO radar is to change the pairing between the antennas and the associated radiated waveforms \cite{hassanien2018dual}. In this case, a permutation matrix, in lieu of the selection matrix, is applied to shuffle the waveforms assigned to the different antennas. Fig. \ref{Fig::IM2} shows the combined IM, through waveform shuffling, and signal phase modulation. 

Code-shift keying (CSK) falls under the category of waveform diversity and it is a type of IM, where each symbol is indexed by a pulse code sequence. Each sequence, which can be in a direct sequence \cite{tedesso2018code} or FH form \cite{hassanien2017dual, xu2022hybrid}, is transmitted over one pulse repetition interval. CSK indexing in FH radars can be combined with antenna indexing \cite{xu2022hybrid, eedara2022dual}. Since the multiplication of each frequency hopping pulse by a complex value does not change the hopping frequency orthogonality, CSK indexing in FH radars can be combined with concurrent transmission of phase-modulated signal. A generalized approach is proposed in [24], where each symbol is represented by a code that modulates the FH waveform and each hop is multiplied by one pulse of the code, achieving a high data rate DFRC system. It is shown that the different codes can be chosen to reduce range sidelobe modulations and to maintain approximately the same sidelobe levels when using non-orthogonal codes. The work in \cite{elbir2024index} includes a table that compares IM-ISAC techniques in terms of the signal processing tools employed, the achieved data rate, citing both the advantages and drawbacks.

\subsubsection{IM Involving Carrier Frequency, Bandwidth, and Polarization}
Radar system parameters such as carrier frequency, bandwidth, and antenna polarization can also serve as domains of IM. Indexing can be achieved by jointly exploiting multiple carrier frequencies and their allocation across antenna elements~\cite{huang2020majorcom, ma2021frac}. This multi-carrier agility introduces additional spectral DoFs, enabling higher data rates through combined frequency–spatial index modulation. The work in~\cite{temiz2023radar} further employs center frequency, bandwidth, and antenna polarization all together as modulation indexes, demonstrating polarization as a viable IM dimension. Furthermore, these IM schemes can be integrated with phase modulation to further increase data throughput compared with IM alone~\cite{ma2021frac, temiz2025fmcw}.

\subsection{Communication-Centric Design: Radar Sensing with Communication Signals} \label{Sec::comm-centric}
In contrast to radar-centric designs, communication-centric ISAC systems exploit the communication signal itself for radar sensing. In this paradigm, the existing communication waveform is reused to enable radar functionality without requiring dedicated sensing resources. 

\vspace{0.5ex}
\subsubsection{Radar Sensing with Pilot and Reference Signals}
A classical communication frame includes pilots and preambles used for channel estimation and synchronization. These deterministic signals, known to both the transmitter and receiver, can also serve as radar sensing waveforms~\cite{bazzi2025mutual, ma2022downlink}. Since their primary role is channel estimation, their properties—such as constant amplitude and impulse-like autocorrelation—are naturally suitable for sensing. A representative example is WiFi-based sensing, where the receiver estimates the channel state information (CSI) from long training symbols and extracts radar parameters such as range, velocity, and motion features~\cite{ma2019wifi}. Similarly, IEEE~802.11ad-based radar systems exploit the preamble of single-carrier physical layer frames~\cite{kumari2017ieee}, leveraging the excellent cross-correlation properties of Golay complementary sequences. Although these reference signals yield favorable ambiguity function (AF) characteristics, their duration within the overall frame is relatively short compared to the data payload, often resulting in limited sensing signal-to-noise ratio (SNR) relative to the total transmitted power.

Nevertheless, sensing with reference signals remains attractive as a communication-standard-compatible approach that minimizes performance compromise. This motivates recent advances in pilot-based ISAC designs, where pilot symbols are optimized to serve both channel estimation and sensing. For instance,~\cite{bazzi2025mutual} employs mutual information (MI) for ISAC pilot symbol design, while~\cite{zhu2023pilot} investigates pilot resource allocation for flexible S\&C trade-offs. Such new designs indicate that pilot and reference signals can play a key role in enabling efficient and low-overhead ISAC implementation within existing wireless standards.

\vspace{0.5ex}
\subsubsection{Radar Sensing with Communication Data Payload} \label{Sec::Const}
\begin{figure}[t!]
    \centering
    {\includegraphics[width=0.48\textwidth]{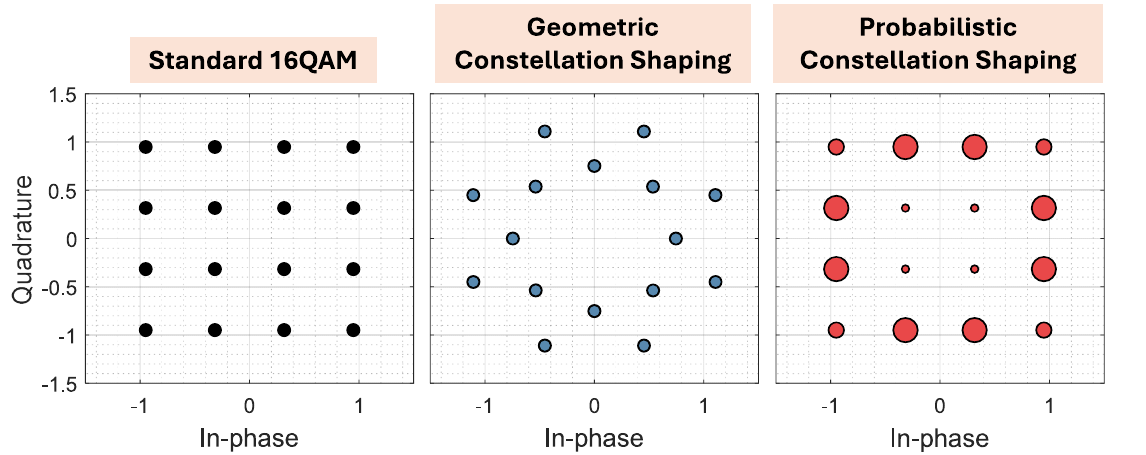}} 
    \caption{ISAC modulation constellation design: Standard 16QAM, geometric constellation shaping, and probabilistic constellation shaping.}
    \label{Fig::Const_concept}
\end{figure}

\begin{figure*}[t!]
    \centering
    {\includegraphics[width=0.8\textwidth]{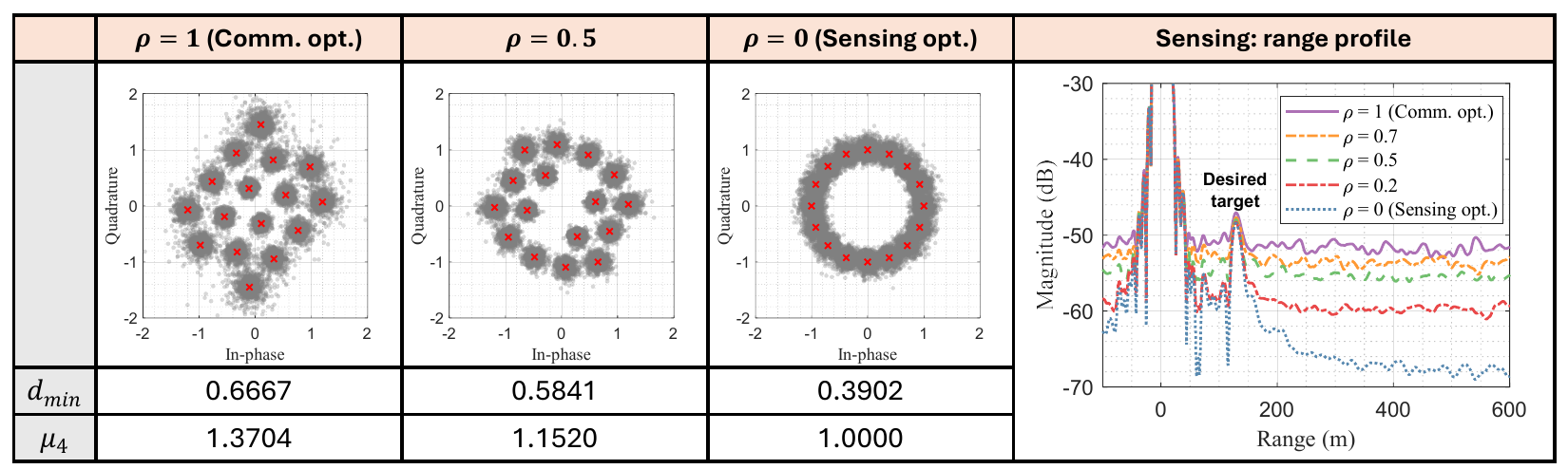}} 
    \caption{Geometric constellation shaping for communication-centric ISAC: (a) 16-ary modulation constellation designed under various S\&C priority ratios, and (b) measured range profiles with matched filtering receiver, showing trade-offs between range sidelobes and MED of constellation symbol.}
    \label{Fig::Const}
\end{figure*}
The sparse nature of pilot transmission, which limits sensing performance, motivates efforts to extend sensing across the whole communication frame, including the data payload. For decades, communication signals such as Digital Video Broadcasting-Terrestrial (DVB-T) have been utilized for opportunistic passive sensing, representing one of the earliest communication-centric approaches that exploit existing communication infrastructure for radar sensing~\cite{chalise2017performance, berger2010signal, falcone2010experimental, palmer2012dvb}. Early research on communication-centric DFRC systems demonstrated that existing communication waveforms,  carrying data payloads, can be repurposed for monostatic radar sensing while maintaining communication performance~\cite{sturm2011waveform}. However, the randomly modulated data payload leads to fluctuations in the AF and increased range-Doppler (RD) sidelobes, which degrade target detection and parameter estimation accuracy compared with dedicated radar waveforms.

The seminal work in~\cite{liu2025cp} demonstrated that the cyclic prefix (CP) OFDM waveform outperforms other candidates such as single-carrier, orthogonal time frequency space (OTFS), and affine frequency division multiplexing (AFDM) in terms of ranging sidelobe levels in single-input single-output (SISO) links, providing a theoretical framework for sensing performance under random signaling. In parallel, several recent studies~\cite{liu2025uncovering, liao2025pulse, du2024reshaping} have analyzed the AF characteristics of modulated communication signals, offering deeper insights into their inherent sensing capabilities. Building on these findings, it has been shown that communication-centric ISAC transmitters utilizing data payloads can be systematically designed to achieve flexible S\&C trade-offs by exploiting available time-frequency domain DoF, including modulation constellation~\cite{du2024reshaping, geiger2025constellation, hu2025learning, yang2024constellation, han2025constellation}, time-domain pulse shaping~\cite{liu2025uncovering, liao2025pulse}, and subcarrier power allocation~\cite{han2025sensing, zhang2025optimal}. These approaches bridge the gap between purely opportunistic sensing and fully integrated ISAC designs, enabling practical signaling adaptability within existing communication frameworks.

\textit{a) Impact of Signal Constellation in OFDM-ISAC:}  
For sensing with data payloads, it makes sense to study the impact of different constellations on performance. Focusing on CP-OFDM ISAC systems, compatible with 5G and 6G standards, it has been shown that their ranging performance under a matched filtering (MF) receiver depends on the kurtosis of the modulation constellation, or equivalently, the fourth-order moment for unit-power, zero-mean constellations.  Assuming the data payloads are modulated by an $M$-ary constellation set $\mathcal{S} = \{s_1 , s_2 , \dots , s_M\}$ with $\sum_{m=1}^M s_m = 0$ and $\frac{1}{M}\sum_{m=1}^M |s_m|^2 = 1$, the fourth-order moment (kurtosis) is given by  
\begin{equation}
    \mu_{4} = \frac{1}{M}\sum_{m=1}^{M} |s_m|^4.
    \label{Eqn::mu}
\end{equation}
The kurtosis directly influences key sensing metrics, including the integrated sidelobe level (ISL) of the auto-correlation function (ACF) (or the Doppler cut of the AF)~\cite{liu2025uncovering}, target detection probability~\cite{geiger2025constellation}, and ranging mean-square error (MSE)~\cite{han2025constellation} in multi-target scenarios with MF receivers. Unit-amplitude constellations from the PSK family, characterized by $\mu_{4} = 1$, yield optimal sensing performance, whereas constellations with $\mu_{4} > 1$ lead to degraded sensing accuracy. Notably, this relationship holds specifically for MF receivers; its impact on sensing performance differs under mismatched filtering (MMF) receivers~\cite{han2025constellation, han2025sensing, keskin2025fundamental, mercier2020comparison}, as will be further discussed in Section~\ref{Sec::RX_F}.

\textit{b) Constellation Shaping:}  
Building on the previous analysis, flexible S\&C trade-offs can be realized by directly designing the modulation constellation. The constellation can be optimized through probabilistic shaping~\cite{yang2024constellation}, geometric shaping~\cite{han2025constellation}, or their joint design~\cite{geiger2025constellation}, as illustrated in Fig. \ref{Fig::Const_concept}. Communication performance can be characterized using metrics such as information entropy~\cite{yang2024constellation}, MI under additive white Gaussian noise (AWGN) channels~\cite{geiger2025constellation}, and minimum Euclidean distance (MED)~\cite{caire2002bit, han2025constellation}. For example, a joint optimization problem can be formulated to minimize the kurtosis for sensing while maximizing the MED for communication, as expressed in~\cite{han2025constellation}:  
\begin{equation}\label{Eqn::P0}
    \begin{aligned}
    & \underset{\{s_m\}_{m=1}^M}{\text{minimize}}
    \qquad (1-\rho) \cdot \mu_4 + \rho \cdot(- d_{\min}) \\
    & \text{subject to} \qquad |s_i - s_j| \geq d_{\min}, \quad \forall s_i \neq s_j \in \mathcal{S},
    \end{aligned}
\end{equation}
where $d_{\min}$ denotes the MED between modulation symbols, and $\rho \in [0,1]$ represents the priority weight between sensing and communication. Example designs with $M = 16$ are illustrated in Fig.~\ref{Fig::Const}, showing that the modulation constellation not only influences the communication performance but also governs the ranging accuracy. This confirms that geometric constellation shaping provides an effective mechanism for flexibly balancing sensing and communication in communication-centric ISAC systems. It is worth noting that the constellation design based on kurtosis applies exclusively to the MF receiver, while receiver-specific ISAC constellation designs are discussed in~\cite{han2025constellation, du2025probabilistic}. Furthermore, ISAC signal modulation based on constellation selection offers a practical alternative, since constellation shaping approaches typically require modifications to the demodulation process at communication receivers.

\vspace{0.5ex}
\subsubsection{Index Modulation in OFDM-ISAC Systems}
Beyond its application in radar-centric ISAC, IM is also an attractive modulation technique in communication-centric ISAC systems, most notably, those employing OFDM. In conventional OFDM, information is conveyed solely through the modulation symbols placed on all active subcarriers. IM, however, introduces an additional information-bearing dimension by exploiting the indices of the active subcarriers themselves \cite{abu2009subcarrier, bacsar2013orthogonal, wen2015achievable}. By activating only a subset of subcarriers and mapping part of the information onto their activation pattern, IM provides a unique mechanism to improve achievable rate and bit error rate (BER) performance without requiring extra bandwidth or transmit power \cite{wen2015achievable}. 

When integrated into OFDM-based ISAC systems, a simple IM approach is to disjointly allocate subcarriers for radar sensing while simultaneously activating or deactivating subcarriers to provide IM for communication \cite{csahin2021ofdm, huang2021index, elbir2024index}. However, this setup of null subcarrier distributions introduces drawbacks for both functionalities. On the communication side, null subcarriers reduce the achievable data rate, whereas on the sensing side, they increase range sidelobes and degrade target detection performance. In particular, the presence of spectral holes leads to null observations of certain sensing reflections, which necessitates advanced receiver processing techniques such as compressed sensing to reconstruct the missing information \cite{huang2021index}. An alternative approach that alleviates this problem is to use two different power levels instead of on-off subcarriers. This way, data are transmitted on all subcarriers without null observations \cite{sui2025multi}.

It has recently been shown \cite{hawkins2024ofdm} that IM provides new DoFs for balancing communication throughput and radar sensing performance by collecting multiple observations of the received signal to fill in the holes in IM-OFDM systems. In so doing, the sparse activation of subcarriers can still leverage the benefits of IM, reducing inter-carrier interference (ICI) and improving symbol error performance, while the index domain carries additional bits with minimal overhead. From the sensing perspective, carefully designed subcarrier activation patterns can improve the AF and influence range–Doppler performance. This creates opportunities to tailor IM schemes for dual functionality. In essence, some subcarriers prioritize robust communication, while others are optimized for accurate sensing. For example, the work in \cite{yang2024superposed} superimposed a dedicated sensing sequence with good auto-correlation properties, improving the sensing performance in IM-OFDM. Nevertheless, the joint design of IM-OFDM for ISAC has not been extensively explored. Most existing works consider IM as an add-on to conventional OFDM, whereas a dedicated co-design that jointly optimizes index patterns, waveform structures, and sensing objectives could enable more flexible trade-offs between communication and sensing, ultimately enhancing overall system performance.

\section{The ISAC Transmitter: MIMO Precoding Design}\label{Sec::TX_B}
Complementary to the modulation and signal designs discussed above, precoding can also offer additional DoF in designing ISAC trade-offs. This section overviews state-of-the-art MIMO-ISAC precoding techniques, outlining the fundamental frameworks for joint radar–communication beamforming. 
\subsection{MIMO Precoding Design for ISAC} \label{Sec::JointSignaling}
\begin{figure}[t!]
    \centering
    {\includegraphics[width=0.48\textwidth]{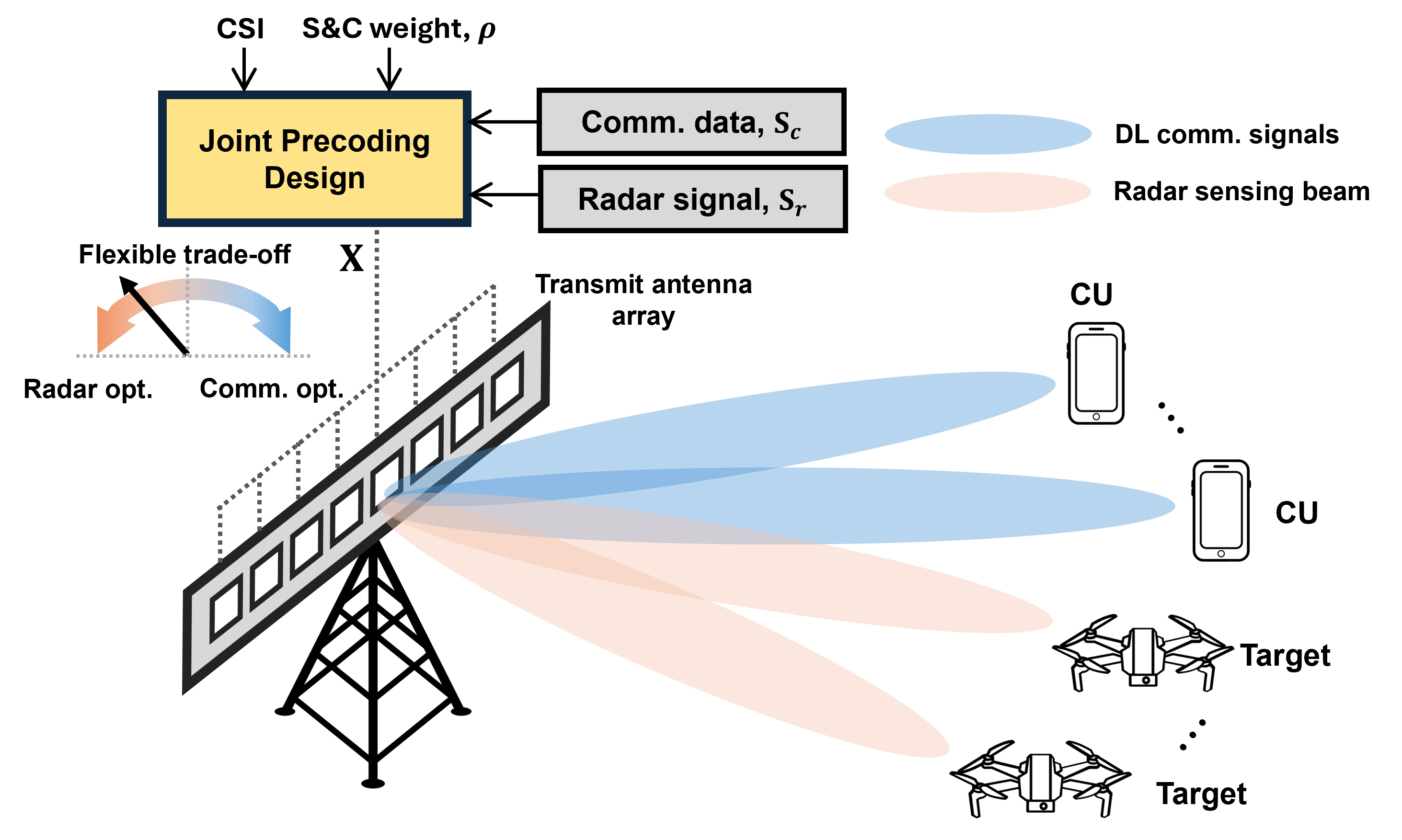}}
    \caption{Illustration of joint MIMO precoding design for multi-user communication and multi-target sensing.}
    \label{Fig::3}
\end{figure}
Unlike modulation approaches that embed one functionality into the platform of the other, joint signaling design treats both as co-primary objectives and balances their performance through multi-objective optimization of time–frequency–spatial resources. The main objective of joint signaling optimization is to design spatial communication precoders and MIMO radar beamforming weights for multiple transmit antennas, as illustrated in Fig.~\ref{Fig::3}. 

To this end, one approach is to express the transmit ISAC signal $\mathbf{X}$ over $L$ blocks as a weighted superposition of communication signals and dedicated radar probing signals \cite{liu2020joint, hua2023optimal, he2022energy, choi2024joint}. Consider a transmitter equipped with $N_t$ antennas that serves $U$ single-antenna communication users (CUs) while simultaneously detecting and estimating the parameters of $K$ targets. The transmit signal with a linear block-level precoding (BLP) is then modeled as  
\begin{align}
    \mathbf{X} = \mathbf{W}_c \mathbf{S}_c + \mathbf{W}_r \mathbf{S}_r, 
    \label{Eqn::Sig_Model}
\end{align}
where $\mathbf{W}_c = [\mathbf{w}_1, \dots, \mathbf{w}_U] \in \mathbb{C}^{N_t \times U}$ denotes the communication precoder for the $U$ users, $\mathbf{S}_c \in \mathbb{C}^{U \times L}$ represents the communication data streams, $\mathbf{W}_r \in \mathbb{C}^{N_t \times N_t}$ is the radar beamforming matrix, and $\mathbf{S}_r \in \mathbb{C}^{N_t \times L}$ corresponds to the radar signals. The above signal model can be viewed as a generalization of the unified signal model $\mathbf{X} = \mathbf{W} \mathbf{S}$, which is also used in the literature to provide ISAC functionality through weighted precoding optimization \cite{liu2021cramer}.

Importantly, the weighted-sum signal model in BLP provides additional DoF compared to the unified model \cite{liu2018mu}, whose covariance matrix becomes rank-deficient when $U < N_t$. The auxiliary sequence $\mathbf{S}_r$, which increases the DoF of the transmitted signal, can be specifically designed to have better cross-correlation and auto-correlation properties to suppress RD sidelobes in the matched filtering output, thereby improving target detection and interference suppression \cite{cui2013mimo, xu2012zero, he2009designing}. For more details of ISAC signaling models, readers are referred to \cite{liu2020joint, liu2021cramer}. 

We remark that the signal model in \eqref{Eqn::Sig_Model} assumes a fully digital MIMO array, where the number of transmit antennas equals the number of radio frequency chains. While this architecture provides maximum flexibility, its hardware cost and power consumption can limit practical deployment in large-scale MIMO systems \cite{han2022high}. To improve hardware efficiency, sparse arrays, hybrid beamforming architectures, or phased-array modules are often considered, depending on the application scenario. In such cases, the signal model can be readily revised or extended to hybrid beamforming \cite{nguyen2024joint, qi2022hybrid, wang2022partially, liyanaarachchi2023joint, liu2019hybrid} and multi-beam analog beamforming \cite{luo2019optimization, suh2023null, zhang2018multibeam}.

The joint precoding design is typically formulated as a multi-objective optimization problem that captures both sensing and communication goals. Such a formulation enables flexible trade-offs between the two functionalities while accounting for MIMO transceiver specifications such as total transmit power, per-antenna power, or peak-to-average power ratio (PAPR). Although many variations of this problem have been studied in the literature, a unified structure can be expressed as  
\begin{equation}\label{Eqn::P1}
    \begin{aligned}
    & \underset{\mathbf{X}}{\text{maximize}}
    \qquad   \rho \tilde{f}_c(\mathbf{X}) \pm (1-\rho) \tilde{f}_r(\mathbf{X})  \\
    & \text{subject to}
    \qquad \quad  c_i(\mathbf{X}) \leq C_i, \ \forall i,
    \end{aligned}
\end{equation} 
where $\tilde{f}_c(\mathbf{X})$ and $\tilde{f}_r(\mathbf{X})$ denote normalized performance functions for communication ${f}_c(\mathbf{X})$ and radar sensing ${f}_r(\mathbf{X})$, respectively, and $c_i(\mathbf{X})$ represents a system-level specification constrained by $C_i$ (e.g., power budget or constant-modulus condition). The parameter $\rho \in [0,1]$ serves the same function as in (2). In practice, this multi-objective problem is often transformed into a single-objective problem by recasting one objective as a constraint while optimizing the other. The formulated problem for ISAC signaling design is generally non-convex, and can be solved using semi-definite relaxation (SDR) \cite{liu2018mu}, successive convex approximation (SCA) \cite{fang2025low}, or learning-based approaches \cite{wu2024efficient, temiz2025deep, zhang2025low}.

In the next section, we provide an overview of the ISAC metrics that are commonly used under the above framework. It is worth noting that a new stream of research investigates network-level ISAC optimization and has introduced corresponding network-level ISAC metrics \cite{han2025network, meng2024cooperative, meng2025network,meng2024network}. However, as the article focuses on link-level ISAC design, it falls outside the scope of this article.

\vspace{0.5ex}
\subsection{Communication Performance Metrics}
From the communication perspective, the main focus of joint signaling design is to account for both multi-user interference (MUI) and radar-induced interference at each CU. To this end, metrics such as per-user signal-to-interference-plus-noise ratio (SINR) \cite{liu2021cramer, liu2020joint, hua2023optimal}, achievable/sum rate \cite{li2024framework, hua2023mimo, ren2023fundamental}, and MI \cite{wei2023waveform, ouyang2023integrated} have been widely adopted, all of which capture communication quality-of-service (QoS) under DFRC operation. Early work employed the total MUI energy as the performance metric \cite{liu2018toward}:

\begin{equation}
    f_{c}(\mathbf{X}) = \|\mathbf{H} \mathbf{X} - \mathbf{S}_c \|_{F}^2,
\end{equation}
where $\mathbf{H} = [\mathbf{h}_1, \dots, \mathbf{h}_U]^H \in \mathbb{C}^{U \times N_t}$ with $\mathbf{h}_u \in \mathbb{C}^{N_t \times 1}$ denoting the channel between the ISAC transmitter and user $u$.
For direct intuition on communication performance, the average per-user SINR of user $u$ can be expressed as  
\begin{equation}
    f_{c,u}(\mathbf{X}) = \frac{|\mathbf{h}_u^H \mathbf{w}_u|^2}{\sum_{i=1,i \neq u}^U|\mathbf{h}_u^H \mathbf{w}_i|^2 + \|\mathbf{h}_u^H \mathbf{W}_r\|^2 + \sigma_c^2}, \; \forall{u}
    \label{Eqn::SINR}
\end{equation}
where $\sigma_c^2$ represents the noise power at the CU. The denominator reflects the interference contributions, including both MUI and radar signals received at the user. Denoting \eqref{Eqn::SINR} as $\gamma_u$, the achievable/sum rate can then be also derived as \cite{li2024framework, hua2023mimo, ren2023fundamental}:
\begin{align}
    f_{c,u}(\mathbf{X}) & = \log_2(1+\gamma_u), \;\; \text{(per-user achievable rate)} \label{Eqn::Rate1} \\
    f_c(\mathbf{X}) &= \sum_{u=1}^U \log_2(1+\gamma_u), \;\; \text{(sum-rate)}
    \label{Eqn::Rate2}
\end{align}
It is important to note that the SINR expression in \eqref{Eqn::SINR} captures only the average performance, with respect to the data stream, over an $L$-block transmission. Consequently, signaling (precoder) designs based on this metric guarantee average communication-symbol SINR performance regardless of the specific realization of $\mathbf{S}_c$. This limitation can be addressed through symbol-level designs, which will be discussed in detail in Section~\ref{Sec::SLP}.

As observed in \eqref{Eqn::SINR}, evaluating communication performance requires instantaneous CSI at the transmitter, which may not always be available in practical implementations. To enhance robustness, imperfect CSI is often addressed by incorporating bounded CSI errors or statistical CSI into the design problem, ensuring that the resulting signaling provides guaranteed worst-case performance \cite{choi2024joint, zhang2023robust, zhao2024robust, dai2024joint}. 

\begin{figure}[t!]
    \centering
    {\includegraphics[width=0.35\textwidth]{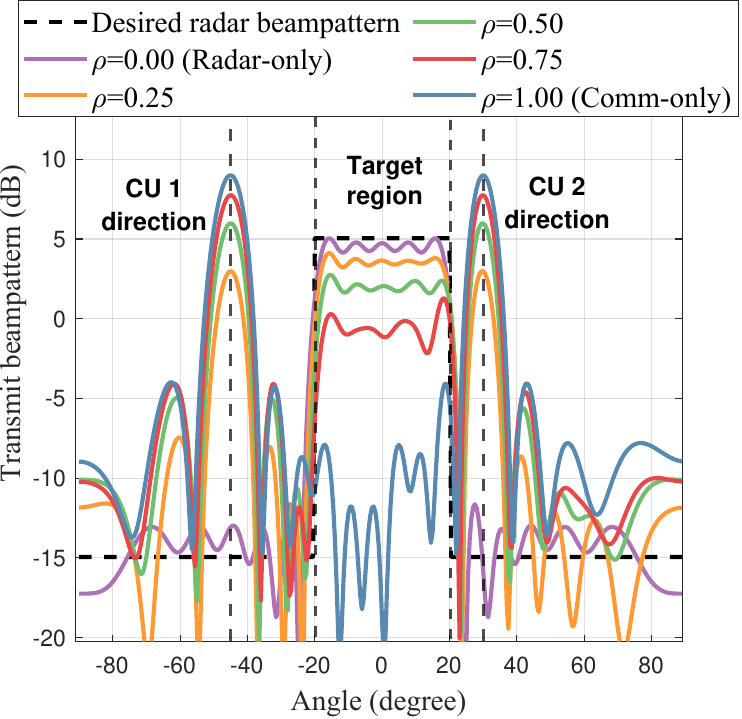}}
    \caption{Beampattern–SINR trade-off design with $N_t = 16$ transmit antennas and $U=2$ users located at $-45^{\circ}$ and $30^{\circ}$. The beampattern matching error varies with the priority weight $\rho$, illustrating that joint signaling design enables a flexible trade-off between sensing and communication performance.}
    \label{Fig::4}
\end{figure}

\vspace{0.5ex}
\subsection{Sensing Performance Metric}
Precoding for the sensing task aims to illuminate targets using multiple antennas, where performance is primarily characterized in the spatial (or angular) domain. Accordingly, the design often leverages classical metrics from MIMO radar beamforming, including beampattern matching \cite{stoica2007probing, cheng2017constant, fuhrmann2008transmit}, angle estimation Cramér–Rao lower bound (CRLB) \cite{li2007range}, SINR and signal-to-clutter-plus-noise ratio (SCNR) \cite{cui2013mimo, li2016mimo}, MI \cite{yang2007mimo}.  

\subsubsection{Beampattern matching}  
The transmit beampattern describes the spatial distribution of radiated power across angles. For a steering vector $\mathbf{a}(\theta) \in \mathbb{C}^{N_t}$ at angle $\theta$, the beampattern of the ISAC signal is given by $\mathbf{a}^H(\theta)\mathbf{R}_{\text{X}}\mathbf{a}(\theta)$,  
where $\mathbf{R}_{\text{X}} = \tfrac{1}{L}\mathbf{X}\mathbf{X}^H$ denotes the transmit covariance matrix. The objective of beampattern matching is to minimize the error between the designed ISAC beampattern and a pre-defined desired beampattern $P(\theta)$. Accordingly, the beampattern matching MSE for $M$ angular samples $\{\theta_i\}_{i=1}^M$ is expressed as  
\begin{equation}
    f_r(\mathbf{X}) = \frac{1}{M}\sum_{i=1}^M \left| \alpha P(\theta_i) - \mathbf{a}^H(\theta_i)\mathbf{R}_{\text{X}}\mathbf{a}(\theta_i) \right|^2,
    \label{Eqn::Beam}
\end{equation}
where $\alpha$ is a scaling factor. This metric has been widely adopted in MIMO radar beamforming for both target search and tracking, since it enables controlled power distribution across multiple spatial directions, even under uncertainty in the target channel. For joint ISAC signaling design with beampattern matching, readers are referred to \cite{liu2018mu, hua2023optimal, liu2020joint, choi2024joint}, which present various optimization algorithms for solving the formulated problem. An illustrative numerical example is shown in Fig.~\ref{Fig::4}, demonstrating the flexible trade-off between sensing and communication performance as the priority weight $\rho$ is varied. It is observed that as the priority weight shifts toward sensing, the resulting beampattern approaches the desired radar beampattern, whereas prioritizing communication focuses the beam toward the CUs, thereby maximizing their QoS.

\subsubsection{CRLB}  
\begin{figure}[t!]
    \centering
    {\includegraphics[width=0.35\textwidth]{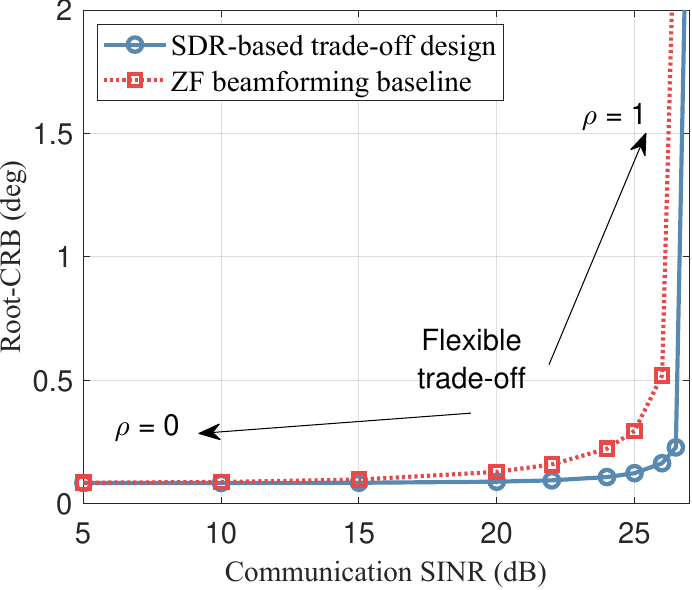}}
    \caption{CRLB–SINR trade-off with $N_t = 16$ transmit antennas and $U=2$ users located at $-45^{\circ}$ and $30^{\circ}$. The SDR-based joint signaling design is compared with the ZF beamforming baseline, demonstrating an extended trade-off region and improved performance.}
    \label{Fig::5}
\end{figure}

In joint signaling design, CRLB can be employed as a sensing performance metric, as it directly characterizes the fundamental limit of parameter estimation accuracy. The objective function can be expressed as  
\begin{equation}
    f_r(\mathbf{X}) = \left[ \mathbf{J}^{-1}(\theta) \right]_{1,1},
    \label{Eqn::CRLB}
\end{equation}
where $\mathbf{J}(\theta)$ is the Fisher information matrix determined by the transmit covariance and the sensing channel. Minimizing CRLB in joint design improves the theoretical accuracy of target parameter estimation under power constraints. The resulting optimization problem is generally non-convex but can be handled using SDR, as demonstrated in \cite{liu2021cramer}. Fig~\ref{Fig::5} illustrates the CRLB–SINR trade-off obtained from SDR-based solution compared with classical zero-forcing (ZF) beamforming. The results show that the joint design based on SDR extends the achievable trade-off region, yielding improved CRLB–SINR performance compared to the conventional baseline.

Evaluating the deterministic CRLB in \eqref{Eqn::CRLB} requires prior knowledge of the true target angle, which is typically unknown in practice. To mitigate this limitation, several works \cite{xiong2023fundamental, xu2024mimo,liu2025joint, su2025secure} have considered the use of Bayesian CRLB as an alternative performance metric, where prior information is incorporated to relax the requirement of knowing the true parameter.

Beyond the presented metrics, sensing SINR/SCNR \cite{choi2024joint}, MI \cite{li2024framework}, Ziv-Zakai bound \cite{chazan1975improved}, and Kullback-Leibler divergence \cite{fei2024revealing} can also be exploited in ISAC signaling. We remark that joint signaling design continues to evolve toward handling increasingly complex and practical scenarios, while offering flexible trade-offs between the two functionalities. This trend ultimately paves the way for realizing ISAC systems that can be effectively deployed in real-world environments.

\vspace{0.5ex}
\section{Interference Exploitation in ISAC Transmission} \label{Sec::SLP}
Traditionally, interference in wireless communication systems has been treated as a harmful factor that degrades QoS and must be mitigated. Conventional transmitter designs with linear BLP handle MUI as detrimental and attempt to suppress it statistically over a block of symbols. As a result, instantaneous performance is not ensured, which limits overall efficiency in medium-to-high SNR regimes. This has motivated research into precoding methods that instead exploit interference at the symbol level on an instantaneous basis rather than cancel it \cite{masouros2010correlation, masouros2009dynamic, masouros2013known, zheng2014rethinking, masouros2015exploiting, li2020tutorial}. In this section, we investigate recent advances in ISAC transmitter design with interference exploitation. By exploiting constructive interference (CI) in joint signaling design, ISAC transmitters can enhance communication reliability while simultaneously improving instantaneous sensing performance.

\vspace{0.5ex}
\subsection{Constructive Interference Exploitation}
\begin{figure}[t!]
    \centering
    {\includegraphics[width=0.5\textwidth]{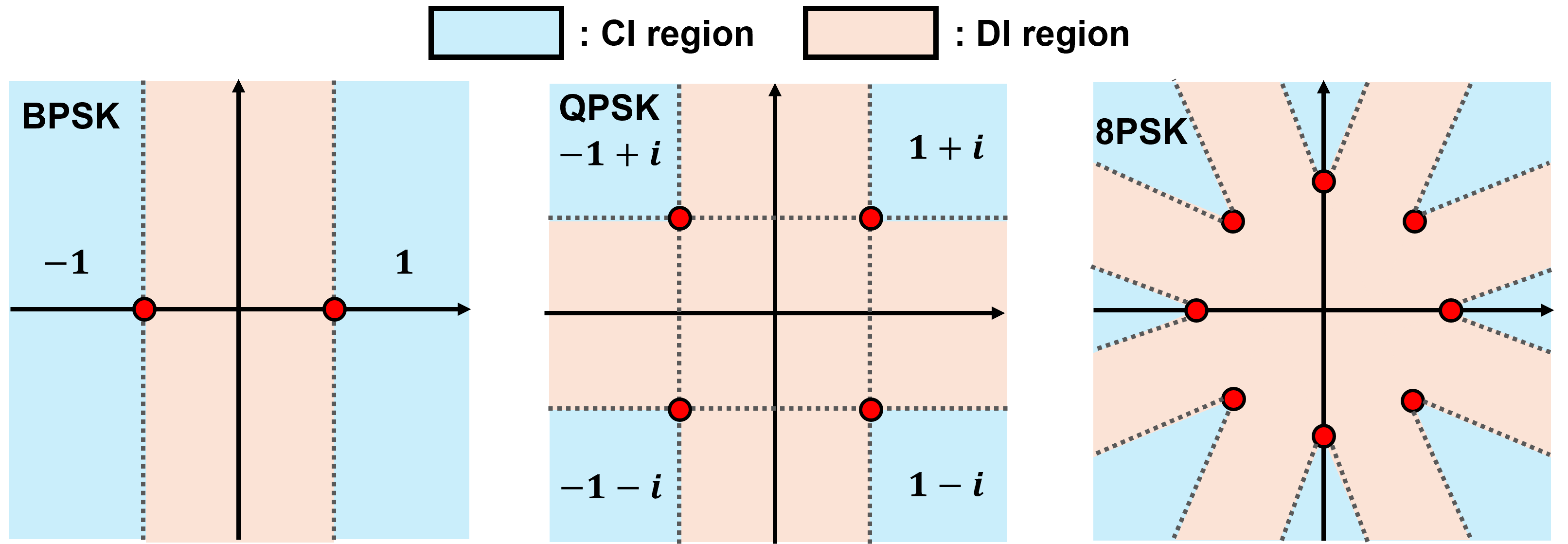}}
    \caption{CI and DI regions in BPSK, QPSK, and 8PSK constellations.}
    \label{Fig::6}
\end{figure}

A breakthrough came with the concept of CI, which refers to interference that pushes received signals at CU further away from the decision boundaries of the modulated symbol constellation, thereby enhancing useful signal power. As the counterpart of CI, destructive interference (DI) is defined as interference that drives the received signal back to the decision boundaries, reducing useful signal power. Exemplary CI and DI regions for BPSK, QPSK, and 8PSK are illustrated in Fig. \ref{Fig::6}. These concepts motivated the development of SLP \cite{masouros2013known, masouros2015exploiting}, which operates on a symbol-by-symbol basis and exploits both CSI and data symbol knowledge to control not only the power but also the direction of interference at CU receivers. 

SLP operates at the symbol timescale. This allows the transmitter to manipulate MUI in a way that makes it constructive. In a downlink (DL) multi-user multiple-input single-output (MU-MISO) system, the transmitted signal is
\begin{equation}
    \mathbf{x} = \sum_{u=1}^{U} \mathbf{w}_u s_u = \mathbf{W}\mathbf{s},
    \label{Eqn::model_slp}
\end{equation}
where $\mathbf{w}_u \in \mathbb{C}^{N_T}$ is the precoder for user $u$, $s_u$ is the modulation symbol, and $\mathbf{s}$ is the symbol vector. The received signal at user $u$, ignoring noise, is
\begin{equation}
    r_u = \mathbf{h}_u^H \mathbf{x} = \lambda_u s_u,
\end{equation}
where $\mathbf{h}_u$ is the channel vector and $\lambda_u \in \mathbb{C}$ captures the effect of interference on the amplitude and phase of symbol $s_u$ after precoding. To make interference constructive, the following CI conditions are considered to design SLP. 

For an $M$-PSK constellation, the constructive region is defined as the angular sector of width $\pm \pi/M$ around each symbol. Accordingly, the CI condition for user $u$ can be expressed as  
\begin{equation}
    \Big[ \Re(\lambda_u) - \sqrt{\Gamma_u \sigma_c^2} \Big] 
    \tan\!\left( \frac{\pi}{M} \right) 
    \geq \big| \Im(\lambda_u) \big|, 
    \label{Eqn::CI}
\end{equation} 
where $\Gamma_u$ denotes the SNR target. $\Re(\cdot)$ and $\Im(\cdot)$ denote the real and imaginary part, respectively. It should be noted that the CI concept extends to multi-level constellations, such as QAM, star-QAM, and amplitude and phase-shift keying (APSK), via symbol-scaling. For details on symbol-scaling for CI exploitation in QAM, see \cite{alodeh2017symbol, li2020interference}. A comprehensive overview of CI exploitation and SLP is given in \cite{li2020tutorial}.

\vspace{0.5ex}
\subsection{Symbol-level Precoding for ISAC Transmitter Design}
Recent advances in MIMO ISAC transmitter design, as discussed in Section~\ref{Sec::JointSignaling}, have primarily focused on BLP. This implies that once multiplied with the data symbols, the properties of the waveform, including radar aspects such as RD sidelobes, will be subject to instantaneous variations. Unlike BLP, SLP enables each symbol to simultaneously satisfy communication constraints while shaping favorable radar characteristics. This property ensures that sensing performance metrics such as beampattern matching or RD sidelobe suppression remain consistent, even with a limited number of snapshots, which is particularly valuable in highly dynamic environments \cite{chen2023symbol, li2024mimo, liu2021dual}.

With the transmit signal model $\mathbf{x} = \mathbf{W}\mathbf{s} \in \mathbb{C}^{N_t}$ in \eqref{Eqn::model_slp}, the optimization problem for SLP-based joint signaling ISAC design can be generally formulated as  

\begin{equation}
    \begin{aligned}
        & \underset{\mathbf{x}}{\text{maximize}}
        \qquad   \rho \tilde{f}_c(\mathbf{x}) \pm (1-\rho) \tilde{f}_r(\mathbf{x})  \\
        & \text{subject to}
        \quad \;\;\; {f}_c(\mathbf{x}) = \Gamma_u, \\
        & \qquad \qquad \quad \quad \Big[ \Re(\mathbf{h}_u^H \mathbf{x} e^{-j \phi_u}) - \sqrt{\Gamma_u \sigma_c^2} \Big] 
        \tan\!\left( \frac{\pi}{M} \right) \\
        & \qquad \qquad \qquad \qquad \qquad \qquad \geq \big| \Im(\mathbf{h}_u^H \mathbf{x} e^{-j \phi_u}) \big|, \;\; \forall{u} \\
        & \qquad \qquad \quad \; \;\; c_i(\mathbf{x}) \leq C_i, \ \forall i,
    \end{aligned}
    \label{Eqn::SLP_ISAC}
\end{equation}
where $\phi_u \in [0, 2\pi]$ is the corresponding phase of $s_u$. $f_r(\mathbf{x})$ denotes a sensing-oriented objective described in Section \ref{Sec::TX_B}. The CI constraints ensure that the received symbols at each CU remain in the constructive region, thereby guaranteeing the required communication QoS. The additional constraints $c_i(\mathbf{x}) \leq C_i$ capture system specifications such as power budget or constant-modulus conditions. 

This formulation highlights two important distinctions from block-level ISAC transmitter design. First, while block-level designs optimize signals only statistically, with respect to the data stream, over $L$ snapshots, SLP guarantees that each transmit vector $\mathbf{x}$ contributes simultaneously to communication and sensing objectives on an instantaneous basis. Second, unlike BLP, where $\mathbf{x}$ is strictly a linear mapping of the data symbols $\mathbf{s}$, SLP directly designs $\mathbf{x}$ with knowledge of $\mathbf{s}$, thereby allowing symbol-by-symbol adaptation. These properties give SLP-based ISAC transceivers much finer control over instantaneous S\&C performance. 

\begin{figure}[t!]
    \centering
    {\includegraphics[width=0.20\textwidth]{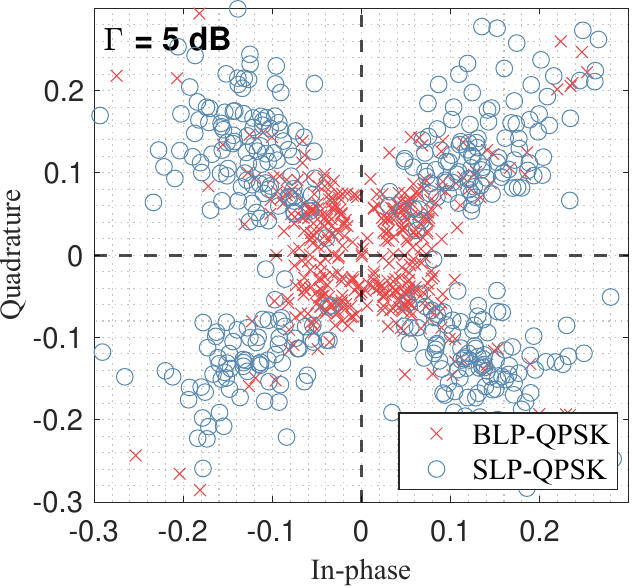}}
    {\includegraphics[width=0.20\textwidth]{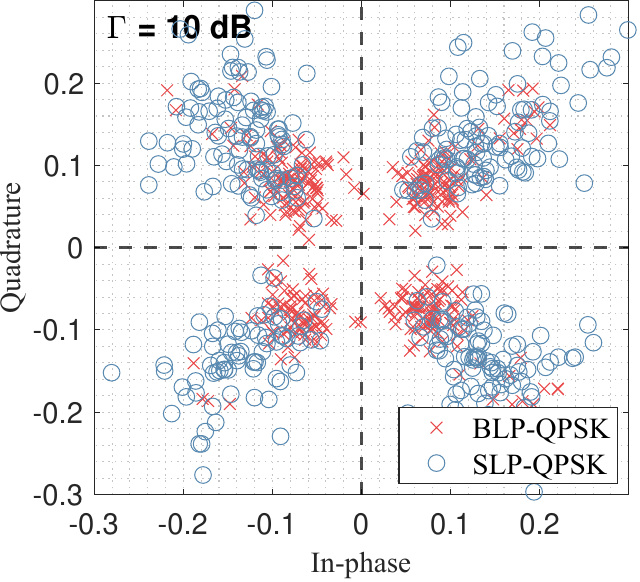}} \\
    {\includegraphics[width=0.20\textwidth]{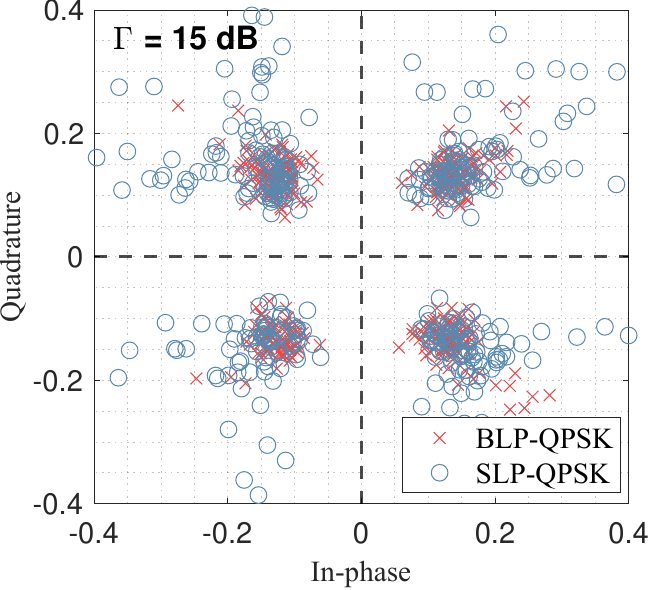}}
    {\includegraphics[width=0.20\textwidth]{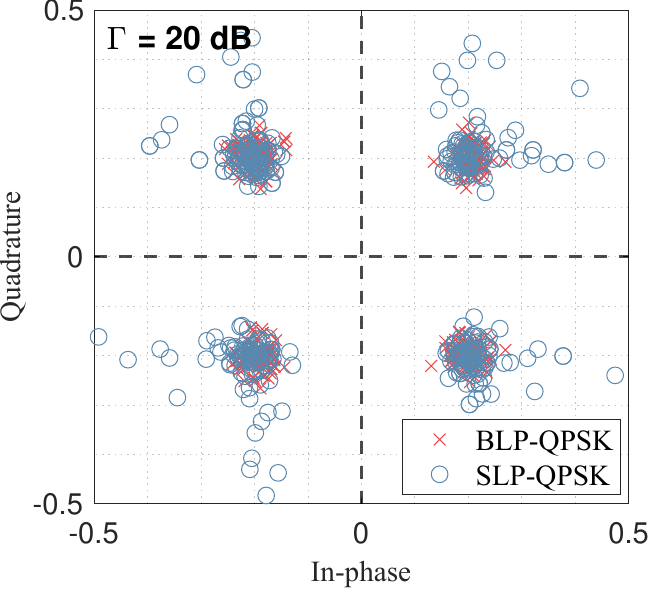}}
    \caption{Received symbols for different level of communication QoS under CRLB-SINR (or SNR) trade-off design with BLP and SLP.}
    \label{Fig::SLP}
\end{figure}

For instance, the works in \cite{liu2021dual, wang2022symbol} design MIMO DFRC transmit beamforming using SLP, thereby providing instantaneous S\&C trade-offs in terms of radar beampattern shaping and CI-based SINR. With the additional consideration of PAPR, these designs also incorporate constant-modulus power constraints. The resulting non-convex problems are either relaxed to SDP formulations \cite{wang2022symbol}, or solved using iterative algorithms such as majorization–minimization or augmented Lagrangian methods \cite{liu2021dual, zhang2025low}. Another line of work extends the BLP-based CRLB–SINR trade-off design into the symbol-level domain \cite{chen2023symbol, babu2024symbol, wang2022symbol}, of which examples for the received symbols are shown in Fig. \ref{Fig::SLP}. In particular, the work in \cite{babu2024symbol} demonstrates the superior performance of SLP compared to BLP when applied to near-field ISAC scenarios. Moreover, instead of CRLB, alternative sensing metrics such as radar estimation minimum mean-square error (MMSE) have been proposed for $f_r(\mathbf{x})$ \cite{liao2023symbol}. These optimization problems can also be addressed by SDR or SCA, similar to their BLP counterparts. The reported results overall consistently show that SLP-based ISAC signaling achieves enhanced joint S\&C performance relative to BLP-based designs, while additionally guaranteeing instantaneous symbol-level performance \cite{liu2021dual, wang2022symbol,chen2023symbol, babu2024symbol, liao2023symbol,wang2025symbol, wang2025constructive, wang2025interference}.

Recent works have also extended SLP in ISAC beyond narrowband settings. In wideband MIMO-OFDM systems, SLP provides additional temporal DoF, enabling direct control of RD sidelobes through symbol-level optimization. For example, \cite{li2024mimo} addresses a key drawback of dual-functional MIMO-OFDM ISAC waveforms, namely, the high RD sidelobes introduced by random data symbols in the MF receiver, which severely degrade target detection and parameter estimation. By incorporating SLP into MIMO-OFDM ISAC, both temporal and spatial DoF are exploited to directly shape the AF of the transmit signal. Specifically, the optimization problem is formulated to minimize the ISL of RD maps while ensuring target illumination power and maintaining CI-based multi-user communication QoS. This demonstrates that symbol-level optimization not only transforms harmful MUI into a communication gain but also significantly enhances radar sensing capability through improved MF output.

\vspace{0.5ex}
\subsection{Overcoming Complexity in Interference Exploitation for ISAC Design}
A major barrier to the practical deployment of SLP in ISAC transmitters is its computational complexity. Because SLP-based transmitter design requires a tailored precoder for each symbol combination, the computational burden increases rapidly with the number of antennas, users, and symbol durations within a coherence interval. This challenge is further intensified in ISAC, where the design must simultaneously satisfy CI constraints for communication and radar sensing-oriented objectives. Several recent studies have therefore proposed algorithmic frameworks to mitigate this symbol-level complexity while retaining most of the performance benefits.

One line of work focuses on optimization-driven methods. Iterative reformulations such as separable and dual optimization \cite{yang2023speeding} and inversion-free alternating direction method of multipliers (ADMM) updates \cite{yang2024low} decompose the SLP problem into parallelizable subproblems that avoid costly matrix inversions. Closed-form CI-based precoder solutions further accelerate an iterative SLP design ~\cite{wen2024low, li2020interference, wei2020secure, li2018interference}. Low-complexity designs for large-scale antenna arrays have also been developed, where hybrid SLP architectures reduce the number of radio frequency chains required while maintaining interference-exploitation gains \cite{domouchtsidis2018symbol}. Although these methods were originally proposed for MU-MISO communication systems, they hold strong potential for extension to SLP-based ISAC designs.

Another promising direction is learning-based design. Model-driven frameworks such as ADMM-SLPNet \cite{yang2023admm} employ deep unfolding to translate iterative optimization steps in SLP into trainable neural network layers, providing both interpretability and fast convergence. The unfolded network derived from the iterative optimization has also been applied to SLP in DFRC systems \cite{zhang2025low}, achieving near-optimal S\&C trade-offs with significantly reduced complexity. In addition, supervised and hybrid learning techniques can approximate dual-functional SLP waveform mappings \cite{jiang2025slp}, while unsupervised methods \cite{mohammad2021unsupervised} learn feasible interference-exploitation solutions without requiring labeled data.

These algorithmic advances indicate that the complexity challenge in SLP-based ISAC can be effectively addressed through a combination of mathematical simplification, problem restructuring, and learning-based approaches. Particularly promising are model-based learning frameworks that unfold optimization-inspired algorithms into neural architectures \cite{yang2023admm, zhang2025low}, striking a practical balance between performance and complexity. Together, these developments are transforming SLP from a theoretically powerful yet computationally prohibitive technique into a practical enabler for real-time ISAC signaling with MIMO transceivers.

\section{ISAC Receiver Design}\label{Sec::RX}
\begin{figure}[t!]
    \centering
    {\includegraphics[width=0.375\textwidth]{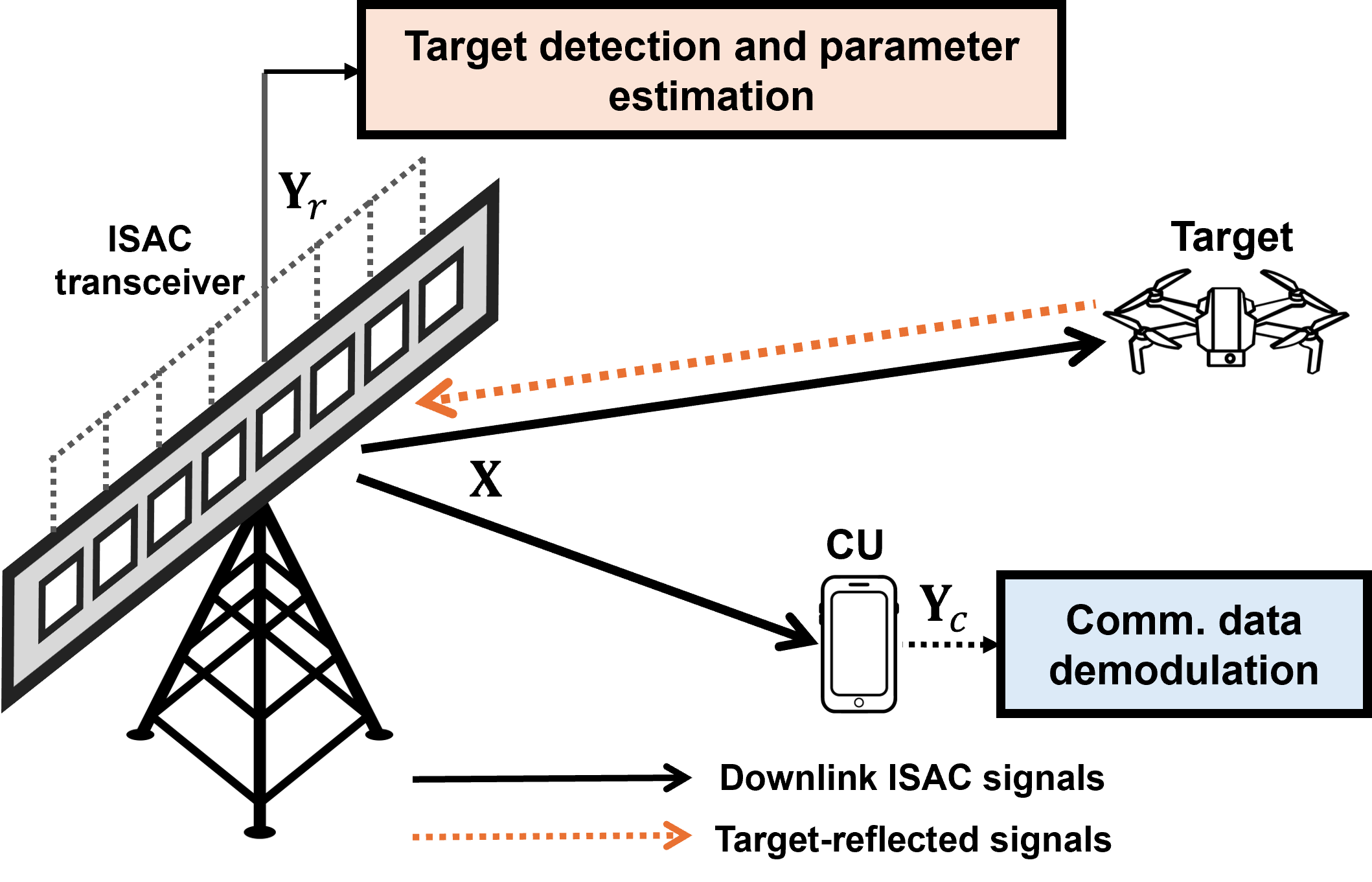}}
    \caption{Illustration of a downlink mono-static ISAC scenario, where the CU demodulates communication data while the sensing receiver detects targets and estimates their parameters.}
    \label{Fig::RX1}
\end{figure}

The ISAC receiver plays a critical role in achieving full dual functionality. The unified ISAC signals directly influence both radar and communication receiver designs, particularly in radar-centric IM-based ISAC and communication-centric ISAC with data payloads. For joint receiver operation, the receiver must simultaneously decode communication data and extract sensing information from the same received signal. Realizing this dual functionality requires advanced signal processing, estimation, and receiver architectures capable of handling the coupled radar–communication tasks. Focusing on the scenario with unified ISAC signal transmission, this section reviews representative ISAC receiver design methodologies, highlighting key architectures, processing techniques, and implementation aspects.

\subsection{Receiver Design for Radar-Centric ISAC}
Considering the ISAC scenario illustrated in Fig.~\ref{Fig::RX1}, this subsection explores communication and sensing receiver designs with a focus on IM-based ISAC systems. In IM-based ISAC, communication data are conveyed through IM bits, requiring reliable IM bit recovery at the communication receiver. 

\subsubsection{Communication Receiver Design}
The communication receiver in IM-based ISAC is designed to recover the active indexing pattern and corresponding data symbols. To this end, least-squares estimation is typically employed to extract the radar parameter values used for indexing and decode the embedded IM bits. In antenna-selection-based IM, the IM bits can be detected using a sparse array dictionary~\cite{wang2018dual}. Let $\mathbf{y}_c$ denote the received signal at the CU for a given pulse. The receiver estimates the active steering vector as
\begin{equation}
    \hat{i} = \underset{i}{\mathrm{argmin}} \; \left\|  \mathbf{y}_c / \alpha - \bar{\mathbf{a}}_i (\theta) \right\|_2,
    \label{eq:rx_IM}
\end{equation}
where $\alpha$ is a scaling factor and $\bar{\mathbf{a}}_i(\theta) = \Phi_i \mathbf{a}(\theta)$, with $\Phi_i \in \mathbb{C}^{N_t \times N_t}$ representing the antenna-selection matrix. The receiver evaluates the distance between the estimated vector and each dictionary element, selecting the index $\hat{i}$ that minimizes it. Receiver designs for other forms of IM can be similarly extended using least-squares estimation. For example,~\cite{ma2021frac} presents a receiver for joint antenna–frequency IM with phase modulation, while~\cite{elbir2024index} provides a comprehensive overview of IM-based ISAC receivers.

Since the complexity of IM-based ISAC receivers increases with the size of the dictionary or IM codebook, exhaustive search across all indexing patterns becomes computationally expensive. To address this issue,~\cite{temiz2025fmcw} proposes a low-complexity receiver design for carrier frequency, bandwidth, and antenna polarization IM combined with phase modulation. The receiver first estimates the IM bits based on the predefined codebook and subsequently demodulates the phase symbols after compensating for the corresponding IM parameters. This two-stage processing effectively decouples IM detection from phase demodulation, significantly reducing receiver complexity while maintaining reliable data recovery.

\subsubsection{Sensing Receiver Design} 
The sensing receiver, having full knowledge of the transmitted waveform, can apply matched or mismatched filtering accordingly. However, IM involving radar system parameters and phase modulation inevitably introduces undesirable RD sidelobes that can degrade sensing accuracy. For instance, IM using chirp bandwidth variations causes fluctuation in range resolution across chirps; thus, applying a fixed-size fast Fourier transform (FFT) leads to range inconsistencies over slow time, distorting Doppler estimation~\cite{temiz2023experimental}. This effect can be mitigated by employing variable-size range FFTs that adapt to the chirp bandwidth~\cite{temiz2023experimental}. Likewise, variations in the chirp center frequency in IM-FMCW induce additional phase shifts beyond those caused by target motion, increasing Doppler sidelobe levels~\cite{temiz2025radar}. These distortions can be compensated by exploiting prior knowledge of the transmitted chirp parameters.

\subsection{Receiver Design for Communication-Centric OFDM-ISAC} \label{Sec::RX_F}
\begin{figure}[t!]
    \centering
    {\includegraphics[width=0.40\textwidth]{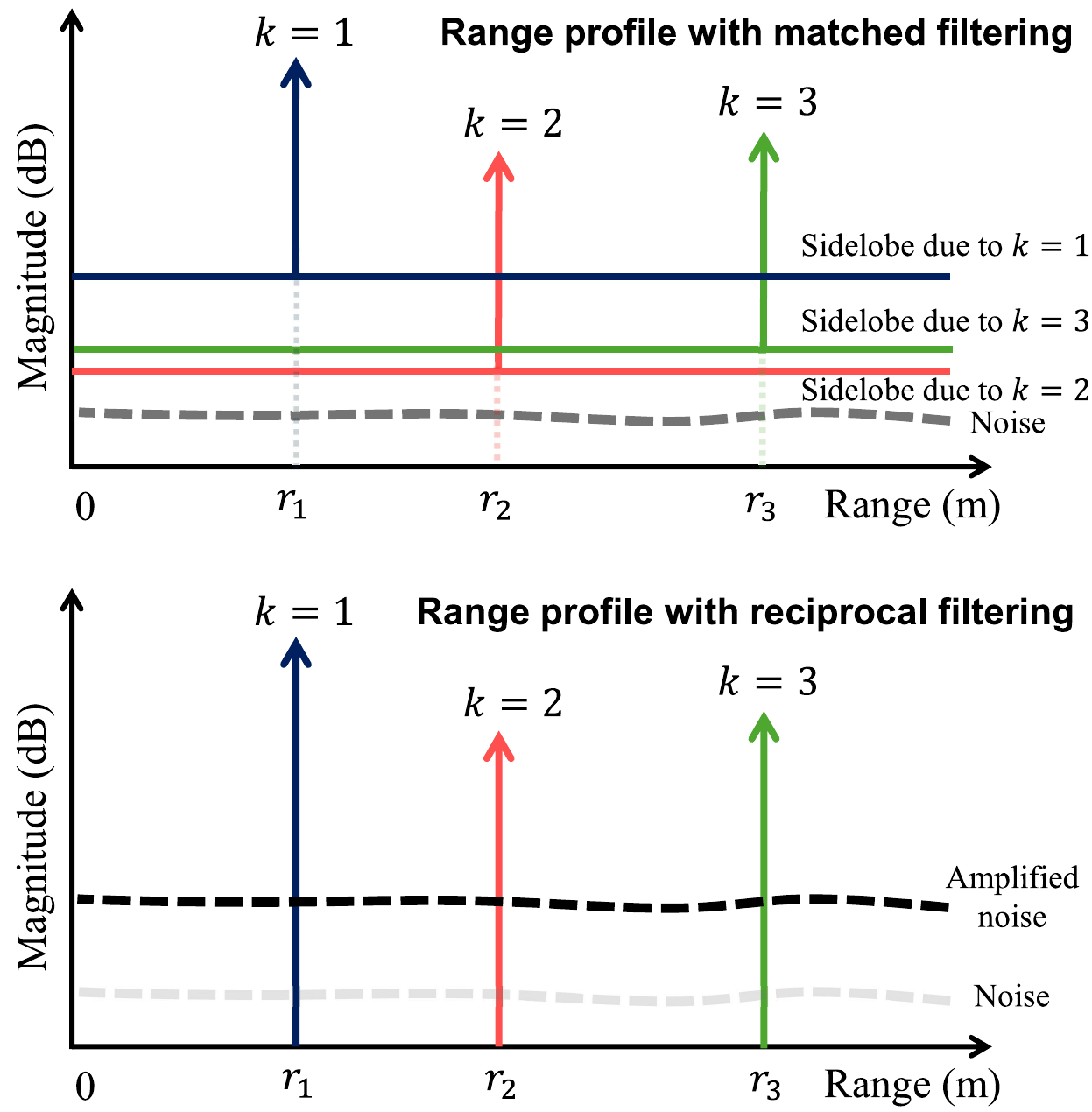}}
    \caption{Comparisons of MF and RF: OFDM-ISAC receiver processing with data payload signals.}
    \label{Fig::MFRF}
\end{figure}
Another important aspect of ISAC receiver design lies in the sensing receiver processing with data payload transmission, where the characteristics of communication payload signals directly influence sensing performance. For the communication-centric ISAC, the design of the receiver processing chain plays a critical role in determining the achievable sensing performance. Focusing on CP-OFDM-based ISAC systems, this section provides an insight on how sensing performance is governed by the receiver architecture when using data-embedded OFDM signals. 

Given that the transmitted OFDM signal $\mathbf{x} \in \mathbb{C}^{N_s}$ with $N_s$ subcarriers is modulated from an $M$-ary constellation set $\mathcal{S} = \{s_1 , s_2 , \dots , s_M\}$ with $\sum_{m=1}^M s_m = 0$ and $\frac{1}{M}\sum_{m=1}^M |s_m|^2 = 1$, a frequency-domain received signal model with $K$ targets after CP removal is given by
\begin{align}
    \mathbf{y}_r  = \mathbf{a}^T \mathbf{H} \mathbf{X} + \mathbf{z}, \label{RX_signal}
\end{align}
where $\mathbf{a} = [\alpha_{1}, \alpha_{2}, \dots, \alpha_{K}]^T \in \mathbb{C}^{K \times 1}$ denotes the complex amplitudes that incorporate the path loss and radar cross-section (RCS) of each target. The delay-channel matrix is expressed as $\mathbf{H} = [\mathbf{h}(\tau_1), \mathbf{h}(\tau_2), \dots , \mathbf{h}(\tau_K)]^T \in \mathbb{C}^{K \times N_s}$, where $\tau_{k}$ is the time-of-flight (ToF) from the ISAC transmitter to target $k$ and back to the receiver. The delay steering vector is defined as $\mathbf{h}(\tau) = \begin{bmatrix} 1, & e^{-j2 \pi \Delta f \tau}, & \dots, & e^{-j2 \pi (N_s-1)\Delta f \tau} \end{bmatrix}^T \in \mathbb{C}^{N_s \times 1},$ with subcarrier spacing $\Delta f = B/N_s$, where $B$ denotes the signal bandwidth. The transmitted signal $\mathbf{X}$ is the diagonal matrix of $\mathbf{x}$, i.e., $\mathbf{X} = \text{diag}(\mathbf{x})$. Finally, $\mathbf{z}$ denotes the AWGN at the sensing receiver, following $\mathbf{z} \sim \mathcal{CN}(0,\sigma^2 \mathbf{I}_{N_s})$.

\vspace{0.5ex}
\subsubsection{Matched Filtering Receiver}
\begin{table}[t!]
\centering
    \fontsize{8.3}{14}\selectfont
    \caption{Values of $\mu_4$ and $\nu_{-2}$ for $M$-QAM, and $M$-APSK modulation schemes. The APSK modulation formats with code rates of 2/3 are defined in the DVB standard~\cite{etsi2020dvbs2x}. } 
        \label{table2}
    \begin{tabular}{>{\centering\arraybackslash} m{3em} | >{\centering\arraybackslash} m{3em}  | >
    {\centering\arraybackslash} m{3em}  | >{\centering\arraybackslash} m{3.5em} | >{\centering\arraybackslash} m{3.5em} | >{\centering\arraybackslash} m{3.5em}}
        \toprule
             & 16QAM & 64QAM & 256QAM & 16APSK & 32APSK \\ 
            \hline
        \midrule
        $\mu_4 $   & 1.32 & 1.38 & 1.40 & 1.25 & 1.41    \\ \hline
        $\nu_{-2} $  & 1.89 & 2.69 & 3.44 & 2.50 & 3.23  \\ 
        \bottomrule
    \end{tabular}
\end{table}

As the classical radar receiver architecture, matched filtering remains the most widely adopted radar processing due to its property to yield the optimal SNR at the output. In its basic form, MF is implemented through time-domain cross-correlation between the received echoes and the reference transmitted signal~\cite{cook2012radar}. Alternatively, in CP-OFDM systems, MF processing can be performed efficiently in the frequency-domain after cyclic prefix removal~\cite{mercier2020comparison}. An important characteristic of MF is that its output follows the AF of the transmitted signal. Consequently, multiple target reflections appear as shifted and scaled replicas of the AF pattern in the RD domain. This property provides a direct mapping between waveform characteristics and sensing performance.

The MF receiver in CP-OFDM is implemented by multiplying the received signal by the conjugate of the reference transmitted signal. The output of MF, $\mathbf{y}_{\text{MF}} = \mathbf{y}_r\mathbf{X}^H$, becomes
\begin{align}
    \mathbf{y}_{\text{MF}} = \mathbf{a}^T \mathbf{H} |\mathbf{X}|^2 + \mathbf{z}_{\text{MF}},
    \label{Eqn::MF}
\end{align}
where $\mathbf{z}_{\text{MF}}$ follows the same noise characteristics as $\mathbf{z}$, assuming a unit-variance constellation. From \eqref{Eqn::MF}, it is observed that the MF receiver preserves the noise power, while each target channel is weighted by the squared magnitude of the corresponding TX subcarrier. This indicates that an instantaneous non-flat transmit spectrum induces sidelobes in the AF, which manifest as unwanted artifacts in the multi-target range profile, as illustrated in Fig.~\ref{Fig::MFRF}. Accordingly, the effective SINR of the MF output for target $k$ can be expressed as \cite{geiger2025constellation, han2025constellation}
\begin{align}
    \mathrm{SINR}_{\text{MF},k} = \frac{N \cdot |\alpha_{k}|^2}
     {(\mu_{4}-1) \cdot \sum_{j\neq k}^K|\alpha_{j}|^2 + \sigma^2}. \label{Theo_eq2}
\end{align}
As discussed in Section~\ref{Sec::Const}, the MF output performance is directly influenced by the fourth-order moment $\mu_{4}$ of the modulation constellation defined in \eqref{Eqn::mu}. A lower $\mu_{4}$ value yields reduced sidelobe interference, thereby improving the effective SINR and enhancing ISAC ranging performance.

\vspace{0.5ex}
\subsubsection{Mismatched Filtering Receiver}
Traditionally, a mismatched filter in radar receivers has been developed to overcome the limitations of the MF, effectively suppressing range sidelobes caused by imperfect TX waveforms~\cite{mcaulay1971optimal}. The MMF receiver also can be employed in OFDM-based sensing, offering significantly improved sidelobe suppression at the cost of some SNR loss compared with the MF receiver.
 
\textit{a) Reciprocal Filtering:}  
Reciprocal filtering (RF) is a representative MMF technique widely adopted in OFDM-based radar sensing~\cite{sturm2011waveform, keskin2025fundamental, wojaczek2018reciprocal, han2023sub}. It is also known as modulation-symbol-based processing~\cite{sturm2011waveform} or a ZF-type receiver~\cite{mercier2020comparison}. The key idea of the reciprocal filtering receiver is to eliminate the data dependency in the received signal through element-wise division by the TX symbols. This operation is implemented as $\mathbf{y}_{\text{RF}} = \mathbf{y}_r\mathbf{X}^{-1}$, yielding
\begin{align}
    \mathbf{y}_{\text{RF}} = \mathbf{a}^T \mathbf{H} + \mathbf{z}_{\text{RF}}.
\end{align}
It is observed that the output of the RF receiver is free from the effects of random signaling in the signal term, implying that it eliminates sidelobes from other delay sources. However, the RF receiver introduces noise amplification, leading to SNR degradation that depends on the modulation constellation, as illustrated in Fig.~\ref{Fig::MFRF}. The post-processed noise $\mathbf{z}_{\text{RF}}$ remains zero-mean but its variance is reshaped due to the reciprocal operation, following $\mathbf{z}_{\text{RF}} \sim \mathcal{CN}(0, \nu_{-2} \cdot \sigma^2 \mathbf{I}_{N_s})$~\cite{wojaczek2018reciprocal, keskin2025fundamental, han2025sensing}, where $\nu_{-2} \geq 1$ denotes the inverse second-order moment of the modulation constellation $\mathcal{S}$, given by
\begin{equation}
    \nu_{-2} = \frac{1}{M}\sum_{m=1}^{M} |s_m|^{-2}.
    \label{Eqn::nu}
\end{equation}
Since the RF receiver eliminates sidelobe interference from multiple targets, the resulting SNR for target $k$ can be expressed as
\begin{align}
    \mathrm{SNR}_{\text{RF},k} = \frac{N \cdot |\alpha_{k}|^2}
     {\nu_{-2} \cdot \sigma^2}.
     \label{Eqn::SNR_RF}
\end{align}
It is worth noting that the RF receiver exhibits a different dependence on the modulation constellation compared with the MF receiver, implying that the constellation design for ISAC differs between the two receiver architectures \cite{han2025constellation}. In Table~\ref{table2}, the values of $\mu_4$ and $\nu_{-2}$ for QAM and APSK modulation schemes are summarized, providing insight into the sensing performance associated with specific receiver processing using data payload signals.

\textit{b) Linear MMSE Receiver:}  
The linear MMSE (LMMSE)-type receiver, also known as Wiener filtering, is another form of MMF used in radar sensing, and it has also been widely adopted for channel estimation in wireless communication systems~\cite{hung2009pilot}. It is generally defined as $\mathbf{y}_{\text{LMMSE}} = \mathbf{y}_r\left(|\mathbf{X}|^2 + (\mathbf{a}^H \mathbf{a}/\sigma^2)\mathbf{I}\right)^{-1}\mathbf{X}.$
Importantly, the LMMSE receiver provides a balanced trade-off between the MF and RF receivers under non-unit-amplitude constellation. While the MF maximizes output SNR but suffers from high sidelobes, and the RF suppresses sidelobes at the cost of significant noise amplification, the LMMSE receiver adjusts its filtering behavior according to the instantaneous input SNR. As a result, it achieves effective sidelobe suppression while minimizing SNR loss, yielding the improved dynamic range compared to the MF and RF receivers~\cite{keskin2025fundamental}. Although this superiority makes the LMMSE receiver a promising solution for OFDM-based ISAC systems operating under varying SNR conditions, its implementation requires prior knowledge of the target SNRs. The work in~\cite{keskin2025fundamental} provides a practical approach to realizing the LMMSE receiver for OFDM-based sensing systems.

As a summary of sensing receiver design, it is important to note that the choice between MF and MMF receivers depends on several factors, including the number of targets and their SNRs, clutter interference level, modulation constellation, and receiver complexity~\cite{keskin2025fundamental, rodriguez2023supervised, quirini2024clutter}.

\subsection{Joint Receiver Design}
\begin{figure}[t!]
    \centering
    {\includegraphics[width=0.40\textwidth]{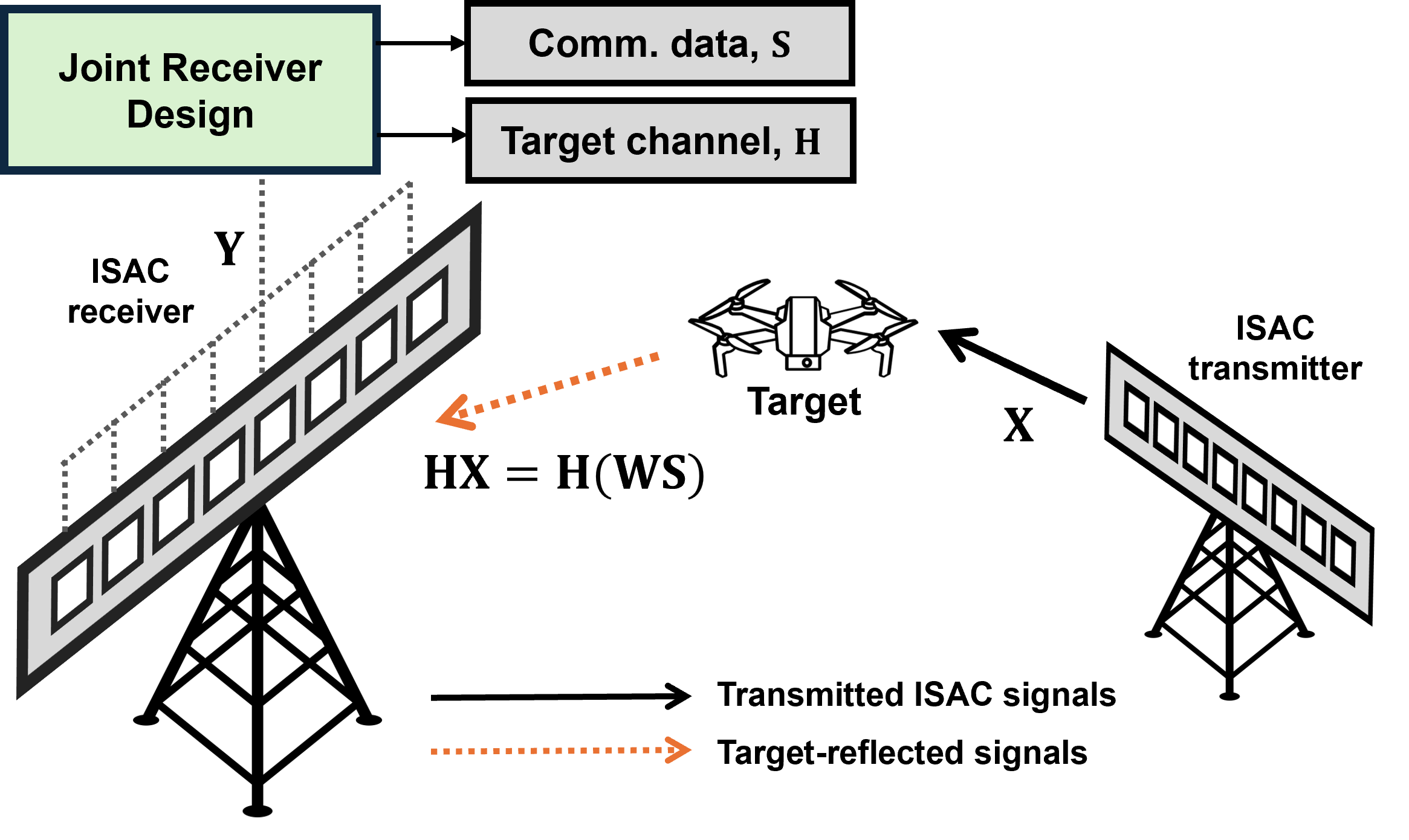}}
    \caption{Illustration of a joint receiver design for a bi-static ISAC scenario without prior knowledge of the transmitted ISAC signal at the receiver.}
    \label{Fig::RX2}
\end{figure}
This subsection examines joint receiver design in a bi-static ISAC scenario where the receiver lacks knowledge of the transmitted ISAC signal, as illustrated in Fig.~\ref{Fig::RX2}. In such a setup, either the transmitter or receiver can be the base-station (BS) or CU, covering uplink (UL), downlink, or BS-to-BS bi-static configurations.

Existing studies have primarily explored joint receiver design under separate radar and communication transmissions, assuming that the two independent signals arrive synchronously~\cite{dong2023joint, yu2024addressing, yu2025framework}. This problem is often addressed using successive interference cancellation: first, the receiver detects communication data based on known communication channels while treating radar echoes as interference, then subtracts the reconstructed communication component to estimate the radar response~\cite{chiriyath2015inner}. From the authors’ perspective, such schemes correspond to separate transmission rather than unified ISAC signaling, representing a special case within the broader joint receiver design framework.

Given that an $N_t$-antenna ISAC transmitter sends a unified ISAC signal $\mathbf{X} \in \mathbb{C}^{N_t \times L}$, which is reflected by $K$ targets, the received baseband signal at an ISAC receiver equipped with $N_r$ antennas over $L$ symbol intervals can be expressed as  
\begin{equation}
    \mathbf{Y} = \mathbf{H}\mathbf{X} + \mathbf{Z},
    \label{eq:rx_model}
\end{equation}
where $\mathbf{H} \in \mathbb{C}^{N_r \times N_t}$ denotes the channel matrix including sensing target parameters and $\mathbf{Z}$ represents additive noise and potential clutter. Assuming the transmitted signal is unprecoded, i.e., $\mathbf{W} = \mathbf{I}_{N_t}$, we have $\mathbf{X} = \mathbf{S} \in \mathbb{C}^{N_t \times L}$. The ISAC receiver aims to jointly detect the communication data $\mathbf{S}$ and estimate the radar channel $\mathbf{H}$. This joint receiver design problem can be formulated as~\cite{hu2024isac, valiulahi2025isac}  
\begin{equation}
    \hat{\mathbf{H}}, \hat{\mathbf{S}} = \underset{\mathbf{H},\,\mathbf{S}}{\mathrm{argmin}} \; \|\mathbf{Y} - \mathbf{H}\mathbf{S}\|_F^2.
    \label{eq:rx_prob}
\end{equation}
It should be noted that the joint estimation problem in~\eqref{eq:rx_prob} does not yield a unique solution unless additional constraints or prior knowledge of $\mathbf{H}$ and $\mathbf{S}$ are incorporated.

Under these conditions, conventional interference-cancellation and pilot-assisted receivers fail, as they rely on predefined training sequences or reference links to decouple sensing and communication components. This limitation has led to the emergence of blind estimation frameworks capable of jointly recovering communication data and radar parameters, such as time delay, Doppler shift, and angles of arrival and departure, directly from the received echo signals~\cite{bigdeli2022noncoherent, valiulahi2025isac, valiulahi2025isac_2}. The resulting inference task is inherently bilinear and nonconvex, since both the transmitted data symbols and the channel responses are unknown and multiplicatively intertwined. Without appropriate structural prior knowledge, such bilinear inverse problems are ill-posed, precluding unique or stable recovery.  

To overcome this challenge, recent advances leverage atomic norm minimization (ANM) and its lifted variants (LANM) to impose low-rank and sparsity-promoting regularization, thereby transforming the blind recovery task into a convex optimization problem with theoretical guarantees. ANM provides a gridless approach to sparse super-resolution, capturing continuous delay-Doppler-angle features without discretization errors. Building on this, LANM introduces a structured model in which the unknown transmit waveform lies within a known low-dimensional subspace, encoded by a dictionary or compression matrix. This framework enables simultaneous estimation of radar scene parameters and communication data, solvable via SDR under certain coherence and separation conditions~\cite{valiulahi2025isac, valiulahi2025isac_2}. By exploring different dictionary structures, LANM-based receivers offer tunable trade-offs between estimation accuracy, sample efficiency, and computational complexity \cite{valiulahi2025isac_2}. Although these blind receivers incur higher algorithmic cost and may exhibit suboptimal accuracy compared with pilot-assisted counterparts, they establish a powerful foundation for pilot-free ISAC operation, particularly in scenarios where pilot signaling is infeasible, contaminated, or spectrally inefficient. Consequently, ANM and related blind recovery approaches represent a critical step toward high-resolution and spectrum-efficient ISAC receiver architectures.

Learning-based ISAC receivers for joint data and target parameter estimation have recently emerged to address computational complexity and performance degradation in time-varying environments. A representative data-driven approach is a two-stage transformer-based receiver that performs sliding-window symbol detection followed by MUSIC-based angle–delay estimation, achieving robust performance with minimal training and strong generalization under dynamic channels~\cite{hu2024isac}. In parallel, a model-driven ISAC receiver proposed in~\cite{jiang2024isac} unrolls classical estimation algorithms for passive sensing and data recovery into trainable layers, enabling end-to-end learning with interpretable structure. This hybrid framework demonstrates significant gains in both data demodulation and sensing parameter estimation compared with conventional signal demodulation methods, leveraging both pilot- and data-assisted sensing.

\section{Full-Duplex ISAC Transceiver Design} \label{Sec::FD-ISAC}
The term (in-band) full-duplex~\cite{smida2024band} refers to wireless systems in which a transceiver simultaneously transmits and receives (STAR) on the same frequency band. At the physical layer, full-duplex (FD) operation introduces the inherent challenge of self-interference (SI)~\cite{Erricolo11145123}, i.e., the FD transceiver’s own transmission interferes with its reception. This phenomenon is reasonably ignored in the previous sections, as in much of the cited ISAC literature, when the focus is on the integrated performance of communication and sensing, as well as the trade-off or synergy between these functionalities.

As illustrated in Fig.~\ref{fig:FDsystems}, full-duplex ISAC pertains only to monostatic scenarios, in which a base station-like transceiver's receiver operates as a radar to extract information about targets or the surrounding environment, while its transmitter simultaneously sends radar and/or communication signals. Thus, the same MIMO antenna system is used simultaneously in downlink and uplink over the same frequency band. A full-duplex MIMO system is usually pseudo-bistatic, in the sense that the transmit and receive arrays may be physically separate but located nearby, or a single array may be divided into transmit and receive sub-arrays. Nevertheless, all pseudo-bistatic configurations, where STAR operation takes place within the same site, are regarded as monostatic, and the direct interference from a CU or another ISAC transceiver in true bistatic setups is not considered as SI. The presence of SI inherently couples the transmit and receive designs, both in terms of its mitigation and the overall ISAC operation. A monostatic setup can, therefore, jointly design both sides in a non-distributed manner, as discussed in this section.

\subsection{Full-Duplex Integrated Sensing and Communications}

Full-duplex ISAC scenarios can be characterized into two classes according to Fig.~\ref{fig:FDsystems} based on whether the sensing function is integrated with downlink or uplink communication function. The fundamental difference between the classes comes from self-interference exploitation in the spirit of Section~\ref{Sec::SLP}: In downlink ISAC, the harmful SI signal is essentially a short-delay multipath component within the useful sensing signal, while the SI is only harmful for sensing in uplink ISAC. In principle, it would be possible to imagine also a hybrid of the classes, which integrates sensing and communications in both downlink and uplink simultaneously, but research on such scenarios is still very limited.

\subsubsection{Monostatic Downlink Sensing}
As illustrated on the scenario (a) of Fig.~\ref{fig:FDsystems}, full-duplex downlink ISAC means using downlink transmissions for sensing in a monostatic manner. The previous received signal models are updated to include self-interference through the self-interference channel $\mathbf{H}_{si}$ as
\begin{equation}
\mathbf{Y} = \mathbf{H}\mathbf{X} = (\mathbf{H}_{r} + \mathbf{H}_{si})\mathbf{X},
\label{eq:FDsignalDL}
\end{equation}
where $\mathbf{X}$ and $\mathbf{Y}$ are transmitted and received signals, respectively, while $\mathbf{H}_{r}$ is the radar channel including target parameters. The system applies BLP to generate the ISAC transmitted waveform as per (\ref{Eqn::Sig_Model}). The system aims at transmitting communication data $\mathbf{S}_{c}$ in $\mathbf{X}$ to downlink CUs, while simultaneously estimating from $\mathbf{Y}$ the target channel $\mathbf{H}_{r}$ within the combined channel $\mathbf{H}$ by either mitigating the effect of $\mathbf{H}_{si}\mathbf{X}$ or taking it into account in joint transceiver design.

\subsubsection{Bistatic Uplink Sensing}
\begin{figure}[t!]
    \centering
    {\includegraphics[width=0.45\textwidth]{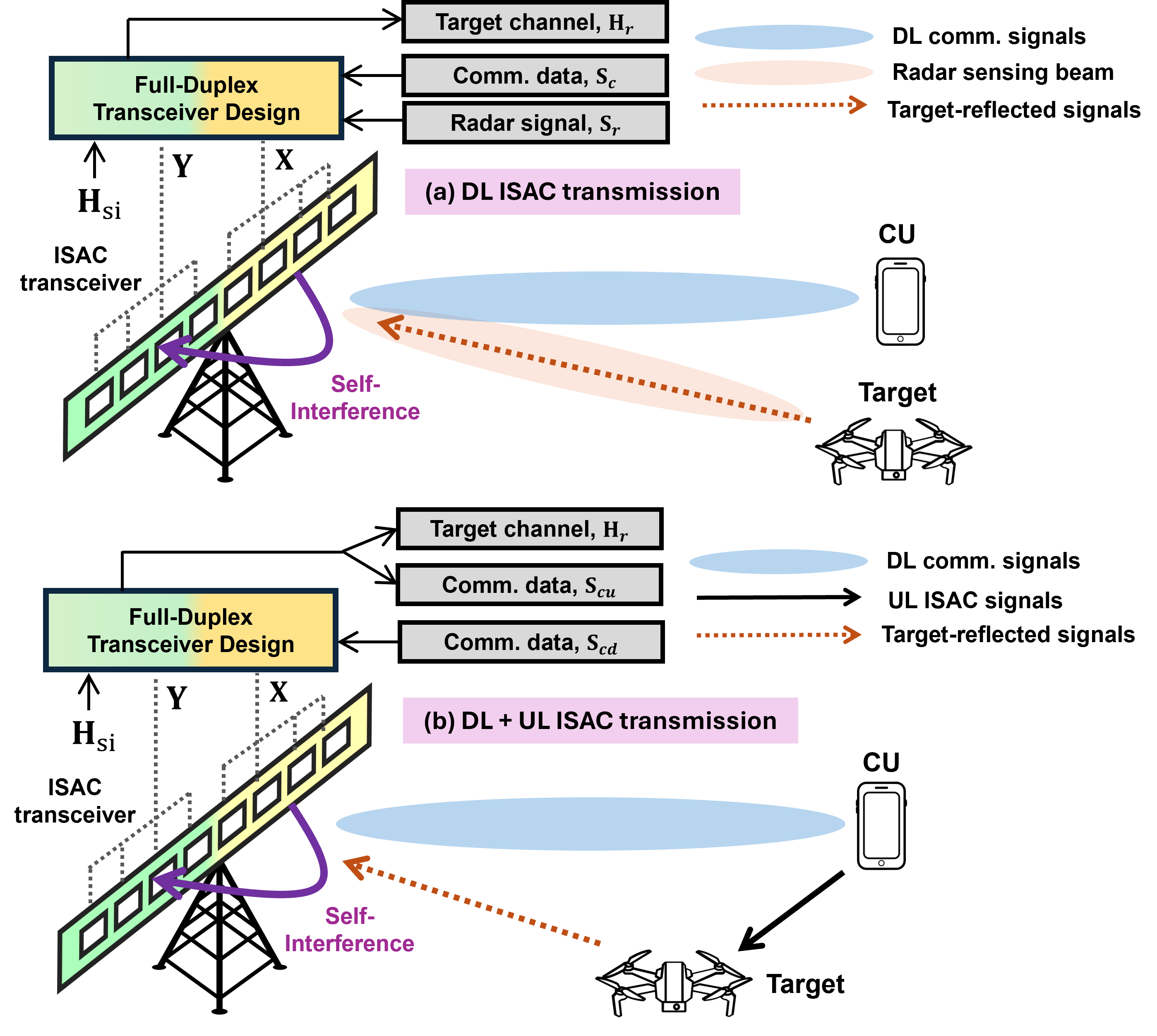}}
    \caption{Scenarios for full-duplex integrated communication and sensing, which are inherently subject to self-interference within the MIMO transceiver.}
    \label{fig:FDsystems}
\end{figure}

As illustrated on the scenario (b) of Fig.~\ref{fig:FDsystems}, full-duplex uplink ISAC means using uplink transmissions for sensing in a bistatic manner while simultaneous downlink communication transmissions cause self-interference through the self-interference channel $\mathbf{H}_{si}$. The model of the received signal is updated as
\begin{equation}
\mathbf{Y} = \mathbf{H}_{r}\mathbf{X}_{cu} + \mathbf{H}_{si}\mathbf{X},
\label{eq:FDsignalUL}
\end{equation}
where $\mathbf{X}_{cu}$ and $\mathbf{X}$ are signals transmitted by the CUs and the ISAC transceiver, respectively. Like above, the system generates the transmitted waveform as $\mathbf{X} = \mathbf{W}_{c}\mathbf{S}_{cd} + \mathbf{W}_{r}\mathbf{S}_{r}$ per (\ref{Eqn::Sig_Model}), where $\mathbf{S}_{cd}$ is downlink communication signals and $\mathbf{S}_{r}$ equals to zero. The system aims at transmitting communication data $\mathbf{S}_{cd}$ in $\mathbf{X}$ to downlink CUs while simultaneously estimating from $\mathbf{Y}$ the target channel $\mathbf{H}_{r}$ under the SI signal $\mathbf{H}_{si}\mathbf{X}$ by either mitigating the effect thereof or taking it into account in joint transceiver design. The system may also aim at simultaneously receiving an uplink communication data signal $\mathbf{S}_{cu}$ transmitted in $\mathbf{X}_{cu}$ so that it operates in a full-duplex manner also from the plain communications perspective.

\subsubsection{Hybrid Downlink--Uplink Sensing}

A comprehensive full-duplex ISAC system might also perform simultaneously both downlink monostatic and uplink bistatic sensing with both downlink and uplink communication data transmission. The signal model would be a combination of the above:
\begin{equation}
\mathbf{Y} = \mathbf{H}_{ru}\mathbf{X}_{cu} + \mathbf{H}\mathbf{X} = \mathbf{H}_{ru}\mathbf{X}_{cu} + (\mathbf{H}_{rd} + \mathbf{H}_{si})\mathbf{X},
\label{eq:FDsignalHybrid}
\end{equation}
which now contains separate radar channels $\mathbf{H}_{ru}$ and $\mathbf{H}_{rd}$ for uplink and downlink sensing, respectively. In such full-duplex hybrid ISAC systems, the monostatic downlink sensing part becomes particularly difficult because it needs to perform under both the self-interference signal $\mathbf{H}_{si}\mathbf{X}$ and the interference signal $\mathbf{H}_{ru}\mathbf{X}_{cu}$ from uplink sensing. Developing the feasibility of the concept is proposed as a quest for future research.

\subsection{Self-Interference Mitigation}

The original classification of self-interference mitigation schemes in full-duplex MIMO relaying~\cite{Riihonen5985554} holds also for full-duplex MIMO-ISAC systems: Physical isolation, time-domain cancellation, and spatial-domain suppression; all these can be passive or active means. In fact, the mitigation schemes surveyed next could be applied with any full-duplex MIMO transceiver, because their purpose is to minimize (the effect of) the self-interference signal $\mathbf{H}_{si}\mathbf{X}$ in (\ref{eq:FDsignalDL})--(\ref{eq:FDsignalHybrid}) by modifying $\mathbf{H}_{si}$ and $\mathbf{X}$ into $\hat{\mathbf{H}}_{si}$ and $\hat{\mathbf{X}}$, respectively, before transmission or by modifying $\mathbf{Y}$ into $\hat{\mathbf{Y}}$ before processing for sensing, while limiting the collateral effect to sensing and communications. Typically, the desirable level of SI suppression in full-duplex operation exceeds 100 dB \cite{smida2024band}.

\subsubsection{Physical Isolation}

It would be highly beneficial to design the full-duplex array architecture to begin with in such a way that $\mathbf{H}_{si} \to \hat{\mathbf{H}}_{si} \approx \mathbf{0}$. However, in practice, physical isolation schemes can only somewhat reduce the SI leakage at best through some $\hat{\mathbf{H}}_{si}$ with lower gain. Using the same antenna element for transmitting and receiving is feasible in full-duplex SISO-ISAC, where a passive circulator or an equivalent active component (of which there are many variants) allows a degree of transmitter-receiver isolation. However, in full-duplex MIMO-ISAC, such components cannot mitigate inter-antenna interference, even if intra-antenna interference is suppressed. Thus, full-duplex MIMO-ISAC arrays are usually implemented with two separate arrays or at least sub-arrays for transmitting and receiving. The (sub-)array separation enables passive isolation through propagation distance, element directivity, placing isolating or destructively resonating materials, and obstacles between the (sub-)arrays as well as active means such as meta-materials~\cite{Gong10520298,Gong10335685} to control the SI coupling.

\subsubsection{Time-Domain Cancellation}

Time-domain cancellation refers to all subtractive means by which
\begin{equation}
\hat{\mathbf{Y}} = \mathbf{Y} - \tilde{\mathbf{H}}_{si}\tilde{\mathbf{X}},
\end{equation}
where $\tilde{\mathbf{H}}_{si}$ and $\tilde{\mathbf{X}}$ are estimates of $\mathbf{H}_{si}$ and $\mathbf{X}$, respectively. Ideally the self-interference signal $\mathbf{H}_{si}\mathbf{X}$ in (\ref{eq:FDsignalDL})--(\ref{eq:FDsignalHybrid}) would then disappear 
from $\hat{\mathbf{Y}}$ without any collateral effect to sensing and communications. However, in practice, $\tilde{\mathbf{H}}_{si}$ is only an imperfect estimate of the true channel. Moreover, although $\mathbf{X}$ is theoretically known as $\tilde{\mathbf{X}}$, transmitter hardware impairments, such as nonlinear distortion, phase noise and offset, and transmitter noise, make $\tilde{\mathbf{X}}$ deviate from the actual transmitted signal $\mathbf{X}$. Subtractive cancellation can be implemented at analog or digital baseband, intermediate frequency band and radio frequency band within the transceiver chains. Nevertheless, implementations outside the digital baseband are generally prohibitively complex for full-duplex MIMO-ISAC systems, as they would require a dedicated electronic cancellation circuit between every pair of transmit and receive antennas.

\subsubsection{Spatial-Domain Suppression}

Spatial-domain suppression refers to schemes that modify the transmit and receive beamforming by transmitting $\hat{\mathbf{X}} = \mathbf{W}_{tx}\mathbf{X}$ and receiving
\begin{align}
\hat{\mathbf{Y}} 
&\overset{\phantom{(\ref{eq:FDsignalDL})}}{=} \mathbf{W}_{rx}\mathbf{Y}\\ 
&\overset{(\ref{eq:FDsignalDL})}{=} \mathbf{W}_{rx}\mathbf{H}_{r}\mathbf{W}_{tx}\mathbf{X} 
+ \mathbf{W}_{rx}\mathbf{H}_{si}\mathbf{W}_{tx}\mathbf{X}
\nonumber\\
&\overset{(\ref{eq:FDsignalUL})}{=} \mathbf{W}_{rx}\mathbf{H}_{r}\mathbf{X}_{cu} 
+ \mathbf{W}_{rx}\mathbf{H}_{si}\mathbf{W}_{tx}\mathbf{X}
\nonumber\\
&\overset{(\ref{eq:FDsignalHybrid})}{=} \mathbf{W}_{rx}\mathbf{H}_{ru}\mathbf{X}_{cu} + \mathbf{W}_{rx}\mathbf{H}_{rd}\mathbf{W}_{tx}\mathbf{X} 
+ \mathbf{W}_{rx}\mathbf{H}_{si}\mathbf{W}_{tx}\mathbf{X}
\nonumber
\end{align}
instead of~(\ref{eq:FDsignalDL})--(\ref{eq:FDsignalHybrid}), where $\mathbf{W}_{tx}$ and $\mathbf{W}_{rx}$ are corresponding spatial filtering matrices. The objective is to spatially suppress the last term in any of the above variations such that $\mathbf{W}_{rx}\mathbf{H}_{si} \approx \mathbf{0}$, $\mathbf{H}_{si}\mathbf{W}_{tx} \approx \mathbf{0}$, $\mathbf{W}_{rx}\mathbf{H}_{si}\mathbf{W}_{tx} \approx \mathbf{0}$. If achieving exact nulls is not feasible, these products should at least be minimized according to an appropriate metric, while simultaneously ensuring that the radar channels $\mathbf{H}_{r}$, $\mathbf{H}_{ru}$, or $\mathbf{H}_{rd}$ are affected as little as possible, such that sensing performance remains intact when accounting for the spatial filtering applied by $\mathbf{W}_{tx}$ and $\mathbf{W}_{rx}$. Here, downlink ISAC and uplink ISAC are fundamentally different in the sense that the receive filter $\mathbf{W}_{rx}$ affects obviously both, whereas the transmit filter $\mathbf{W}_{tx}$ affects only the former, so that $\mathbf{H}_{si}\mathbf{W}_{tx} \approx \mathbf{0}$ is reasonable.

\subsection{Full-Duplex ISAC Transceiver Design}

The joint design of full-duplex MIMO transceivers is one of the least researched branch of ISAC studies. It is the next step from the spatial-domain suppression described above into designs that optimize communications and sensing performance explicitly taking into account the self-interference. Such multi-objective optimization problems can be expressed with the same unified structure as in (\ref{Eqn::P1}), but the normalized sensing metric $\tilde{f}_r(\mathbf{X})$ needs to model the self-interference per~(\ref{eq:FDsignalDL})--(\ref{eq:FDsignalHybrid}) or additional constraints $c_i(\mathbf{H}_{si}\mathbf{X}) \leq C_i$ need to be introduced for limiting the effect of self-interference.

A prominent solution~\cite{Barneto9933894} is based on spatial-domain suppression to transmit $\hat{\mathbf{X}} = \mathbf{W}_{tx}\mathbf{X}$ and beampattern matching at $M$ angular samples per Section~\ref{Sec::TX_B}, for which the sensing performance metric in \eqref{Eqn::Beam} is updated as follows:
\begin{equation}
    f_r(\mathbf{X}) = \frac{1}{M} \sum_{i=1}^M \left|\alpha P(\theta_i) - \mathbf{a}^H(\theta_i) \mathbf{W}_{tx}\mathbf{R}_{\mathrm{X}} \mathbf{W}_{tx}^H \mathbf{a}(\theta_i) \right|^2,
\end{equation}
while $\mathbf{W}_{tx}$ is chosen such that $\mathbf{H}_{si}\mathbf{W}_{tx} = \mathbf{0}$. The SI suppression is achieved using the Moore--Penrose pseudoinverse $\mathbf{H}_{si}^+$ of the SI channel $\mathbf{H}_{si}$ as $\mathbf{W}_{tx} = \mathbf{I} - \mathbf{H}_{si}^H(\mathbf{H}_{si}^+)^H$ due to the identity $\mathbf{H}_{si} = \mathbf{H}_{si}\mathbf{H}_{si}^H(\mathbf{H}_{si}^+)^H$. Accordingly, the solution matches the designed full-duplex beampattern with the pre-defined desired beampattern $P(\theta)$ while eliminating SI. Nevertheless, the original solution in~\cite{Barneto9933894} is a bit more involved due to the considered hybrid analog--digital array architecture at mm-wave frequencies.

\subsection{Recent Advances in Full-Duplex MIMO-ISAC}
The full-duplex capability has been recognized as an enabler for ISAC~\cite{Barneto9363029}, although it is already a necessity for the downlink ISAC, where the SI is unavoidable as explained above. On the other hand, the full-duplex capability in itself is considered an essential technology in 6G evolution~\cite{Smida10158724}. Full-duplex ISAC is further surveyed in~\cite{Du11050889} and, with MIMO transceivers, in \cite{Smida10769781}. The development of general full-duplex MIMO transceivers~\cite{Kolodziej10460255} also facilitates full-duplex MIMO-ISAC, e.g., to reduce complexity and facilitate distributed processing~\cite{Kolodziej10159257} as well as to implement full-duplex wideband MIMO~\cite{Islam9789966} and massive MIMO systems~\cite{Alexandropoulos9933358,Le-Ngoc10483101,Mohammadi10684260}. Some state-of-the-art works consider also ISAC or, in other words, multi-function systems explicitly~\cite{Kolodziej9363026}, although there is much room for original research.

The applications of full-duplex ISAC are currently rapidly emerging. The research in~\cite{Chen11173435} develops a full-duplex MIMO BS for near-ground precipitation sensing. Some state-of-the-art works develop spatial-domain suppression and joint transceiver design using the aforementioned sub-array configuration~\cite{Lin10813460,Zhou11059381}, while most works on MIMO-OFDM ISAC~\cite{liyanaarachchi2023joint,Xiao10634583} still presume that the self-interference problem is implicitly solved for downlink ISAC BSs. Vehicular applications~\cite{Talha11169427} are also timely for full-duplex ISAC, where the development is progressing towards MIMO systems. As a very recent concept, fluid-antenna system (FAS) is a promising solution for a full-duplex ISAC system, where a BS communicates simultaneously with downlink and uplink users while performing target sensing~\cite{Tang11142587}.

\section{Secure ISAC Transceiver: Data Security} \label{Sec::PLS}
\begin{figure}[t!]
    \centering
    \subfigure[]{\includegraphics[width=0.24\textwidth]{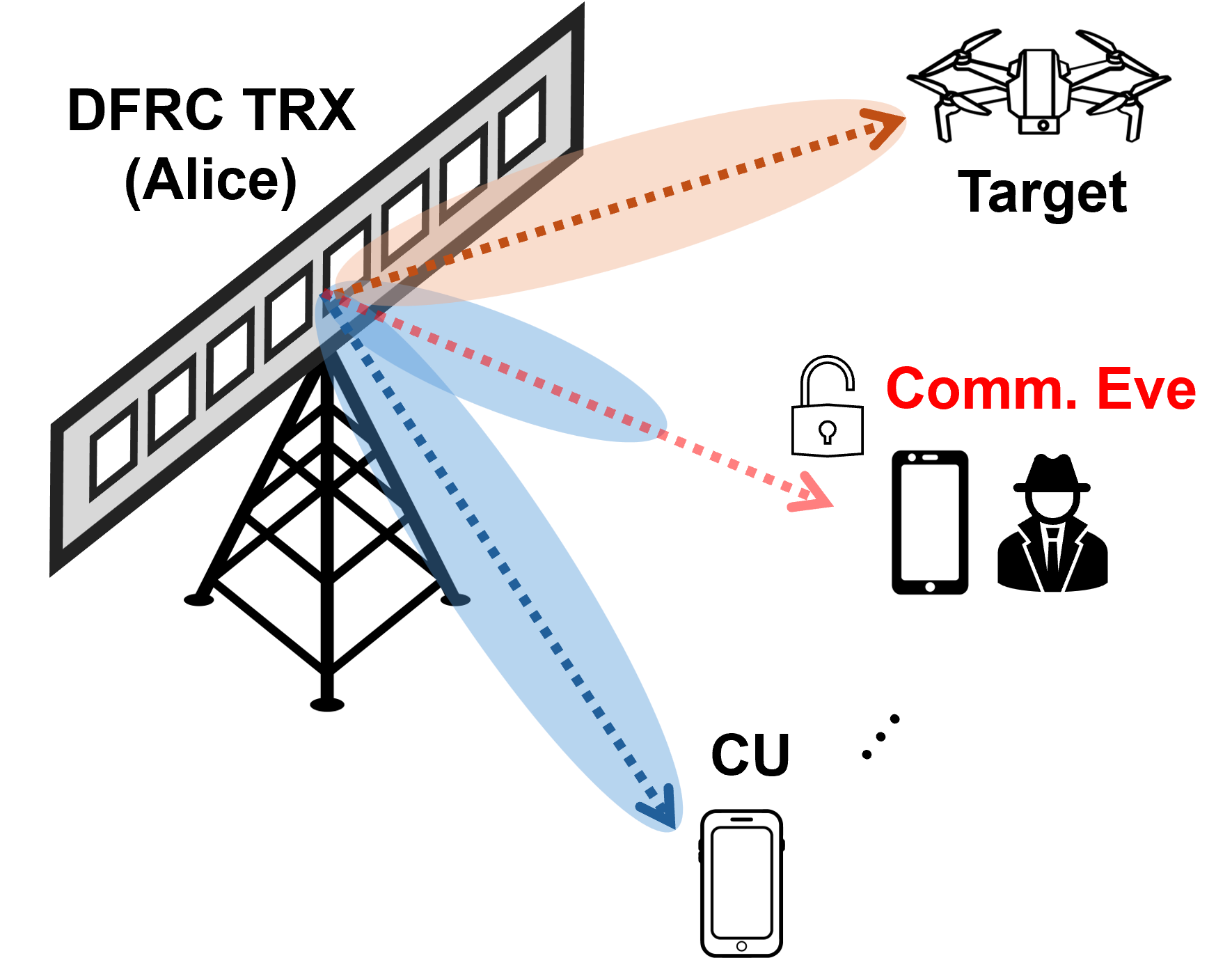}}
    \subfigure[]{\includegraphics[width=0.24\textwidth]{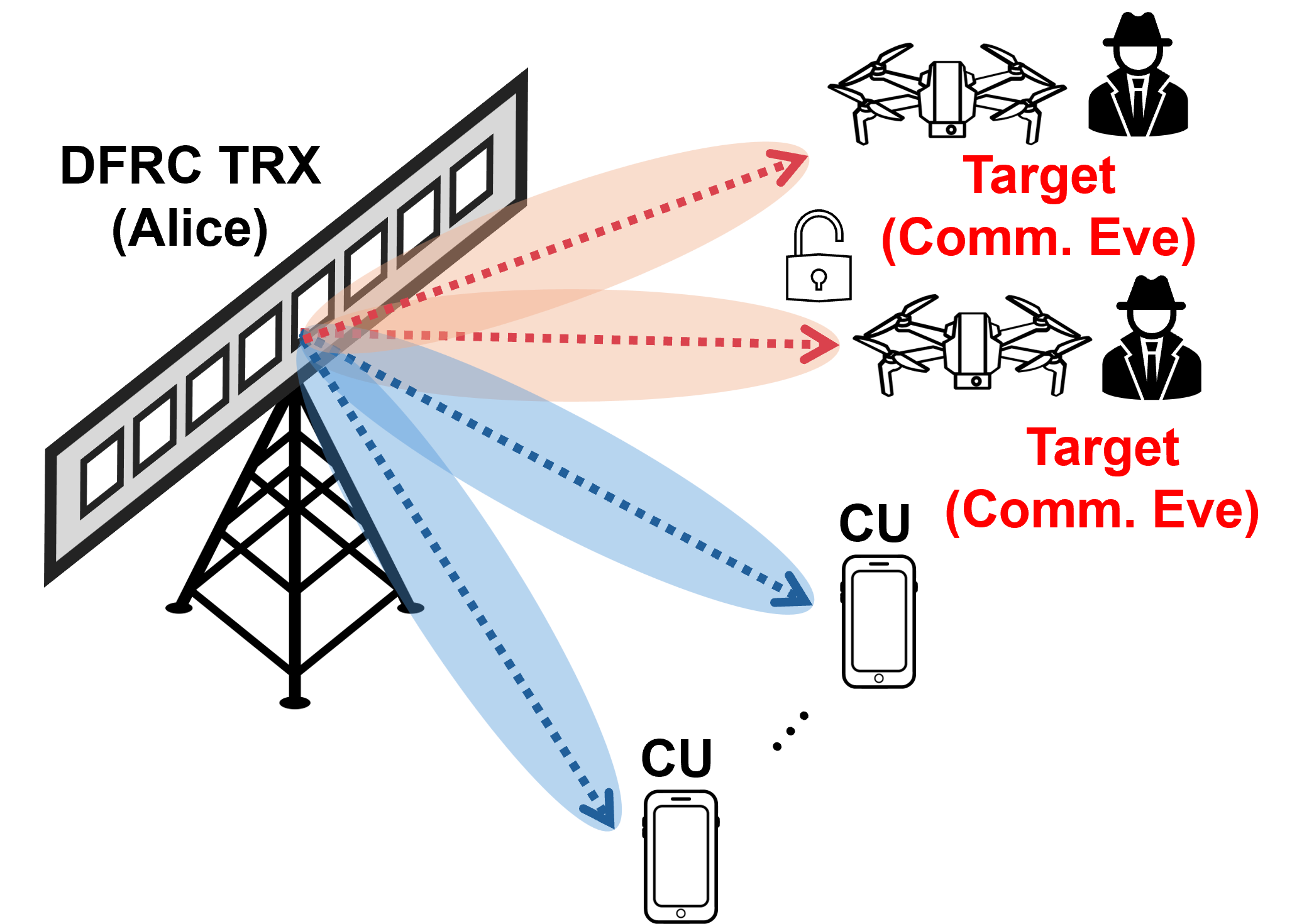}}
    \caption{ISAC security threat scenarios with communication data eavesdroppers. (a) External communication eavesdropper in the ISAC coverage area. (b) Malicious sensing target acting as an data eavesdropper.}
    \label{Fig::Sec_comms}
\end{figure}
The security of information transfer in wireless communication systems has been a long-standing challenge~\cite{shiu2011physical}. PLS techniques have been extensively studied as a built-in defense mechanism complementing upper-layer encryption and authentication techniques~\cite{yang2015safeguarding}. However, unlike conventional wireless systems, ISAC introduces new data security vulnerabilities at the physical layer due to its inherent dual-functionality of communication and radar sensing. Beyond the classical scenario where an external eavesdropper (Eve) resides within the ISAC coverage area, as illustrated in Fig.~\ref{Fig::Sec_comms}(a), a unique threat arises when the sensing target itself acts as a malicious eavesdropper. In ISAC scenarios, target illumination is carried out with a data-carrying probing signal. This presents the opportunity to the target to behave as an unauthorized receiver, attempting to extract information embedded in the transmitted waveform, as shown in Fig.~\ref{Fig::Sec_comms}(b). This dual-functionality complicates the application of conventional PLS strategies, since it must simultaneously support target illumination and secure data transmission. Therefore, this ISAC data-security scenario excludes the use of large classes of classical PLS technologies such as secure beamforming and null steering, since steering nulls toward the target/Eve would result in no illumination of the target and thus inhibit the sensing functionality. Instead, the aim of the ISAC transmitter is to illuminate the target with a high-power beam while using a signal that prevents eavesdropping of the data. In this section, we discuss recent advances in data-secure ISAC transceiver design, focusing on directional modulation (DM), MIMO signaling design, and jamming functionality to counteract eavesdropping threats while maintaining reliable sensing functionality.

\subsection{ISAC Data Security with Artificial Noise}
Artificial noise (AN)-aided transmission has been established as one of the most effective PLS techniques~\cite{negi2005secret, chu2015secrecy, nguyen2016joint}. In such designs, the transmitter injects carefully structured AN into the transmitted waveform, acting as a jamming component that degrades Eve’s reception. Unlike conventional PLS, AN in data-secure ISAC transmission simultaneously supports radar sensing within the same spectral, temporal, and spatial resources, ensuring that its presence does not compromise sensing accuracy or target detection performance as well as the communication performance of the CUs~\cite{deligiannis2018secrecy, su2020secure, chen2025secure, he2024joint, li2025secure, dong2023joint_sec, ren2023robust, xu2022robust, ren2022optimal}. 

The AN-aided MU-MISO transmit signal serving $U$ single-antenna users can be expressed as
\begin{equation}
    \mathbf{X} = \mathbf{W}_c\mathbf{S}_c + \mathbf{N},
    \label{Eqn::AN}
\end{equation}
where $\mathbf{N} \in \mathbb{C}^{N_t \times L}$ denotes the AN component, and its covariance matrix is given by $\mathbf{R}_{N} = \tfrac{1}{L}\mathbf{N}\mathbf{N}^H$. Assuming a sufficiently large block length $L$, the communication data and AN are considered statistically independent. When the sensing target acts as an eavesdropper, the received signal at Eve can be modeled as
\begin{equation}
    \mathbf{y}_{E} = \beta_{E} \mathbf{a}^H(\theta)\mathbf{X} + \mathbf{z}_{E},
    \label{Eqn::Eve_comm}
\end{equation}
where $\beta_E$ denotes the path-loss coefficient, $\mathbf{a}(\theta)$ is the transmit steering vector toward direction $\theta$, and $\mathbf{z}_{E}$ represents AWGN following $\mathbf{z}_{E} \sim \mathcal{CN}(0, \sigma_E^2 \mathbf{I})$.  
For the legitimate users, the received signal model follows that presented in Section~\ref{Sec::JointSignaling}.

\vspace{0.5ex}
\subsubsection{Data Security Performance Metric}
The secrecy rate (SR) has been widely adopted as a data security metric~\cite{telatar1999capacity, hero2004secure, negi2005secret}. The SR quantifies the rate gap between the legitimate user and Eve, directly reflecting the confidentiality of information transmission. Similarly, data-secure ISAC signaling design employs the SR as the main measure of communication secrecy. From the received signal model of Eve in~\eqref{Eqn::Eve_comm}, the achievable rate at Eve is expressed as
\begin{equation}
    R_E = \log_2 \left( 1 + \frac{|\beta_E|^2 \mathbf{a}^H(\theta)\mathbf{R}_{c}\mathbf{a}}{|\beta_E|^2\mathbf{a}^H(\theta)\mathbf{R}_{N}\mathbf{a} + \sigma_E^2} \right),
    \label{Eqn::Eve_rate}
\end{equation}
where $\mathbf{R}_{c}$ denotes the covariance matrix of the communication signal. 

Using the achievable rate of the legitimate user $u$, denoted by $R_{B,u}$ in~\eqref{Eqn::Rate1}, the worst-case achievable SR is defined as~\cite{su2020secure, dong2023joint_sec}
\begin{equation}
    R_s(\mathbf{X}) = \min_u \big[ R_{B,u} - R_E \big]^+,
\end{equation}
where $[\cdot]^+$ denotes the operator $\max(\cdot,0)$. The sum SR also can be exploited to describe the overall data security measure of the DFRC system \cite{jia2023physical, deligiannis2018secrecy}. With this SR metric, various optimization problems for data-secure ISAC signaling can be formulated by jointly considering the radar sensing and communication objectives described in Section~\ref{Sec::JointSignaling}. 

\vspace{0.5ex}
\subsubsection{Optimization for Data-Secure ISAC Signaling}
The optimization problem for data-secure ISAC signaling is generally formulated by adding the SR constraint to the joint signaling design problem in~\eqref{Eqn::P1}. The goal is to achieve desired sensing and communication performance while guaranteeing a required level of security. This formulation inherently introduces a new trade-off among sensing, communication, and data security. Alternatively, the problem can be reformulated as a reciprocal optimization framework, where one of the objectives, sensing, communication, or data security, is selected as the objective function, while the remaining metrics are enforced as constraints. Such a formulation enables flexible prioritization depending on system requirements, available DoF, and target performance levels.

One representative example in~\cite{su2020secure} considers the joint design of data-secure ISAC signaling by maximizing the SR while guaranteeing both communication and sensing performance. Specifically, the problem aims to minimize the Eve’s SNR under per-user SINR constraints for legitimate CUs, radar beampattern matching accuracy, and total transmit power constraints. On the other hand, the work in~\cite{hou2024optimal}, employing CRLB as the sensing metric, maximizes the SR under similar communication and sensing constraints. Accordingly, the general joint signaling design problem for maximizing the SR can be expressed as
\begin{equation}
    \begin{aligned}
        & \underset{\mathbf{X}}{\text{maximize}} 
        && R_s(\mathbf{X}) \\
        & \text{subject to} 
        && f_c(\mathbf{X}) \ge \Gamma_u, \; \forall u, \\
        &&& f_r(\mathbf{X}) \le \epsilon_r, \\
        &&& c_i(\mathbf{X}) \le C_i, \; \forall i,
    \end{aligned}
    \label{Eqn::secure_opt}
\end{equation}
where $\Gamma_u$ denotes the performance threshold for each CU (e.g., SINR or achievable rate), $f_r(\mathbf{X})$ represents the sensing metric as discussed in Section~\ref{Sec::JointSignaling}, and $\epsilon_r$ is the sensing tolerance level. The additional constraints $c_i(\mathbf{X}) \le C_i$ capture system specifications such as total power, per-antenna power, or constant-modulus conditions. This formulation jointly characterizes the three-way trade-off among communication QoS, sensing accuracy, and data security, illustrating how the spatial DoFs of the transmitter can be adaptively allocated to achieve data-secure and efficient ISAC operation. The resulting optimization problems are typically non-convex, and solutions have been developed using SDR, SCA, and iterative approaches~\cite{deligiannis2018secrecy, su2020secure, dong2023joint_sec}.

\vspace{0.5ex}
\subsubsection{Uncertainty on Target Eavesdropper}
\begin{figure}[t!]
    \centering
    {\includegraphics[width=0.45\textwidth]{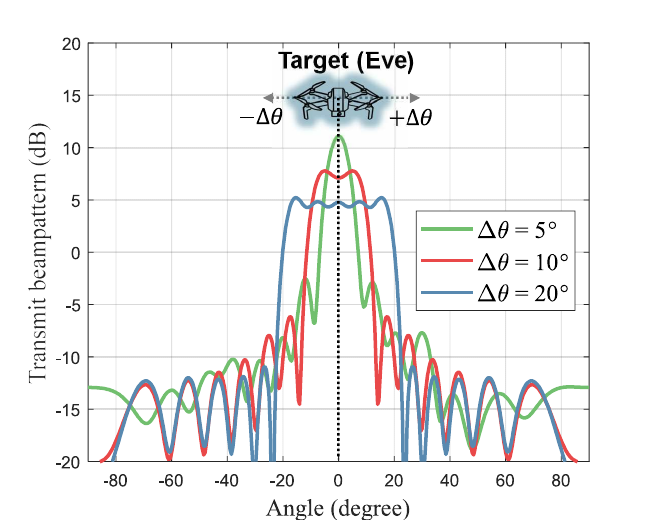}}
    \caption{AN-aided data-secure ISAC design \cite{su2020secure} with $N_t = 16$ transmit antennas and $U=2$ users with SINR threshold $\Gamma_u = 10$dB and various target location uncertainty $\Delta \theta$.}
    \label{Fig::Sec_beampattern}
\end{figure}
One of the key challenges in data-secure ISAC transmitter design lies in the uncertainty of the Eve’s location, as the target Eve may act as a non-cooperative or even mobile object. Consequently, assuming perfect knowledge of Eve’s position or channel may lead to misleading or overly optimistic results in practical ISAC PLS implementations. To address this issue, recent advances in data-secure ISAC transmitter designs incorporate target uncertainty into the optimization framework through robust or probabilistic formulations~\cite{su2020secure, dong2023joint_sec, xu2022robust, hou2024optimal}.

In the context of beampattern-based data-secure ISAC design, one intuitive approach is to broaden the mainlobe width to cover a wider angular region of potential target positions while maintaining low sidelobe power to avoid degrading communication performance~\cite{su2020secure}. Specifically, by defining the angular region of interest as $[\theta_0 - \Delta\theta, \, \theta_0 + \Delta\theta]$, where $\Delta\theta$ denotes the angular uncertainty centered at $\theta_0$, the desired beampattern $P(\theta)$ in~\eqref{Eqn::Beam} can be adaptively shaped to ensure robust illumination and reliable sensing of uncertain targets while maintaining secrecy against potential eavesdroppers. The example beampatterns of AN-aided data-secure ISAC with various uncertainty region is illustrated in Fig. \ref{Fig::Sec_beampattern}. 

It is also worth noting that the work in~\cite{xu2022robust} extends this concept beyond angular uncertainty by incorporating the effects of multi-path fading uncertainty at Eve. By jointly modeling both angular and fading variations, a tractable bound for the combined uncertainty region is developed. On the other hand, CRLB-based data-secure ISAC design, which targets a more fundamental representation of sensing performance, must also account for target uncertainty~\cite{jia2023physical, hou2024optimal, su2025secure}. The posterior CRLB~\cite{hou2024optimal} and Bayesian CRLB~\cite{su2025secure}, both incorporating prior knowledge of the target distribution, have been adopted as sensing metrics to effectively mitigate the impact of target uncertainty.

\subsection{ISAC Data Security Exploiting Subcarrier Interference}
An emerging approach for ISAC data security is waveform-defined PLS, which leverages non-orthogonal waveforms to enhance data transmission confidentiality~\cite{xu2021waveform, xu2024signal, xu2025reliable, zhang2024non}. The core idea is to employ spectrally efficient frequency-division multiplexing (SEFDM)~\cite{darwazeh2018first}, which intentionally introduces ICI by compressing subcarrier spacing below that of conventional OFDM. This deliberate interference prevents Eve from recovering data, while the legitimate CU, aware of the waveform structure, can successfully demodulate it. Accordingly, SEFDM enhances both spectral efficiency and data security.

SEFDM waveforms can be extended to secure ISAC signaling by adopting them within joint precoding design frameworks~\cite{xu2024signal}. The time-domain sample of the SEFDM waveform is expressed as
\begin{equation}
    X_k = \frac{1}{\sqrt{Q}} \sum_{n=0}^{N_s-1} s_n e^{\frac{j2 \pi n k \alpha}{Q}},
\end{equation}
where $k = 0, 1, \dots, Q-1$ and $Q = \kappa N_s$. Here, $\kappa$ denotes the oversampling factor and $\alpha \leq 1$ is the bandwidth compression factor. When $\alpha = 1$, the waveform reduces to standard OFDM. 

The use of SEFDM for data-secure ISAC offers two key advantages. First, it does not require CSI at the transmitter, enabling secure operation even without Eve’s CSI. Second, waveform-defined security remains effective even when Eve is spatially close to the legitimate CU, where AN-aided precoding typically fails due to spatial correlation ~\cite{xu2024signal}. Owing to these properties, SEFDM-based ISAC represents a promising and practical direction for achieving robust data security without additional signaling overhead.

\subsection{ISAC Data Security with Directional Modulation}
Directional modulation (DM) has been extensively studied as a physical-layer security technique for wireless communications \cite{daly2009directional, ding2014establishing, ding2015orthogonal, kalantari2016secure, kalantari2016directional, xie2017artificial, wei2020secure}. DM is inherently a secure approach, as unlike classical modulation that shapes the data constellation at the transmitter, DM aims to create the desired constellation at the intended receiver, thereby preventing data detection at a receiver with an uncorrelated channel. As such, the symbols are directly embedded into the beamforming process by manipulating beamforming weights. In multi-beam transmission scenarios, DM provides the flexibility to securely deliver multiple signals across the user spatial directions with higher SNR than those in other directions. The main difference between DM in communications and DM in ISAC is that the latter typically has spatial DoF tied in producing a desirable radar beam.

\subsubsection{Radar-Centric Directional Modulation}
\begin{figure}[t!]
    \centering
    {\includegraphics[width=0.45\textwidth]{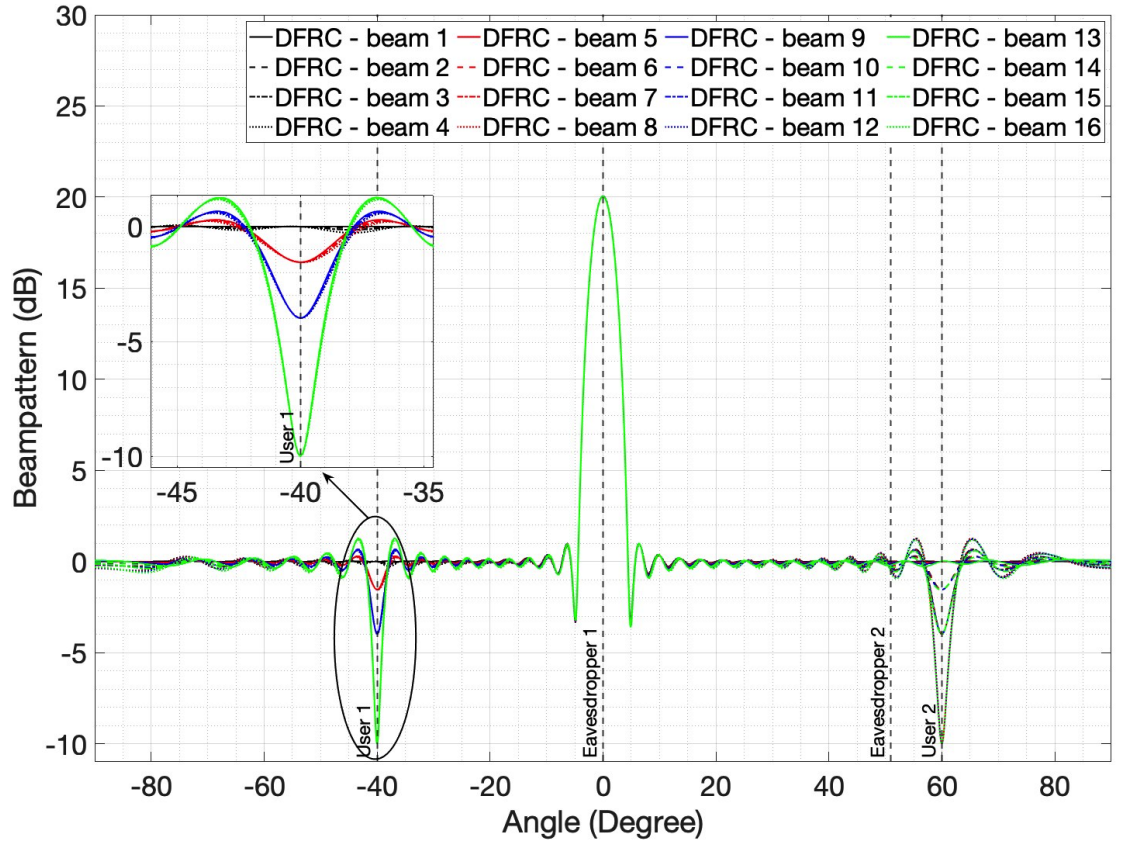}}
    \caption{DM for data-secure DFRC systems.}
    \label{Fig::DM}
\end{figure}
The role of DM in radar has recently gained attention in the context of radar-centric DFRC architecture. Essentially, DFRC systems that set the radar beam complex values, or the spatial response, equal to the communication symbols for intended users can be viewed as performing DM. Since the radar main beam must remain intact, providing the highest attainable gain, DM can be readily applied in the sidelobes. A broader generalization arises when the communication symbols and beam values are not identical but, instead, are related through a dictionary which constitutes a form IM, as discussed in Section \ref{Sec::Radar-centric}.

In radar-centric DFRC systems, data security against eavesdropping is not typically a primary design objective but instead emerges as a byproduct of maintaining close and low sidelobe levels across the field of view, except in the directions of intended users, over multiple beams. In this context, data security is enhanced when Eve observes a more compact sidelobe constellation—both in magnitude and potentially in phase—compared with a rather spread constellation points designed for the intended users to satisfy a preset probability of error. 

The design of a security-driven radar-centric DFRC transmitter of $N_t$ antennas can be achieved by solving the following optimization problem \cite{amin2024similarity}:
\begin{equation}
    \begin{aligned}
        & \underset{\mathbf{w}_{i_1, \dots, i_U}}{\text{minimize}} 
        && || \mathbf{w}_{rad} - \mathbf{w}_{i_1, \dots, i_U}|| \\
        & \text{subject to} 
        && \mathbf{w}_{i_1, \dots, i_U}^H \mathbf{a}(\theta_u) = \mathcal{S}_{u,i_u}, \; \forall u, \\
        &&& \text{Data Security constraints for Eve,}
    \end{aligned}
    \label{Eqn::secure_DM}
\end{equation}
where $U$ is the number of users, $\mathbf{a}(\theta_u)$ and $\theta_u$ are, respectively, the $N_t$-dimensional steering vector of the $u$-th user and the corresponding angle. The first term of the above cost function, $\mathbf{w}_{rad}$, is the desired radar beamformer weight vector, whereas the second term, $\mathbf{w}_{i_1, \dots, i_U}$, is the designed weight vector associated with the $U \times 1$ vector of symbols, $\mathbf{s}_{i_1, \dots, i_U} = [\mathcal{S}_{1,i_1}, \dots, \mathcal{S}_{U,i_U} \in \psi_U]^T$. The symbol associated with the $u$-th user is selected from the dictionary of size $L_u$, which is given by
\begin{equation}
    \psi_u = \{\mathcal{S}_{u,1}, \dots, \mathcal{S}_{u,L_u}\}, \; u = 1, \dots U.
\end{equation}

Without data security constraints, the constellation seen by Eve is given by the complex sidelobe gains at its direction. These values are different in magnitude and phase from those set at the user directions, most notably the former are closer in magnitudes. This evident across all directions in Fig. \ref{Fig::DM}, which shows the designed transmit power radiation pattern with 32 antennas for two users at $-40^{\circ}$ and $60^{\circ}$, each has four symbols, leading to sixteen designed beams. It is important to observe from Fig. \ref{Fig::DM} that the main beam remains intact, whereas most variations are exhibited in the sidelobes. 

Since data security would benefit from more compact constellation at the non-user directions, the work in \cite{cao2023directional} introduced a data security constraint that forces the complex sidelobe gains towards Eve to be equal and, as such, making it more difficult to decipher the information. Such constraint, however, consumes additional degrees of freedom. It also requires either knowledge of the Eve's direction, or channel, otherwise, it enforces such constraint at different presumed directions.

\subsubsection{Constructive–Destructive Interference Exploitation}
\begin{figure}[t!]
    \centering
    {\includegraphics[width=0.24\textwidth]{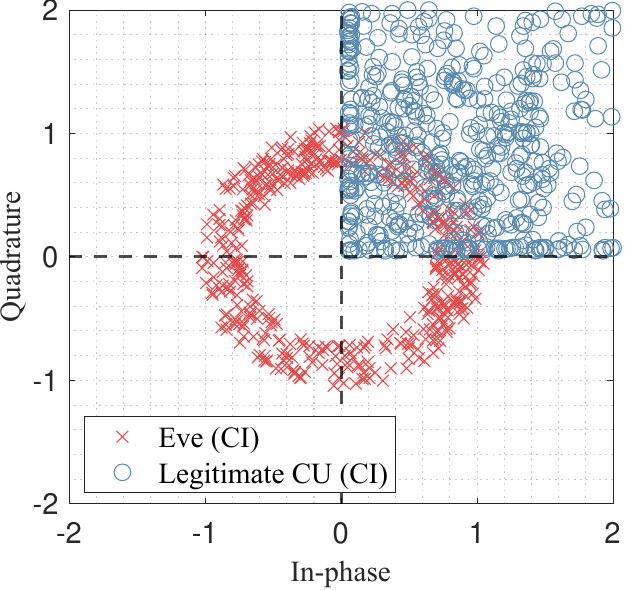}}
    {\includegraphics[width=0.24\textwidth]{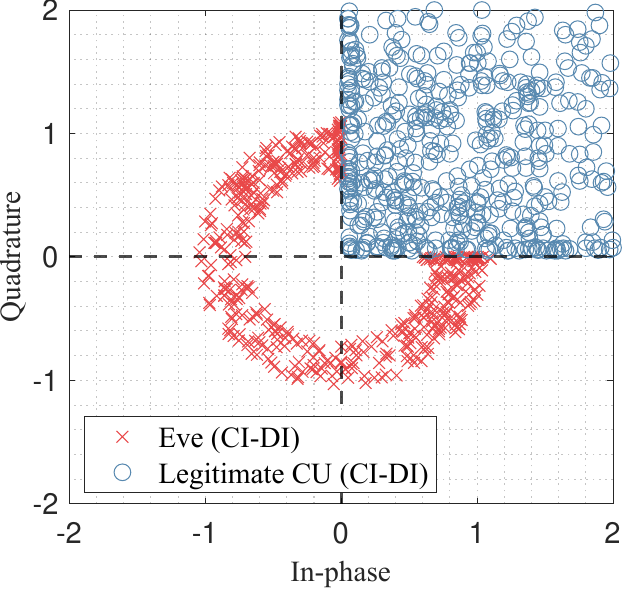}}
    \caption{Received constellations at the legitimate CU and Eve using CI and CI-DI techniques under QPSK modulation for ISAC data security with DM \cite{su2022secure}.}
    \label{Fig::sec_CI}
\end{figure}
Beyond DM in radar-centric ISAC for data security, DM schemes exploiting CI and DI can be integrated into the joint precoding design of data-secure ISAC systems. The key idea is to utilize CI to push the received signals at the legitimate CU farther away from the decision boundaries of modulated symbols, thereby improving communication QoS. Meanwhile, DI can be leveraged to degrade the information-carrying signals received by potential target Eves while maintaining effective target illumination~\cite{su2022secure}. As discussed in Section~\ref{Sec::SLP}, SLP can intentionally force the received symbols into CI or DI regions of the modulation constellation~\cite{kalantari2016directional}. Accordingly, interference exploitation enables more power-efficient secure transmission compared to AN-aided designs, achieving an enhanced trade-off between secure communication and sensing performance.

Unlike conventional DM in PLS, interference exploitation for ISAC security jointly considers both secure communication and sensing performance. Although the CI-based joint precoding design in~\eqref{Eqn::SLP_ISAC} scrambles Eve’s received signals due to the nature of DM, it cannot fully guarantee data security when Eve’s channel is correlated with that of the legitimate CU. To address this issue, the seminal work in~\cite{su2022secure} introduced DI constraints on Eve’s received signals, enabling more secure transmission than the CI-only precoding scheme, referred to as the CI–DI technique.

As illustrated in Fig.~\ref{Fig::6}, SLP aims to force Eve’s received symbols into the DI regions. For QPSK, this DI region can be divided into three distinct zones, where Eve’s received signal may fall into any of them. Readers are referred to equation~(35) in~\cite{su2022secure} for the detailed formulation of these DI constraints. Fig.~\ref{Fig::sec_CI} illustrates exemplary received signal constellations at the legitimate CU and Eve using CI and CI–DI techniques. One key advantage of the CI–DI technique is its ability to achieve more secure data transmission even when the target Eve and CU are spatially correlated. It is worth noting that the impact of target location uncertainty in data-secure ISAC with interference exploitation can be mitigated through Bayesian CRB optimization~\cite{su2025secure}, or by adopting approaches similar to those used in AN-aided data-secure ISAC transmission.

\subsection{Sensing-Assisted Data-Secure ISAC Design}
Apart from introducing new security threats, ISAC also offers unique opportunities to enhance communication data security. Owing to the sensing functionality of the DFRC transceiver, PLS techniques become more feasible, as the system can exploit sensing to detect potential Eves and, at a minimum, estimate their channels or directions, thereby improving communication secrecy through more accurate CSI of the Eves~\cite{liu2025sensing, liu2023securing, su2023sensing, xu2023sensing, xu2025sensing}. The main idea of sensing-assisted, data-secure ISAC design is to first transmit probing signals to detect potential Eves, both active and passive, and estimate their channel parameters. Based on this information, data-secure ISAC signaling is then designed to achieve the desired level of data security while ensuring satisfactory sensing and legitimate communication performance. This establishes a new synergy between radar sensing and secure communication, providing mutual benefits for both functionalities.

A representative work in~\cite{su2023sensing} developed an iterative design framework with sensing-assisted data-secure ISAC. In the initial stage, the transmitter sends an omni-directional probing signal to estimate the parameters of potential eavesdroppers. The sensing results are then used to extract the Eve's direction and enable the characterization of the SR. In the next stage, the transmission aims to maximize the SR while refining the transmit beampattern search region based on the CRLB, which characterizes the variance of the angle estimation error of the Eves. To this end, a weighted-sum optimization problem is formulated to achieve a trade-off between sensing and secure communication performance. The key effect of this design is that by iteratively refining the transmit beampattern using updated estimation results, the transmitter effectively reduces the uncertainty in Eve’s channel, thereby enhancing the achievable SR compared with designs that lack Eve’s channel information.

A similar approach in~\cite{cao2024sensing} divides the process into two distinct stages: first, searching for potential eavesdroppers, and then focusing on secure communication. This protocol is optimized with respect to the number of searching beams and the beamforming design for secure data transmission. More recently, target-tracking capability using extended Kalman filtering has been integrated with ISAC PLS design to handle moving eavesdroppers, jointly optimizing power consumption, legitimate user scheduling, and target-tracking performance~\cite{xu2025sensing}. These studies consistently report that sensing-assisted schemes for PLS significantly enhance secure communication performance compared with conventional AN-aided data-secure signaling. This observation highlights that radar sensing can be further leveraged to enhance the overall data security of future wireless networks.

\section{Secure ISAC Transceiver: Sensing Security}\label{Sec::SensingSecure}
\begin{figure}[t!]
    \centering
    \subfigure[]{\includegraphics[width=0.24\textwidth]{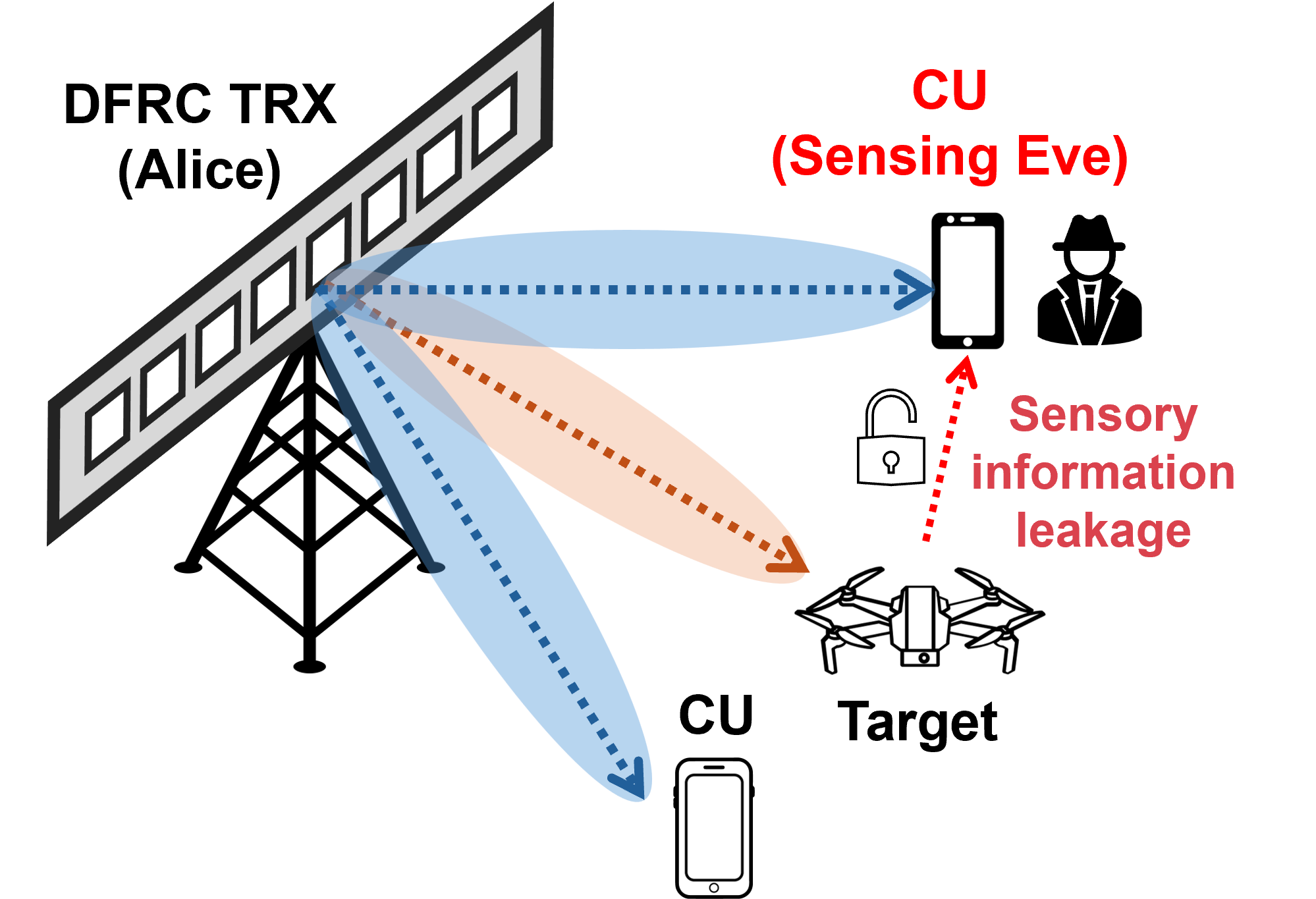}}
    \subfigure[]{\includegraphics[width=0.24\textwidth]{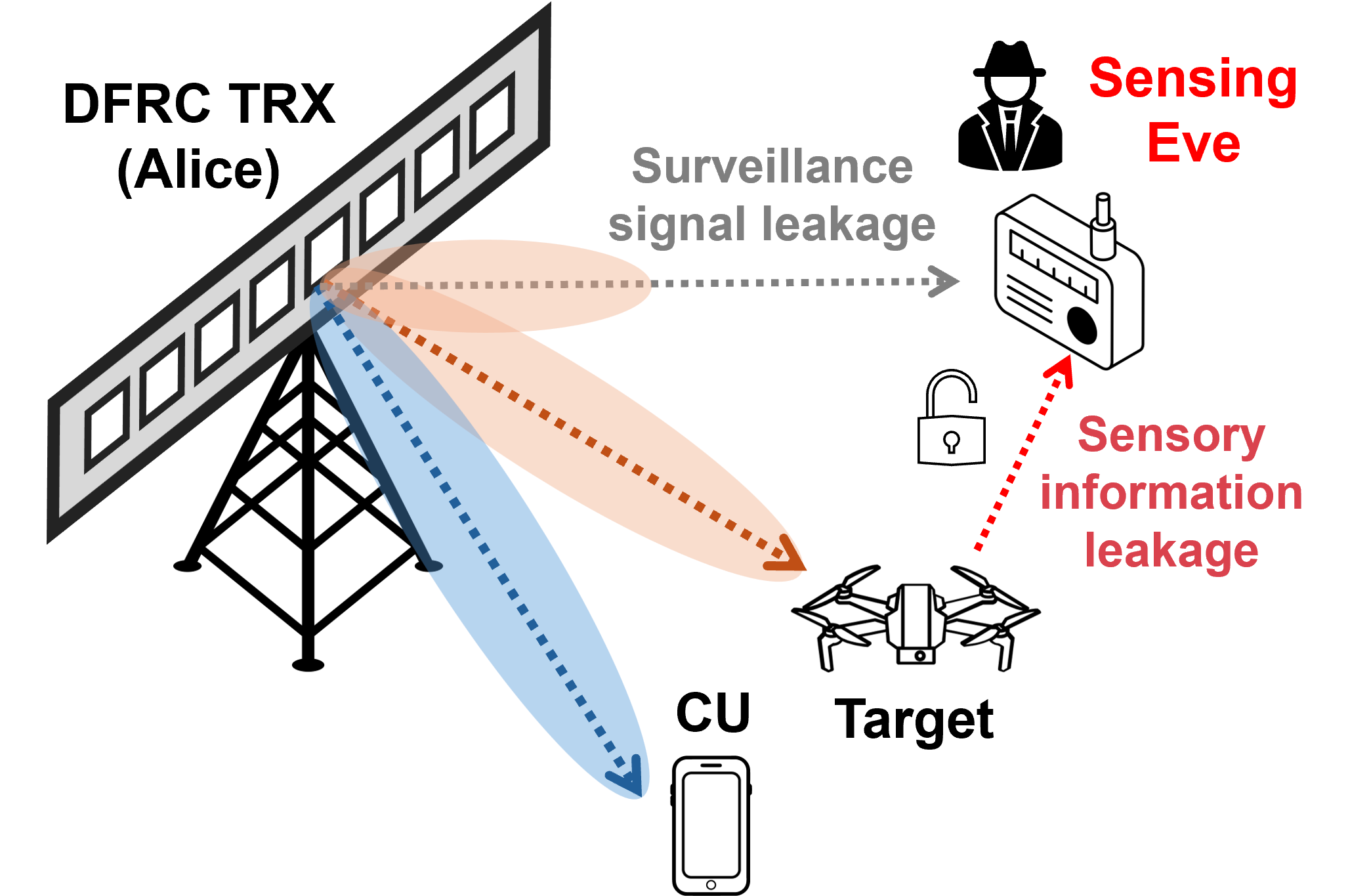}}
    \caption{ISAC security threat scenarios with passive sensing eavesdropper: (a) a CU of the network acts as Eve, and (b) a passive radar acts as Eve, while remaining silent.}
    \label{Fig::Sec_sens}
\end{figure}

Integrating sensing capabilities into wireless networks exposes new vulnerabilities in sensing security. The high-power illumination used for environmental sensing makes opportunistic sensing signals widely accessible \cite{colone2022passive}, allowing unauthorized third parties to exploit these signals to independently infer information about targets and surrounding environments without being compelled to transmit and thereby, exposed ~\cite{qu2024privacy, griffiths2022introduction}. Unlike data transmission, sensing does not involve an encrypted information link, which means that a passive radar eavesdropper can exploit the same ISAC signal as both a reference and a surveillance waveform. In this regard, safeguarding sensing functionality must be achieved through physical layer strategies that intentionally distort or mask the target-related channel information, thereby misleading the eavesdropper~\cite{cigno2022integrating}. Table \ref{tab::sensing_sec} summarizes sensing security vulnerabilities in ISAC and possible solutions that will be discussed in the following subsections.

\begin{table}[t!]
    \centering
    \caption{Overview of sensing security vulnerabilities in ISAC systems and representative countermeasures.}
    {\includegraphics[width=0.45\textwidth]{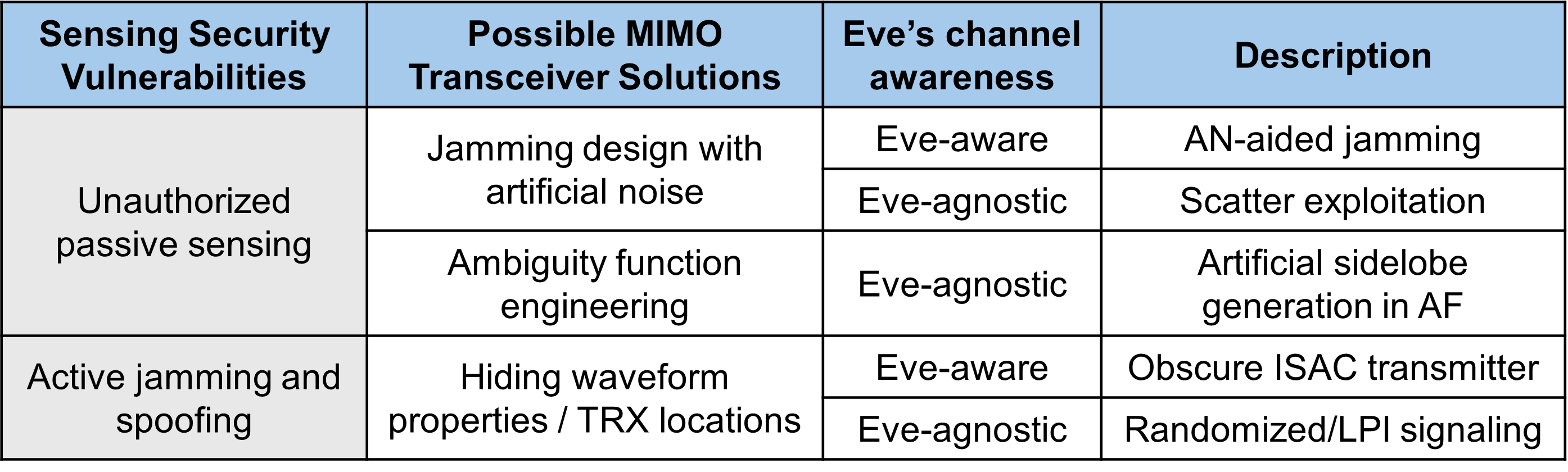}}
    \label{tab::sensing_sec}
\end{table}

\subsection{Jamming Design for Sensing Security}
Jamming has long been employed as an electronic countermeasure to disrupt unwanted receivers by intentionally transmitting interference signals \cite{lothes1990radar}. In sensing-secure ISAC transceiver design, this jamming functionality can be integrated alongside sensing and communication operations \cite{su2020secure}. Similar to PLS techniques for data security, AN can again be utilized, this time to protect the sensing functionality in ISAC by acting as a form of controlled jamming. However, since radar sensing inherently seeks reliable detection and parameter estimation, the design methodologies used for data-secure ISAC cannot be directly applied. Instead, sensing security requires careful consideration of appropriate performance metrics that reflect both the protection level and sensing accuracy. 

\subsubsection{Eavesdropper-Aware Design}
When the CSI or spatial location of Eve is available at the transmitter, the design of secure sensing becomes analogous to that of data security. This scenario may arise when a CU of the network, whose CSI and location is typically known, also attempts to act as an Eve, as illustrated in Fig.~\ref{Fig::Sec_sens}(a). An early work in~\cite{zou2024securing} investigates this case by employing AN. In the proposed secure-sensing ISAC framework, sensing MI is adopted as the performance metric for both the legitimate receiver and Eve. Based on the signal model in~\eqref{Eqn::AN}, the key mechanism enabling sensing security lies in the knowledge disparity of the embedded AN between the legitimate sensing receiver and Eve. The corresponding optimization problem maximizes the legitimate sensing MI while constraining Eve’s sensing MI and the communication SINR, thereby ensuring both sensing reliability and confidentiality. 

This concept is further extended in~\cite{jia2024illegal}, where the sensing target of the ISAC BS itself is regarded as a potential sensing Eve that attempts to detect and estimate other targets using DL ISAC signals. More recently, reconfigurable intelligent surface (RIS)-assisted ISAC designs for sensing security have been proposed, aiming to degrade the target SINR at Eve while guaranteeing the legitimate sensing SINR and communication QoS~\cite{magbool2025hiding}. The key insight in this approach is that the RIS-reflected signal toward Eve acts as interference relative to the target-reflected signal. Hence, by controlling the power of these components, the legitimate transmitter can deny target detection by Eve, particularly when Eve has no knowledge of the RIS location. Although these initial works open new directions for sensing security in ISAC, they rely on a strong assumption that the legitimate ISAC transmitter has perfect knowledge of both Eve’s CSI and the target–Eve channel. In practice, this assumption is rarely valid. Therefore, future research should pursue more practical solutions that account for target uncertainty and imperfect or stochastic Eve CSI, which remain largely unexplored.

\subsubsection{Eavesdropper-Agnostic Design}
In the eavesdropper-agnostic approach, the ISAC transmitter must ensure sensing confidentiality without any prior knowledge of potential eavesdroppers’ locations or channels. Although this is inherently a difficult problem that presents significant design challenges, several unique opportunities in ISAC systems can be exploited to achieve practical and robust sensing security.  
From the perspective of jamming design at the ISAC transmitter, environmental scatterers can serve as natural enablers for deceptive jamming against Eve~\cite{chen2025sensing}. The core idea is to intentionally illuminate selected scatterers together with the intended target so that the resulting reflections act as clutter interference at Eve. These additional echoes distort Eve's sensing observations, degrading its ability to detect or localize the target—particularly when Eve lacks prior knowledge of the presence or geometry of the scatterers. Note that deliberately illuminating scatterers consumes transmit power and spatial DoFs, which may reduce resources available for legitimate sensing and communication. Moreover, clutter-based jamming mainly induces angle-of-arrival (AoA) deception and has limited effectiveness in obscuring range or Doppler, underscoring the need for complementary waveform- or modulation-level countermeasures.

\begin{figure}[t!]
    \centering
    {\includegraphics[width=0.30\textwidth]{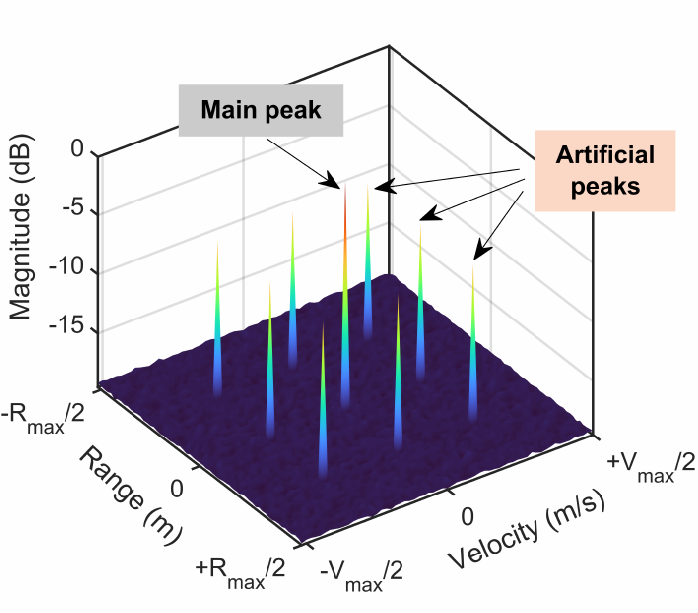}}
    \caption{Ambiguity function engineering with artificial peaks to secure sensing functionality in ISAC.}
    \label{Fig::AF1}
\end{figure}

\begin{figure}[t!]
    \centering
    {\includegraphics[width=0.215\textwidth]{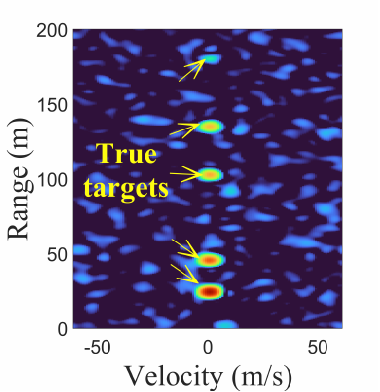}}
    {\includegraphics[width=0.265\textwidth]{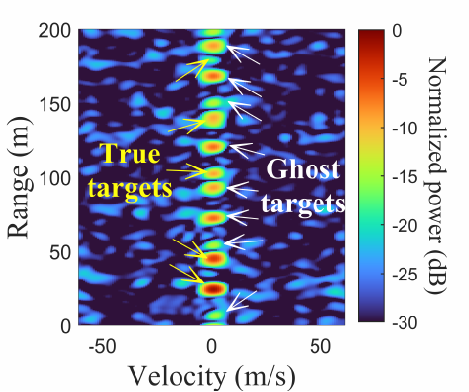}}
    \caption{RD map at passive sensing eavesdropper without (left) and with (right) AF control.}
    \label{Fig::AF2}
\end{figure}

\subsection{Ambiguity Function Engineering}
Complementary to the above signal- and beamforming-level designs, sensing-secure approaches can also be pursued at the ambiguity-function level. A passive radar typically exploits two signals of opportunity: a direct-path reference signal and a target-reflected signal. By cross-correlating these two signals, the passive radar constructs RD-domain measurements to detect and estimate target parameters~\cite{berger2010signal}. Building on this observation, an Eve-agnostic sensing security technique can be developed by deliberately shaping or distorting the AF of ISAC waveforms~\cite{han2025sensing}. Through proper AF engineering, the legitimate receiver can preserve reliable sensing performance, whereas an unauthorized passive radar eavesdropper experiences degraded RD resolution or misleading parameter estimation, effectively concealing the true target information.  

The motivation behind AF engineering arises from the inherent information asymmetry between the legitimate receiver and Eve. Since a passive radar eavesdropper exploits the surveillance signal leakage to infer sensory information, it must rely on MF or time-domain cross-correlation using an imperfect reference signal, suffering from degraded SNR due to incomplete knowledge of the ISAC transmit signal structure~\cite{cui2014target, bkaczyk2015impact}. In contrast, the legitimate sensing receiver, having full knowledge of the transmit waveform, can employ MMF or advanced receiver processing to suppress high sidelobes and maintain accurate target estimation.

\subsubsection{AF Engineering in Communication-Centric ISAC} 
The recent work in~\cite{han2025sensing} introduced the concept of AF engineering in communication-centric ISAC systems employing OFDM waveforms. The key idea is to intentionally design the AF with artificial sidelobes, as illustrated in Fig.~\ref{Fig::AF1}, thereby generating ghost targets at the eavesdropper’s MF receiver as shown in Fig.~\ref{Fig::AF2}, while allowing the legitimate receiver to mitigate them using the MMF receiver.  

Recalling the OFDM-ISAC signal model with $N_s$ subcarriers, the method exploits subcarrier power allocation to shape the ACF, which corresponds to the zero-Doppler cut of the AF. Specifically, the $k$th bin of the frequency-domain ACF is expressed as  
\begin{align}
    \Lambda[k] = \sum_{n=1}^{N_s} |p_n|^2 |s_n|^2 e^{j\frac{2\pi}{N_s}k(n-1)}, 
    \label{eq18}
\end{align}
where $p_n$ denotes the allocated subcarrier power and $s_n$ is the modulated symbol on the $n$th subcarrier. By properly designing $\{p_n\}$ for all subcarriers, the squared secure ACF can be shaped as \cite{han2025sensing} 
\begin{align}
    \mathbb{E}\left[|{\Lambda}[k]|^2\right]
    = N_s^2 \delta[k] + \alpha^2 \sum_{l=1}^{L} \delta[k - l\lambda],
\end{align}
where $\lambda$ represents the periodicity of the artificial peaks, $L = N_s/\lambda - 1$ denotes their total number, and $\alpha$ controls the amplitude of each artificial peak. This design effectively preserves the mainlobe structure required for legitimate sensing, while misleading an unauthorized passive radar by introducing artificial range ambiguities that appear as ghost targets in its RD map. Importantly, it should be noted that the SNR loss in the legitimate sensing receiver is also determined by the design of $p_n$ as discussed in Section~\ref{Sec::RX_F}.

Using this structured AF, a sensing-secure ISAC signaling design has been proposed to balance the three-way trade-off among legitimate sensing performance, communication reliability, and sensing security. The corresponding optimization problem can be formulated as
\begin{equation}\label{Eqn::P2}
    \begin{aligned}
    & \underset{\{p_n\}_{n=1}^{N_s}}{\text{maximize}}
    \qquad   -(1-\rho)\mathcal{L}_{A} + \rho R_c  \\
    & \text{subject to}
    \qquad \qquad   \Delta_{\text{ISL,E}} \geq \epsilon_{\text{ISL}},    \\
    &  \qquad \qquad \qquad \qquad  \Delta_{\text{PSL,E}} \geq \epsilon_{\text{PSL}},
    \end{aligned}
\end{equation}
where $\mathcal{L}_{A}$ denotes the normalized SNR loss at the legitimate sensing receiver, $R_c$ is the achievable rate of the CU, and $\Delta_{\text{ISL,E}}$ and $\Delta_{\text{PSL,E}}$ represent the ISL and PSL of the ACF observed at Eve, respectively. The parameters $\epsilon_{\text{ISL}}$ and $\epsilon_{\text{PSL}}$ specify the required thresholds that determine the desired level of sensing security. This formulation highlights that by properly tuning the power allocation $\{p_n\}$ and weighting factor $\rho$, the ISAC transmitter can flexibly trade off between sensing accuracy, communication throughput, and resistance against sensing eavesdroppers. For more details of the solution and results, we refer the readers to \cite{han2025sensing}.

\subsubsection{AF Engineering in Radar-Centric ISAC} 
The same philosophy of AF engineering for sensing-secure ISAC can also be applied to radar-centric ISAC systems. As discussed in Section~\ref{Sec::Radar-centric}, IM embedded in radar waveforms not only conveys communication data but can also be leveraged to shape the AF for sensing security. In particular, IM implemented through variations in chirp bandwidth and center frequency, together with phase modulation in FMCW radar, enables AF shaping with intentionally introduced artificial sidelobes.

A recent study in~\cite{temiz2025radar} proposes an IM-FMCW-based secure-sensing ISAC framework enhanced with phase modulation, which further increases the DoFs in waveform design. This jointly facilitates Doppler ambiguity control and improves the communication rate. Importantly, the sensing-secure AF is optimized by minimizing the MSE between the desired AF and that of the designed signal, in a manner analogous to beampattern matching. While this deliberately degraded AF distorts target detection and parameter estimation at Eve, the legitimate sensing receiver, having full knowledge of the transmitted IM and phase coding, can compensate for these effects, successfully recovering the target range and velocity without significant performance degradation.  

In summary, AF engineering provides an effective and Eve-agnostic means of designing sensing-secure waveforms for both communication- and radar-centric ISAC systems. Although further research is needed in areas such as MIMO AF design, joint transceiver optimization, and standard-compatible signaling, existing studies have clearly demonstrated the feasibility of achieving secure ISAC through deliberate AF manipulation.

\subsection{Active Security Attack on ISAC: Jamming and Spoofing}
\begin{figure}[t!]
    \centering
    {\includegraphics[width=0.35\textwidth]{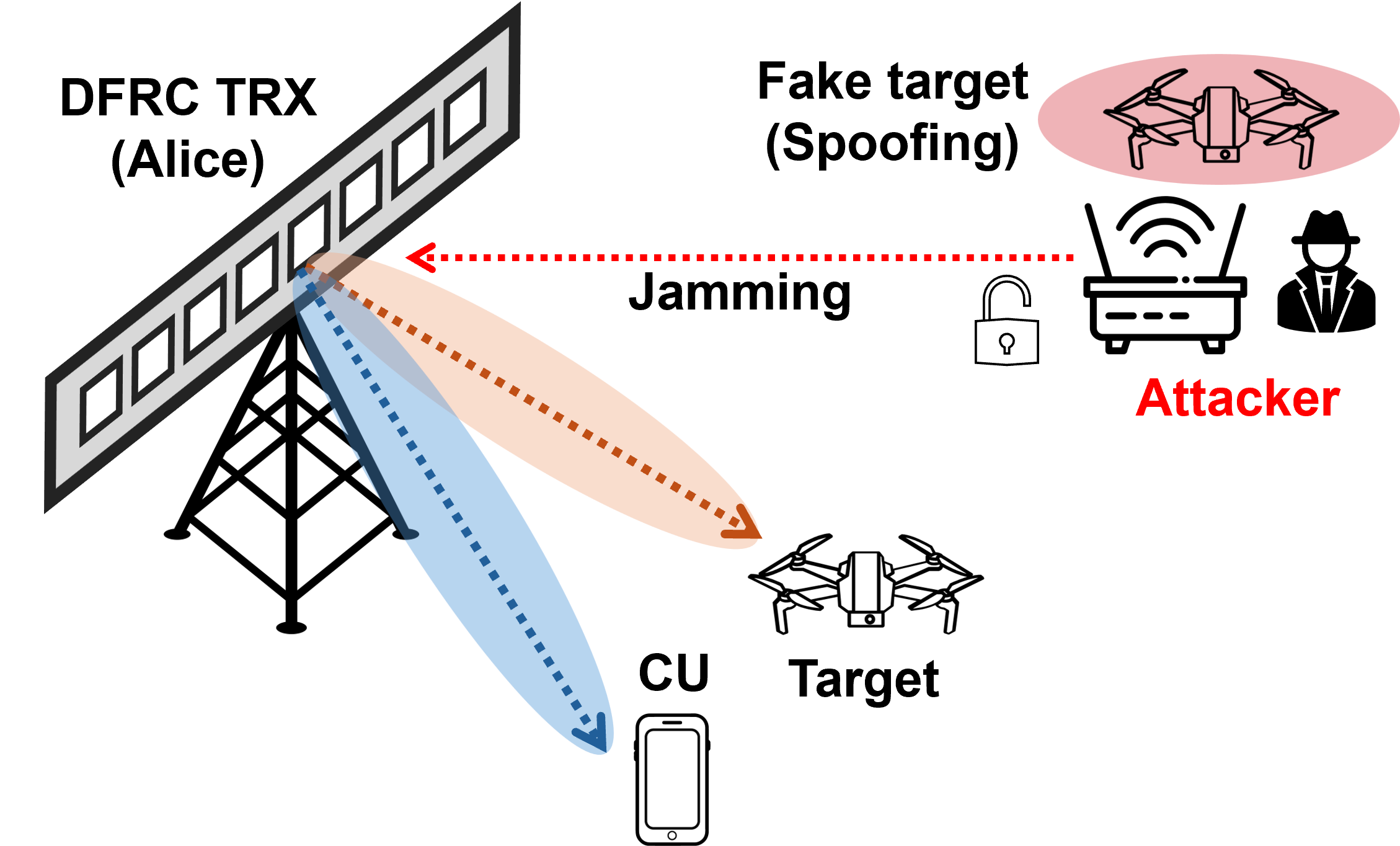}}
    \caption{ISAC security threat scenarios with an active attacker.}
    \label{Fig::Sec_Attack}
\end{figure}
In radar systems, resilience against active attacks such as jamming and spoofing is a fundamental requirement for secure and reliable operation~\cite{greco2008radar, komissarov2021spoofing, chen2007detecting}. The same requirement extends naturally to ISAC, where shared spectral and hardware resources increase vulnerability to electronic countermeasures~\cite{yildirim2025ofdm, chrysanidis2024replay, li2025securing, ma2024sensing}. Consequently, active attacks that target the radar sensing function, for example, intentional jamming that saturates the legitimate receiver or spoofing that injects false echoes to mimic targets, as illustrated in Fig.~\ref{Fig::Sec_Attack}, pose serious threats to ISAC operation and must be addressed in both transceiver design and system deployment.

Recent work in~\cite{yildirim2025ofdm} studies practical active-attack scenarios in WiFi-based sensing and develops jamming models, in which an attacker transmits signals that create artificial targets or corrupt the surveillance channel. Complementary research from the attacker’s perspective shows that sensing-resistant jamming strategies can be designed to be difficult for a legitimate ISAC transceiver to estimate or mitigate~\cite{ma2024sensing}, underscoring the need for robust detection and mitigation techniques in ISAC transceiver design.

A practical defense direction is to deny adversaries accurate knowledge of waveform properties or BS geometry \cite{pace2009detecting, wicks2011principles, dong2019dual}. For example, \cite{li2025securing} proposes a randomized OFDM waveform that dynamically varies subcarrier spacing and carrier-frequency offset across transmissions, making it difficult for an attacker to replicate the legitimate waveform and generate effective adversarial signals. Another promising approach exploits secure transmission strategies that conceal the BS directionality from potential Eve or attackers~\cite{ma2025sensing}. By keeping the BS’s transmit direction and beamforming strategy confidential, the legitimate system reduces the attacker’s ability to accurately orient jamming or spoofing resources, thereby mitigating the risk of successful active attacks. Nevertheless, secure ISAC transceivers resilient to active attacks remain largely underexplored. Developing a foundational framework to address this challenge, particularly by connecting low-probability-of-intercept ISAC waveform design~\cite{shi2025low} with ISAC security, is an important future direction.

\section{ISAC Proof-of-Concept Demonstration}\label{Sec::PoC}
While theoretical research and signal processing advancements have greatly accelerated the integration of communication and radar functionalities, the practical implementation of ISAC systems remains in its early stages. Experimental validation plays a vital role in translating theoretical concepts into real-world ISAC deployments, while simultaneously offering new insights that guide transceiver design. It is worth noting that while several efforts have demonstrated radar sensing capabilities using communication waveforms such as OFDM and OTFS \cite{barneto2019full, correas2023mimo}, the core objective in ISAC demonstration lies in realizing true dual-functionality, achieving both communication and sensing within a unified platform, and validating the trade-offs between the two. This section reviews recent proof-of-concept (PoC) developments and demonstrations of ISAC, with particular emphasis on hardware implementation and practical system realization.

\subsection{Radar-Centric ISAC Demonstration}
\begin{figure}[t!]
    \centering
    {\includegraphics[width=0.45\textwidth]{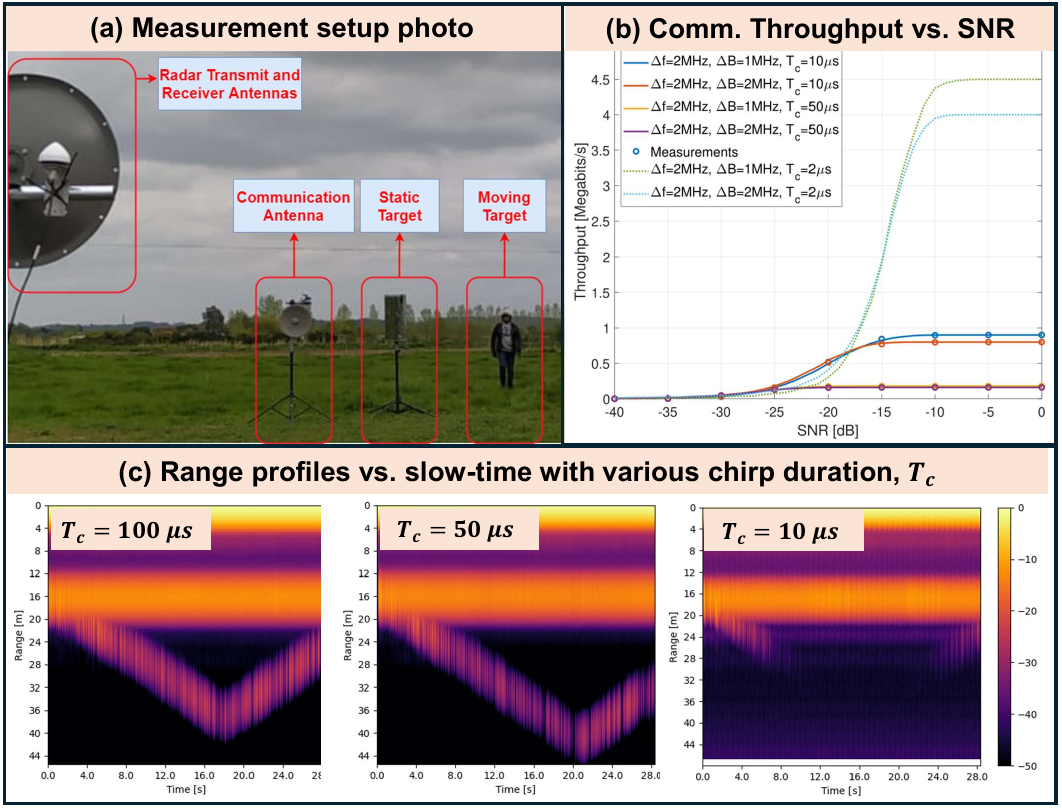}}
    \caption{The PoC demonstration of radar-centric ISAC with IM embedded in the bandwidth and center frequency of FMCW chirps and in the antenna polarization \cite{temiz2023experimental}.}
    \label{Fig::Meas1}
\end{figure}

For radar-centric ISAC systems, the work in~\cite{temiz2023experimental} experimentally demonstrated the dual functionality of radar and communication through IM embedded in the bandwidth and center frequency of FMCW chirps, as well as in the antenna polarization domain. The PoC hardware was implemented on the ARESTOR platform based on a Xilinx RFSoC FPGA~\cite{peters2022arestor}, operating at 2.4~GHz. Notably, the prototype revealed a clear trade-off between communication throughput and radar target SNR as a function of the chirp rate. As illustrated in Fig. \ref{Fig::Meas1}, shorter FMCW chirp durations allow higher communication data rates but reduce the coherent processing gain, thereby degrading the range estimation accuracy of the radar functionality. 

Another IM-based DFRC prototype was experimentally validated in~\cite{ma2021spatial}, which employed generalized spatial modulation via antenna selection. The PoC system was implemented using FPGA boards for both TX and RX at a 5.1~GHz carrier frequency. To emulate moving radar targets in over-the-air measurements, a radar echo generator was developed using a spectrum analyzer and a vector signal generator, which received radar pulses and retransmitted delayed echoes. Experimental results confirmed that the IM-based DFRC with adaptive antenna selection outperformed fixed antenna allocation schemes in both radar sensing and communication performance, achieving improved angle estimation and BER under identical data rate conditions.

\subsection{Communication-Centric ISAC Demonstration}
\begin{figure}[t!]
    \centering
    {\includegraphics[width=0.45\textwidth]{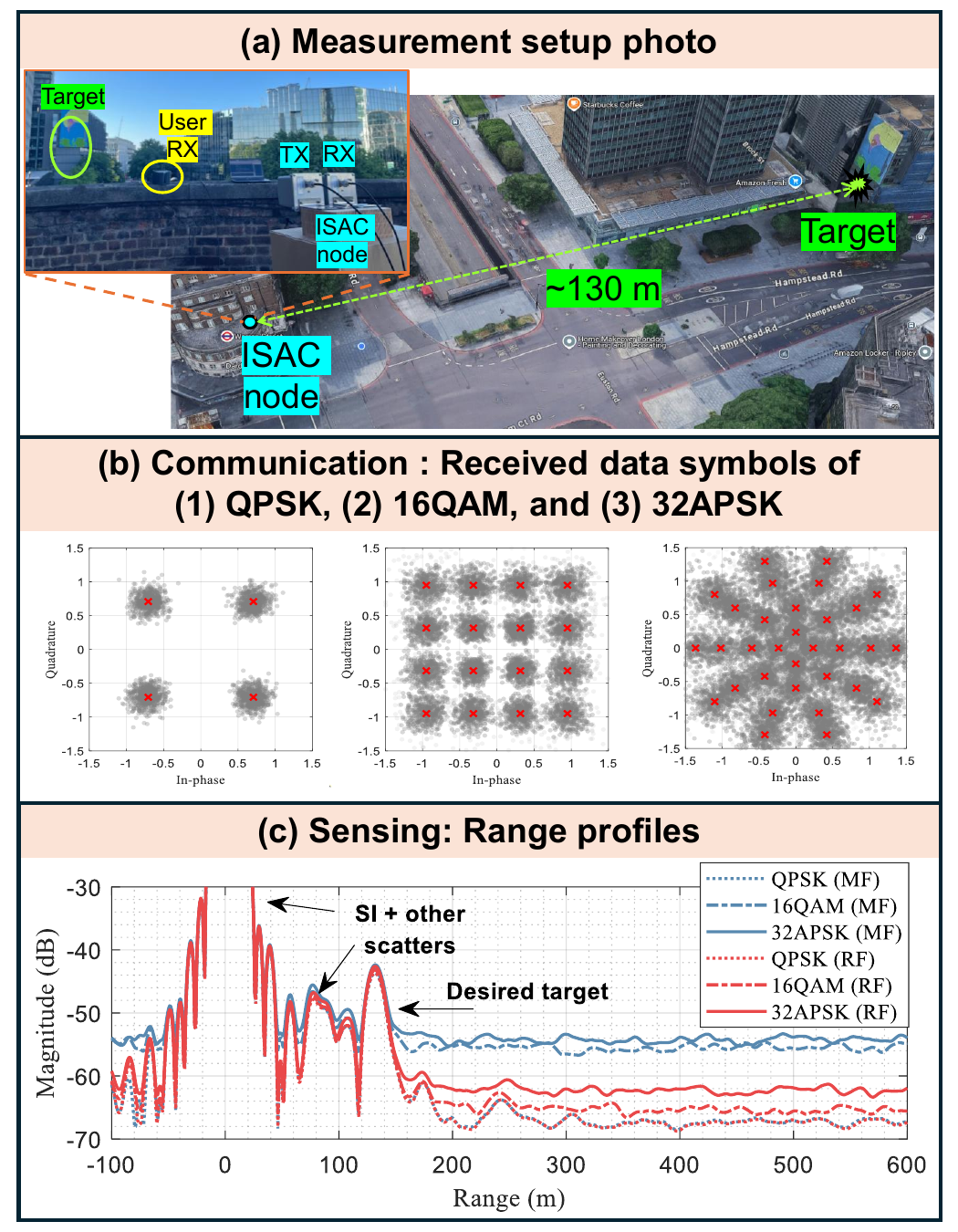}}
    \caption{The PoC demonstration of communication-centric ISAC under various modulation constellations, showing the impact of the constellation geometry on the receiver-specific ranging performance \cite{han2025constellation}.}
    \label{Fig::Meas2}
\end{figure}
PoC systems for communication-centric ISAC have been more actively investigated, particularly focusing on the demonstration of radar sensing using communication signals. Beyond functional implementations of radar sensing, deeper insights into communication-centric ISAC can be obtained by experimentally validating the inherent S\&C performance trade-offs. As illustrated in Fig.~\ref{Fig::Meas2}, an OFDM-ISAC prototype demonstrates the effect of signal modulation on sensing performance under specific receiver processing configurations~\cite{han2025constellation}. In this setup, data payloads transmitted by the ISAC node are received by the single-antenna CU while simultaneously estimating the range of a desired target. The prototype employs a software-defined radio operating at a 2.4~GHz center frequency with a 20~MHz bandwidth. 

This PoC demonstration validates the theoretical analysis of OFDM-based sensing with random signaling presented in Section~\ref{Sec::RX_F}, showing that sensing performance varies with the employed receiver processing, MF or RF, under urban propagation environments with multiple scatters. The ISAC constellation shaping was also validated using the same PoC setup~\cite{han2025constellation}, as illustrated in Fig.~\ref{Fig::Const}, demonstrating the controllable trade-off between S\&C performance based on the constellation geometry. It is noteworthy that the limited transmit power of the software-defined radio can be effectively compensated by exploiting the coherent processing gain across multiple OFDM symbols, enabling reliable detection and estimation of distant targets. Such PoC demonstrations effectively bridge ISAC theory and practical implementation, providing valuable insights into ISAC deployment within existing communication infrastructures.

\subsection{MIMO-ISAC Beamforming Demonstration}
This subsection highlights the PoC demonstration of MIMO-ISAC beamforming, which enables dual functionality for target sensing and communication when these two operations are spatially separated. The seminal demonstration of multifunction waveforms in~\cite{mccormick2017simultaneous} employs a shared antenna array to transmit both radar and communication signals. The software-defined radio platform developed in~\cite{mealey2016beemer}, known as the BEEMER system, operates at 3.5~GHz and serves as the experimental testbed. In this setup, an up-chirped LFM signal is used as the radar waveform, while communication data modulated using a QPSK constellation are transmitted under different shaping filters. The experimental results validate the influence of radar signal interference on communication BER, showing a clear dependency on the radar beam direction and the occupied communication spectrum. 

The joint waveform design was experimentally demonstrated in~\cite{xu2022experimental}, showcasing flexible trade-offs between S\&C performance. The experimental MIMO-OFDM ISAC system was implemented using six Universal Software Radio Peripheral (USRP) devices operating at a 2.4~GHz center frequency. The results revealed a performance trade-off between communication BER and the measured radar beampattern, confirming the feasibility of flexible ISAC operation enabled by the joint precoding design discussed in Section~\ref{Sec::JointSignaling}. Furthermore, to achieve higher sensing accuracy and resolution required by advanced applications, the capabilities of MIMO-ISAC transceivers have been further demonstrated in~\cite{sakhnini2022near, ozkaptan2023mmwave}, validating near-field sensing and mmWave high-resolution imaging performance.

\subsection{Data-Secure ISAC Demonstration}
Beyond the DFRC demonstration, it is noteworthy to see the practical realization of secure ISAC systems. Although still largely underexplored,~\cite{xu2024signal} experimentally validated data-secure ISAC using the SEFDM framework. The prototype, implemented on a USRP platform with six TX antennas, evaluated data security by comparing the error vector magnitude (EVM) of the legitimate CU and Eve placed only 4~cm apart. Notably, while the conventional OFDM signal fails to ensure data confidentiality as Eve’s EVM remains comparable to that of the CU, the SEFDM signal, known only to the legitimate user, significantly degrades Eve’s demodulation accuracy. The results confirm that SEFDM integrated with joint precoding effectively secures ISAC transmission even under strong spatial correlation between CU and Eve. For detailed experimental configurations, readers are referred to~\cite{xu2024signal}.

\section{Conclusion and Future Outlook} \label{Sec::Conclusion}
MIMO transceiver technologies form the foundation of ISAC, providing time, frequency, and spatial degrees of freedom that enable the dual functionality of radar and communication. This article examined the evolution of MIMO transceiver designs for ISAC, establishing the fundamental frameworks for dual-functional radar–communication systems. It also highlighted new physical-layer vulnerabilities introduced by integration and summarized recent MIMO transceiver solutions that jointly address sensing, communication, and security.  

As ISAC moves toward sustainable and large-scale deployment in perceptive mobile networks, new challenges arise in efficient implementation and unprecedented security threats. Addressing these challenges requires continued efforts to explore uncharted problems and exploit emerging opportunities. The following outlook outlines promising directions for future ISAC development.

\subsubsection{Hardware-Efficient MIMO-ISAC Transceivers} 
Next-generation wireless networks demand both high data rates and high-resolution sensing, requiring extremely large antenna arrays and wide bandwidths. However, most existing MIMO transceivers rely on fully digital or fully-connected hybrid architectures, which become impractical for large apertures due to excessive power consumption and complex hardware design. Furthermore, wideband or multi-band sensing with communication signals necessitates high sampling rates in analog-to-digital converters (ADCs), further increasing the power budget. The use of sparse arrays, while extensively explored in radar systems, remains largely untapped in ISAC transmission, missing out on potentially significant hardware gains. Without advances in signal processing and hardware-efficient design, large-scale ISAC deployment will face severe sustainability challenges.

\subsubsection{MIMO Transceivers for Near-Field ISAC}
Extremely large antenna arrays and wide bandwidths developments have two key implications in ISAC transceiver designs. First, the problem becomes near-field, introducing distance-dependent effects that enable joint range and direction-of-arrival (DoA) estimation and facilitate beamfocusing for improved interference mitigation. Operating in the near-field thus opens opportunities for beamfocusing and distance-aware localization. Second, while most existing ISAC studies focus on narrowband signals, future systems will adopt wideband transmission to enhance communication capacity and range resolution. These changes necessitate, in addition to high sampling rate, a redesign of ISAC frameworks to address model mismatches between traditional far-field, narrowband assumptions and realistic near-field, wideband EM environments.

\subsubsection{Theoretical Framework on ISAC Secrecy}
Despite existing secure ISAC designs, a unified theoretical framework for ISAC secrecy remains largely unexplored. In particular, sensing secrecy lacks rigorous foundations for characterizing target detection and parameter estimation under adversarial attacks. Estimation- and information-theoretic analyses are needed to quantify achievable secrecy in radar sensing and to develop unified models that define new secrecy metrics for joint sensing and communication. Such frameworks would guide secure MIMO transceiver design, provide performance benchmarks, and establish fundamental limits on secure ISAC operation.

\subsubsection{Secure Network-Level ISAC Design}
Beyond link-level transceiver design, network-level ISAC offers new opportunities for secure operation through coordinated and cooperative sensing and communication. Developing secure coordination protocols, distributed transmission schemes, and multi-node information fusion strategies will be essential to mitigate emerging security risks in large-scale ISAC networks. Scalable network-level security frameworks should jointly address communication secrecy and sensing privacy while accounting for practical constraints such as limited backhaul capacity.

\subsubsection{ISAC for Artificial Intelligence}
While artificial intelligence (AI) enhances ISAC transceiver optimization, ISAC in turn provides rich sensory data that can empower AI models for perception, localization, and resource allocation in wireless networks. Realizing this synergy requires trustworthy and privacy-preserving learning frameworks, where AI models are trained on ISAC data without compromising communication security or sensing information leakage. Future MIMO-ISAC transceivers should therefore be designed to natively support secure data acquisition, distributed learning, and semantic information extraction for intelligent and autonomous network operation.

\ifCLASSOPTIONcaptionsoff
  \newpage
\fi



%
\bibliographystyle{IEEEtran}
\bibliography{IEEEabrv,references}

%








\end{document}